\documentclass[aps, prd, preprint, preprintnumbers, showpacs, amsfonts, nofootinbib, floatfix]{revtex4}
\usepackage{graphicx}
\usepackage{epsfig}
\usepackage{rotate}
\usepackage{rotating}
\usepackage{multirow}
\usepackage{longtable}
\usepackage[T1]{fontenc}

\def\fo{\hbox{{1}\kern-.25em\hbox{l}}}

\def\beq{\begin{equation}}
\def\eeq{\end{equation}}
\def\eq{\end{equation}}
\def\to{\rightarrow}

\def\bsg{\ifmmode B\to X_s\gamma\else $B\to X_s\gamma$\fi}
\def\bsll{\ifmmode B\to X_s\ell^+\ell^-\else $B\to X_s\ell^+\ell^-$\fi}
\def\bstt{\ifmmode B\to X_s\tau^+\tau^-\else $B\to X_s\tau^+\tau^-$\fi}
\def\shat{\ifmmode \hat{s}\else $\hat{s}$\fi}

\def\binsize{{\rm bin size}}

\def\s2b{s_{2\beta}}

\def\u1{G_{SM}\otimes U(1)^{\prime}}

\def\ry{{\cal R}_{Y^\prime}}
\def\ryy{{\cal R}_{YY^\prime}}

\newcommand{\newc}{\newcommand}

\newc{\lcal}{\int {\cal L}dt}

\newc{\LSP}{{\widetilde{\chi}^0_1}}
\newc{\niki}{{\widetilde{\chi}^0_2}}
\newc{\nuc}{{\widetilde{\chi}^0_3}}
\newc{\ndort}{{\widetilde{\chi}^0_4}}
\newc{\nbes}{{\widetilde{\chi}^0_5}}
\newc{\nalti}{{\widetilde{\chi}^0_6}}
\newc{\mnbir}{{M_{\widetilde{\chi}^0_1}}}
\newc{\mniki}{{M_{\widetilde{\chi}^0_2}}}
\newc{\mnuc}{{M_{\widetilde{\chi}^0_3}}}
\newc{\mndort}{{M_{\widetilde{\chi}^0_4}}}
\newc{\mnbes}{{M_{\widetilde{\chi}^0_5}}}
\newc{\mnalti}{{M_{\widetilde{\chi}^0_6}}}
\newc{\stauR}{{\widetilde{\tau}_R}}
\newc{\stau}{{\widetilde{\tau}_1}}
\newc{\mstop}{m_{\widetilde{t}}}
\newc{\mHpm}{m_{H^\pm}}
\newc{\simgt}{\lower.7ex\hbox{$\;\stackrel{\textstyle>}{\sim}\;$}}
\newc{\simlt}{\lower.7ex\hbox{$\;\stackrel{\textstyle<}{\sim}\;$}}
\newc{\ie}{{\it i.e.}}
\newc{\etal}{{\it et al.}}
\newc{\eg}{{\it e.g.}}
\newc{\kev}{\hbox{\rm\,keV}}
\newc{\mev}{\hbox{\rm\,MeV}}
\newc{\gev}{\hbox{\rm\,GeV}}
\newc{\tev}{\hbox{\rm\,TeV}}
\newc{\xpb}{\hbox{\rm\, pb}}
\newc{\xfb}{\hbox{\rm\, fb}}

\newc{\mtop}{m_t}
\newc{\mbot}{m_b}
\newc{\mz}{m_Z}
\newc{\mw}{M_W}
\newc{\alphasmz}{\alpha_s(m_Z^2)}
\newc{\swsq}{\sin^2\theta_W}
\newc{\tw}{\tan\theta_W}
\newc{\cw}{\cos\theta_W}
\newc{\sw}{\sin\theta_W}
\newc{\BR}{\hbox{\rm BR}}
\newc{\zbb}{Z\to b\bar}
\newc{\Gb}{\Gamma (Z\to b\bar b)}
\newc{\Gh}{\Gamma (Z\to \hbox{\rm hadrons})}
\newc{\rbsm}{R_b^\hbox{\rm sm}}
\newc{\rbsusy}{R_b^\hbox{\rm susy}}
\newc{\drb}{\delta R_b}

\newc{\sgn}{\mbox{sgn}}
\newc{\tbeta}{\tan\beta}
\newc{\uL}{{\widetilde{u}_L}}
\newc{\uR}{{\widetilde{u}_R}}
\newc{\cL}{{\widetilde{c}_L}}
\newc{\cR}{{\widetilde{c}_R}}
\newc{\tL}{{\widetilde{t}_L}}
\newc{\tR}{{\widetilde{t}_R}}
\newc{\dL}{{\widetilde{d}_L}}
\newc{\dR}{{\widetilde{d}_R}}
\newc{\sL}{{\widetilde{s}_L}}
\newc{\sR}{{\widetilde{s}_R}}
\newc{\bL}{{\widetilde{b}_L}}
\newc{\bR}{{\widetilde{b}_R}}
\newc{\eL}{{\widetilde{e}_L}}
\newc{\eR}{{\widetilde{e}_R}}
\newc{\mhp}{m_{H^\pm}}
\newc{\mhalf}{m_{1/2}}
\newc{\emt}{{e/\mu /\tau}}

\newc{\lR}{\widetilde{l}_R}
\newc{\lL}{\widetilde{l}_L}
\newc{\nL}{\widetilde{\nu}_L}
\newc{\naa}{\widetilde{\chi}^0_1}
\newc{\nbb}{\widetilde{\chi}^0_2}
\newc{\ncc}{\widetilde{\chi}^0_3}
\newc{\ndd}{\widetilde{\chi}^0_4}
\newc{\nee}{\widetilde{\chi}^0_5}
\newc{\nff}{\widetilde{\chi}^0_6}
\newc{\caa}{\widetilde{\chi}^{\pm}_1}
\newc{\cbb}{\widetilde{\chi}^{\pm}_2}
\newc{\phit}{\phi_t}
\newc{\phib}{\phi_b}
\newc{\phiew}{\phi_{ew}}
\newc{\htz}{h^0_t}
\newc{\hbz}{h^0_b}
\newc{\hewz}{h^0_{ew}}
\newc{\hsmz}{h^0_{sm}}
\newc{\huz}{h^0_u}
\newc{\hsusyz}{h^0_{susy}}

\def\slashchar#1{\setbox0=\hbox{$#1$}           
   \dimen0=\wd0                                 
   \setbox1=\hbox{/} \dimen1=\wd1               
   \ifdim\dimen0>\dimen1                        
      \rlap{\hbox to \dimen0{\hfil/\hfil}}      
      #1                                        
   \else                                        
      \rlap{\hbox to \dimen1{\hfil$#1$\hfil}}   
      /                                         
   \fi}                                         %
\catcode`@=11
\long\def\@caption#1[#2]#3{\par\addcontentsline{\csname
  ext@#1\endcsname}{#1}{\protect\numberline{\csname
  the#1\endcsname}{\ignorespaces #2}}\begingroup
    \small
    \@parboxrestore
    \@makecaption{\csname fnum@#1\endcsname}{\ignorespaces #3}\par
  \endgroup}
\catcode`@=12

\def\mfig#1#2#3#4{
 \begin{figure}[#1]
 \centering
 \epsfysize=7.5in
 \hspace*{-0.6in}
 \epsffile{#3}
 \caption{#4}
 \label{#2}
 \end{figure}}

\def\sfig#1#2#3#4{
 \begin{figure}[#1]
 \centering
 \epsfysize=2.8in
 \hspace*{0in}
 \epsffile{#3}
 \caption{#4}
 \label{#2}
 \end{figure}}

\begin{document}

\preprint{ CUMQ/HEP 152,  DESY 09-024,  IZTECH-P/0901}

%
\vspace{1.2cm}
%
%
%
%
%
%
%
%
%

\title{\Large  Search for Gauge Extensions of the MSSM at the LHC}
\author{Ahmed Ali$^{a}$}
\email[]{ahmed.ali@desy.de}
\author{Durmu\c{s} A. Demir$^{a,b}$}
\email[]{demir@physics.iztech.edu.tr}
\author{Mariana Frank$^c$}
\email[]{mfrank@alcor.concordia.ca}
\author{Ismail Turan$^c$}
\email[]{ituran@physics.concordia.ca}
\affiliation{$^a$Deutsches Elektronen - Synchrotron DESY,
Notkestr.~85, D-22603 Hamburg, Germany.}
\affiliation{$^b$Department of
Physics, Izmir Institute of Technology, IZTECH, TR35430 Izmir,
Turkey,}
\affiliation{$^c$Department of Physics, Concordia
University, 7141 Sherbrooke St.
 West, Montreal, Quebec, Canada H4B 1R6.}


\begin{abstract}
The extensions of the minimal supersymmetric model (MSSM),
driving mainly from the need to solve the $\mu$ problem,
involve novel matter species and gauge groups. These extended
MSSM models can be searched for
at the LHC via the effects of the gauge and Higgs bosons or
their fermionic partners. Traditionally, the focus has been on
the study of the {\it extra forces} induced by the new gauge and Higgs bosons
present in such models.
An alternative way of studying such effects is through
the superpartners of matter species and the gauge forces. We thus
consider a $U(1)^\prime$ gauge extension of the MSSM, and
perform an extensive study of the signatures of the model through
the production and decays of the scalar quarks and gluino,
which are expected to be produced copiously at the LHC. After a
detailed study of the distinctive features of such models with
regard to the signatures at the LHC, we carry out a
detailed Monte Carlo analysis of the signals from the
process $pp \to n\, leptons + m\, jets + \slash\!\!\!\! \slashchar{E}_T$,
and compare the resulting distributions with those
predicted by the MSSM. Our results show that the searches for the extra
gauge interactions in the supersymmetric framework can
proceed not only through the forces mediated by the gauge and Higgs
bosons but also through the superpartner forces mediated by the gauge
and Higgs fermions. Analysis of the events
induced by the squark/gluino decays presented here
is complementary to the direct $Z^{\prime}$ searches at the LHC.
\end{abstract}

\pacs{\vspace*{-0.9cm}12.60.Cn, 12.60.Jv,14.80.Ly}
\vspace*{-0.9cm}
\maketitle

\tableofcontents

\section{Introduction and Motivation}
Any anticipated model of `new physics', which must obligatorily
rehabilitate the unnatural ultraviolet sensitivity of the standard
model (SM), generically involves new matter species and
interactions beyond the SM. These non-SM
features, if discernable in the ${\rm TeV}$ domain, will be probed by
experiments at the LHC. The search for the non-SM gauge interactions
is of particular importance since non-SM gauge forces at the weak
scale can give important hints about the symmetries of  Nature
at short distances. The search can be carried out by measuring the
anomalies in the rates of scattering processes that involve solely the
SM particles. For instance, $2\rightarrow 2$ scatterings can
receive contributions from the exchanges of the extra gauge bosons
$Z^{\prime}$ or $W^{\prime}$, or extra Higgs bosons, and their
effects can be disentangled by measuring the deviation of the
scattering rate from its SM expectation. However, the effects of the
non-SM gauge interactions are not limited to such processes since
they necessarily participate in interactions of the non-SM particles,
too. This feature extends the search procedure for extra gauge forces
into non-SM particle sector, and can prove useful in establishing
the inner consistency of the model of `new physics'.

The search strategies for, and the signatures of, the extra gauge interactions
depend crucially on the structure of the model of `new physics'. Indeed,
possible selection rules, and correlations among observables
can give rise to distinctive signatures for certain scattering processes.
These observations can be made explicit by considering a specific model of
`new physics'. To this end, ${\rm TeV}$--scale gravity, made possible
by large extra dimensions, and ${\rm TeV}$--scale softly-broken supersymmetric
theories stand up as two main avenues for constructing realistic models.
Supersymmetric theories offer a viable framework for elucidating
these observations, as in these theories the
entire particle spectrum is paired to have the boson--fermion symmetry, and thus,
quadratic divergences that destabilize the scalar field sector
are naturally avoided. In particular, gauge bosons themselves
are paired with the corresponding gauge fermions, and this feature guarantees
that any scattering process involving the gauge bosons possesses a partner
process proceeding with the gauge fermions (along with the exchange of fermions
and scalar fermions). This implies that the search for extended
gauge structures can be performed via both {\it gauge bosons} and {\it gauge
fermions}, and the correlations between the two can reveal the underlying supersymmetric
structure. The theories in higher dimensions, unless endowed with supersymmetry,
do not possess this partnership structure, that is, their forces (induced by the
extended gauge sector or the Kaluza-Klein modes of the known gauge
fields in the bulk) do not acquire contributions from any partner.

In this paper we perform a phenomenological study of the extra
gauge interactions in the context of an extended low-energy softly-broken
supersymmetric model. The minimal supersymmetric model (MSSM)
is based on the SM gauge group $G_{SM}=SU(3)_c\otimes
SU(2)_L\otimes U(1)_Y$. In general, provided that the existing
bounds are respected, this gauge structure can be extended in
various ways motivated by high-energy (SUSY GUTs or strings) or
low-energy ( the $\mu$ problem of the minimal supersymmetry,
the neutrino masses or the cold dark matter) considerations.
The simplest option would be to consider an extra Abelian
symmetry orthogonal to $G_{SM}$ so that the gauge structure at the
${\rm TeV}$ scale takes the form $G_{SM}\otimes U(1)^{\prime}$.
For extending the gauge structure there are other possibilities
as well. For example, one can consider a left-right symmetric
setup $SU(3)_c\otimes SU(2)_L\otimes SU(2)_R\otimes U(1)_{B-L}$
or a more general embedding $SU(3)_c\otimes SU(3)_L\otimes
U(1)^{\prime}$. Each gauge structure comes with its associated
(neutral and charged) gauge bosons and the corresponding gauginos,
and their searches will help
establish the underlying supersymmetric structure.

In this work we attempt to answer the following
question: {\it What are the basic collider signatures of an
extended gauge structure within a supersymmetric framework?}
The answer involves both the {\it forces} mediated by the gauge
bosons and the {\it superpartner forces} mediated by the gauge
fermions. We will answer this question within the following
framework:
\begin{itemize}
\item We will consider $G_{SM}\otimes U(1)^{\prime}$ gauge
    group for definiteness (more general gauge structures
    can be analyzed along the lines of reasoning employed
    for $U(1)^{\prime}$).

\item We will analyze the production and decay processes
    pertaining to the LHC (processes at other colliders like
    Tevatron or the ILC can be analyzed accordingly).
\end{itemize}
This setup might seem too specific to investigate at first sight; however, it will be seen
at the end of this analysis, that the results obtained here are sufficiently generic.

This paper is organized as follows: In Sec.II, we give a description of
the features of the $G_{SM}\otimes U(1)^{\prime}$ model. As several model presentations exist in the literature, we review the features essential for our analysis, relegating the rest to the Appendix for completeness. In Sec. III, we provide a
general discussion of the LHC processes characteristic of the
$G_{SM}\otimes U(1)^{\prime}$ model. In Sec. IV, we analyze these
scattering processes via Monte Carlo
simulations. We summarize and conclude in Sec. V. The Lagrangian of the
$G_{SM}\otimes U(1)^{\prime}$ model is detailed in Appendix A - Appendix D.
For the
remainder of this work, we will refer to our model simply as
the $ U(1)^{\prime}$ model.

\section{The $U(1)^{\prime}$ Model}
There are various reasons for extending the MSSM by an
additional $U(1)$ group. From the point of view of high
energies, an extra $U(1)$ symmetry broken at  the ${\rm TeV}$ scale
frequently arises in grand unified theories and strings
\cite{gut-string}. Seen from the low
energy point of view, introduction of an extra $U(1)$ is motivated by the
need to solve the $\mu$ problem \cite{muprob} of the MSSM.
Indeed, if the $U(1)_{Y^{\prime}}$ charges of the MSSM Higgs
doublets do not sum up to zero it then becomes possible to
promote the $\mu$ parameter to a SM-singlet chiral superfield
$\widehat{S}$ charged solely under the $U(1)_{Y^{\prime}}$
group. This setup, as encoded in the superpotential
\begin{eqnarray}
\label{superpot} \widehat{W} = h_s \widehat{S} \widehat{H}_u \cdot
\widehat{H}_d + h_u \widehat{Q}\cdot \widehat{H}_u \widehat{U} +
h_d \widehat{Q} \cdot \widehat{H}_d \widehat{D} + h_e \widehat{L}
\cdot \widehat{H}_d \widehat{E}\,,
\end{eqnarray}
then induces an effective $\mu$ parameter, $\mu_{eff} = h_s
\langle S \rangle$, below the $U(1)_{Y^{\prime}}$ breaking scale.
The extra chiral field $\widehat{S}$ extends $(i)$ the MSSM Higgs sector via the
additional Higgs field $S$, and $(ii)$ the MSSM neutralino sector via
the additional neutral fermion $\widetilde{S}$ \cite{cvetic}.

The other source of deviation from the MSSM stems from the
presence of the extra gauge boson and its superpartner. Indeed,
the kinetic terms of the gauge superfields in electroweak
sector are given by \cite{kolda,zerwas}
\begin{eqnarray}
\label{gauge} {\cal{L}}_{gauge} =  \frac{1}{32} \left[
\widehat{W}^{a} \widehat{W}^{a} +  \widehat{W}_Y \widehat{W}_Y +
\widehat{W}_ {Y^{\prime}} \widehat{W}_{Y^{\prime}} + 2 \sin\chi
\widehat{W}_Y \widehat{W}_{Y^{\prime}}\right]_{F}\,,
\end{eqnarray}
where $\widehat{W}^{a}$, $\widehat{W}_Y$ and
$\widehat{W}_{Y^{\prime}}$ are, respectively, the gauge
superfields of $SU(2)_L$, $U(1)_Y$ and $U(1)_{Y^{\prime}}$
groups with the gauge couplings $g_2$, $g_Y$ and $g_{Y^{\prime}}$.
The last term in (\ref{gauge}) accounts for the kinetic mixing
(with the angle $\chi$) between the $U(1)_{Y}$ and the $U(1)_{Y^{\prime}}$
gauge superfields.
Eliminating the kinetic mixing in (\ref{gauge}), while  maintaining the hypercharge sector as in the MSSM,
changes the $U(1)_{Y^{\prime}}$ invariance to a new one
$U(1)_{Q^{\prime}}$ with the charge
\begin{eqnarray}
\label{Qfprime}
Q^{\prime}_{f} = \frac{1}{g_{Y^{\prime}} \cos \chi} \left(
 g_{Y^{\prime}} Y^{\prime}_{f} - {g_{Y}} Y_{f} \sin \chi
 \right)\,,
\end{eqnarray}
from which it follows that even if $f$ is neutral under
$U(1)_{Y^{\prime}}$ it still possesses a non-vanishing charge
$Q^{\prime}_{f}$ proportional to its hypercharge times $\tan
\chi$. As our analysis is concerned with the superpartner fermion forces, we present that sector next. In Appendix A we describe the  particle spectrum and the
Lagrangian and analyze the gauge and Higgs boson sectors.

\subsection{Gauge and Higgs Fermions}
The $U(1)^{\prime}$ model possesses no new charged Higgsinos
and gauginos. On the other hand, in the neutral sector it
possesses two new fermion fields: the $U(1)^{\prime}$ gauge
fermion $\widetilde{Z}^{\prime}$ and the singlino
$\widetilde{S}$. In total, there
are 6 neutralino states $\widetilde{\chi}_i^0$ ($i=1,\dots,6$)
\cite{zerwas,langacker-neutral}:
\begin{eqnarray}
\label{neutralino-def}
\widetilde{\chi}_i^0 = \sum_{a} N^0_{i a} \widetilde{G}_a\,,
\end{eqnarray}
where the mixing matrix $N^0_{i a}$ connects the gauge-basis neutral fermion states
$\widetilde{G}_a \in \Big\{\widetilde{B},$ $\widetilde{W}^3,$ $\widetilde{H}^0_d,$
$\widetilde{H}^0_u,$ $\widetilde{S},$ $\widetilde{Z}^{\prime}  \Big\}$ to the physical
neutralinos $\widetilde{\chi}_i^0$. The neutralino masses $M_{\widetilde{\chi}_i^0}$
and the mixing matrix $N^0_{i a}$
are determined via the diagonalization condition
$N^0 {\cal{M}} N^{0\ T} = \mbox{Diag}$ $\Big\{M_{\widetilde{\chi}_1^0},$ $\dots,$ $M_{\widetilde{\chi}_6^0}\Big\}$
for the neutral fermion mass matrix
\begin{eqnarray}
\label{mneut}
{\cal{M}}=\left(\begin{array}{cccccc} M_{\widetilde{Y}} & 0 & - M_{\widetilde{Y}\, \widetilde{H}_d} & M_{\widetilde{Y}\, \widetilde{H}_u} & 0 &
M_{\widetilde{Y}\, \widetilde{Z}^{\prime}}\\
0 & M_{\widetilde{W}} & M_{\widetilde{W}\, \widetilde{H}_d} & - M_{\widetilde{W}\, \widetilde{H}_u} & 0 & 0\\
- M_{\widetilde{Y}\, \widetilde{H}_d} & M_{\widetilde{W}\, \widetilde{H}_d} & 0 & - \mu  & - \mu_{H_u} & \mu^{\prime}_{H_d}\\
M_{\widetilde{Y}\, \widetilde{H}_u} & M_{\widetilde{W}\, \widetilde{H}_d} & - \mu & 0 & - \mu_{H_d} & \mu^{\prime}_{H_u}\\
0 & 0 & - \mu_{H_u} & - \mu_{H_d} & 0 & \mu^{\prime}_S \\
M_{\widetilde{Y}\, \widetilde{Z}^{\prime}} & 0 & \mu^{\prime}_{H_d} & \mu^{\prime}_{H_u} & \mu^{\prime}_S &
M_{\widetilde{Z}^{\prime}}\end{array}\right)\,,
\end{eqnarray}
where certain entries are generated by the soft-breaking sector while others follow from the $SU(3)_c
\otimes SU(2)_L\otimes U(1)_Y\otimes U(1)_{Q^{\prime}}$ breaking. The $U(1)_Y$ gaugino mass
$M_{\widetilde{Y}}$, the $SU(2)_L$ gaugino mass $M_{\widetilde{W}}$, and the $U(1)_{Q^{\prime}}$ gaugino mass
\begin{eqnarray}
M_{\widetilde{Z}^{\prime}} &=& \frac{M_{\widetilde{Y}^{\prime}}}{\cos^2\chi}   - 2 \frac{\tan\chi}{\cos\chi} M_{\widetilde{Y} \widetilde{Y}^{\prime}}
+ M_{\widetilde{Y}} \tan^2\chi\,,
\end{eqnarray}
as well as the mixing mass parameter between $U(1)_Y$ and $U(1)_{Q^{\prime}}$ gauginos
\begin{eqnarray}
M_{\widetilde{Y} \widetilde{Z}^{\prime}} &=& \frac{M_{\widetilde{Y} \widetilde{Y}^{\prime}}}{\cos\chi} - M_{\widetilde{Y}} \tan\chi\,,
\end{eqnarray}
all follow from the soft-breaking sector (See Appendix A). Through the mixing of the gauge bosons,
$M_{\widetilde{Z}^{\prime}}$ and $M_{\widetilde{Y}
\widetilde{Z}^{\prime}}$ exhibit an explicit dependence on the
masses of the $U(1)_Y$ and $U(1)_{Y^{\prime}}$ gauginos, and their
mass mixing.  $M_{\widetilde{Y}
\widetilde{Y}^{\prime}}$ is the soft-breaking mass that mixes the
$U(1)_Y$ and $U(1)_{Y^{\prime}}$ gauginos.

The remaining entries in (\ref{mneut}) are generated by the soft-breaking masses in the Higgs sector via
the $SU(3)_c\otimes SU(2)_L\otimes U(1)_Y\otimes U(1)_{Q^{\prime}}$ breaking. Their explicit expressions
are given by
\begin{eqnarray}
M_{\widetilde{Y}\, \widetilde{H}_d} &=& M_Z \sin\theta_W \cos\beta\,,\;
M_{\widetilde{Y}\, \widetilde{H}_u} = M_Z \sin\theta_W \sin\beta\,,\nonumber\\
M_{\widetilde{W}\, \widetilde{H}_d} &=& M_Z \cos\theta_W \cos\beta\,,\;
M_{\widetilde{W}\, \widetilde{H}_u} = M_Z \cos\theta_W \sin\beta\,,\nonumber\\
\mu &=& h_s \frac{v_s}{\sqrt{2}}\,,\; \mu_{H_d} = h_s \frac{v_d}{\sqrt{2}}\,,\; \mu_{H_u} = h_s \frac{v_u}{\sqrt{2}}\,,\nonumber\\
\mu^{\prime}_{H_d} &=& g_{Y^{\prime}} Q_{H_d}^{\prime} v_d\,,\; \mu^{\prime}_{H_u} = g_{Y^{\prime}} Q_{H_u}^{\prime} v_u\,,\;
\mu^{\prime}_{S} = g_{Y^{\prime}} Q_{S}^{\prime} v_s\,,
\end{eqnarray}
out of which only $\mu$ and $\mu^{\prime}_{S}$ involve $v_s$. These entries
scale with $M_{Z^{\prime}}$, and thus, the heavier the $Z^{\prime}$ boson, the larger the
$\widetilde{S}$--$\widetilde{Z}^{\prime}$ mixing.

The lightest neutralino $\widetilde{\chi}_1^0$ is absolutely stable, and therefore,
it is a natural candidate for cold dark matter in the universe. The singlino $\widetilde{S}$
does not couple to fermions. The other two Higgsinos $\widetilde{H}_{u,d}^{0}$ couple
very weakly to fermions, except for the top quark (and to the bottom quark and the tau lepton to a
lesser extent). Consequently, the scattering processes involving (s)fermions of the first
and second generations are expected to be dominantly sensitive to the gaugino components of
neutralinos.

\section{The LHC Signatures of the $U(1)^{\prime}$ Model}
The CMS and the ATLAS experiments at the LHC, a proton--proton
collider with center-of-mass energy $\sqrt{s}=14\ {\rm TeV}$,
will be searching for physics beyond the SM.
The $U(1)^{\prime}$ model would show up in experiments at the
LHC via the U(1)$^{\prime}$ gauge boson and gauge fermion as well
as the singlet chiral field in its superpotential. These
fermionic and bosonic fields give rise to characteristically
distinct yet not necessarily independent signatures at the LHC
energies. These effects are discussed and contrasted in this section with the ones in the MSSM
by employing the {\it gauge basis} instead of the physical
(mass-eigenstate) basis, for simplicity and clarity of the
discussions.

We first briefly summarize those effects which are
genuine to the $U(1)^{\prime}$ model by considering its bosonic
sector only. These effects have been studied in detail in the
literature \cite{langacker-review}; bounds on various model
parameters will be tightened as more and more experimental data
accumulate. In this work we will not reanalyze these effects,
but  will take into account the  implied constraints.

The bosonic sector of the $U(1)^{\prime}$ model shows up
through the $Z^{\prime}$ gauge boson and the singlet Higgs boson $S$.
The cleanest and the most direct signal of a $Z^{\prime}$ gauge boson, if accessible at the
LHC, will be  a new resonance, centered at $M_{\ell\ell} =
M_{Z^{\prime}}$, in the dilepton spectrum ($\ell = e$ or $\mu$ unless otherwise stated)
\cite{collid,langacker-kang}
\begin{eqnarray}
p\, p \rightarrow Z^{\prime} + X \rightarrow \ell^+ \ell^- + X\,,
\end{eqnarray}
This proceeds through $q\, \overline{q}$ annihilation followed by an $s$-channel $Z^{\prime}$
exchange. The existing bounds from LEP \cite{LEP} and
Tevatron \cite{cdf} require $Z^{\prime}$ to weigh near a ${\rm TeV}$ or higher, depending on
the details of the model which determine the $Z^{\prime}$ couplings to the quarks and leptons
\cite{collid}.

The extra Higgs boson,
$H^{\prime}$ weighs close
to $M_{Z^{\prime}}$ and it is typically the heaviest Higgs
boson in the spectrum \cite{everett,higgs,higgsp}. The $S$
field (which gives rise to the physical $H^{\prime}$ boson
after diagonalization of the Higgs mass-squared matrix) is
produced via
\begin{eqnarray}
\label{Sproduc}
p\, p \rightarrow Z^{\prime} + X \rightarrow  S\, S^{\star} + X\,,
\end{eqnarray}
whereupon the $S$ field subsequently decays into lighter fields in the model:
\begin{eqnarray}
S \rightarrow H_{u}^{0} H_d^{0}\;,\; H_u^{+} H_d^{-}\;,\; H_{d}^{0} \widetilde{t}_L \widetilde{t}^{\star}_R
\;,\; H_{d}^{+} \widetilde{b}_L \widetilde{t}^{\star}_R\;,\; H_{u}^{0} \widetilde{b}_L \widetilde{b}^{\star}_R
\;,\;H_{u}^{-} \widetilde{t}_L \widetilde{b}^{\star}_R\;,\; H_{u}^{0} \widetilde{\ell}_L \widetilde{\ell}^{\star}_R
\;,\;H_{u}^{-} \widetilde{\nu}_L \widetilde{\ell}^{\star}_R\,,
\end{eqnarray}
The
phenomenological implications of these decays have already been
analyzed in \cite{higgs,higgsp}.

There are also effects at the LHC which would involve both the $Z^{\prime}$ and the $S$ fields
in an interacting fashion. One such process is the Higgs production via the Bjorken mechanism
\begin{eqnarray}
p\, p \rightarrow \left(Z, Z^{\prime}\right) + X \rightarrow \left(Z, Z^{\prime}\right) + \mbox{CP-even Higgs bosons} + X\,,
\end{eqnarray}
which differs from its MSSM counterpart by the presence of both the $Z^{\prime}$ and
the $S$ contributions \cite{higgs}. It is
because of these effects, in conjunction with (\ref{Sproduc}), that the Higgs boson discovery limits
 can be modified significantly in the $U(1)^{\prime}$ model.

\subsection{$U(1)^{\prime}$ Effects Through Gauge and Higgs Fermions}
The non-MSSM neutral fermions $\widetilde{S}$ and
$\widetilde{Z}^{\prime}$, which mediate the superpartner
forces, are part of the neutralino sector
(\ref{neutralino-def}), and thus, extraction of the $U(1)^{\prime}$
effects from the collider data can also be accomplished via those
processes involving the neutralinos. At hadron colliders, such
as the LHC, neutralinos ($\widetilde{\chi}_i^{0}$, $i=1,\dots,
6$) can be produced directly in pairs or in association with the
charginos ($\widetilde{\chi}_r^+$, $r=1,2$), gluinos
$\widetilde{g}$ or squarks $\widetilde{q}$ \cite{pair-produce}
\begin{eqnarray}
\label{direct-production}
p\ p \rightarrow \widetilde{\chi}_i^0 \widetilde{\chi}_j^0\;,\;\widetilde{\chi}_i^0 \widetilde{\chi}_r^+
\;,\; \widetilde{\chi}_i^0 \widetilde{g}\;,\;\widetilde{\chi}_i^0 \widetilde{q}\,,
\end{eqnarray}
via the $s$-channel gauge boson exchange (the first two
channels above) or the $t$-channel squark exchange (all the channels).
The trilinear gauge boson couplings are completely
antisymmetric for the $SU(2)_L$ group and do not exist
for the Abelian ones, and hence, $Z$ and $Z^{\prime}$ gauge
bosons do not couple to  the neutral gauginos
$\widetilde{W}^3$, $\widetilde{B}$ and
$\widetilde{Z}^{\prime}$. Instead, they couple  only to the
neutral Higgsinos $\widetilde{H}_{u,d}^{0}$ contributing to the
$\widetilde{\chi}_i^0 \widetilde{\chi}_j^0$ production. On the
other hand, the $W^{\pm}$ boson couples to $\widetilde{W}^3
\widetilde{W}^{\pm}$ as well as  to $\widetilde{H}_{u,d}^{0}
\widetilde{H}_{u,d}^{\pm}$, and thus, the $s$-channel $W^{\pm}$
exchange gives rise to $\widetilde{\chi}_i^0
\widetilde{\chi}_r^+$ final states  containing both the
gauginos and the Higgsinos. In addition, the  $Z^{\prime}$ exchange
(dominantly $Z_2$ exchange for small $Z-Z^{\prime}$ mixing)
causes pair-production of the singlino
$\widetilde{S}$. In fact, this channel is the only mode
which leads to $\widetilde{S}$ production since the $t$-channel
squark exchange produces only the gaugino components of the
neutral fermions. In consequence, while the $s$-channel gauge
boson exchanges generate the $\widetilde{H}_{u,d}^{0}$ and the
$\widetilde{S}$ components of neutralinos, the $t$-channel
squark exchange gives rise to the $\widetilde{W}^3$,
$\widetilde{B}$ as well as the $\widetilde{Z}^{\prime}$ components.
In this sense, the two amplitudes exhibit complementarity in
producing the neutral Higgsinos and the gauginos. Besides,
the neutralino mass matrix (\ref{mneut})
enables the production of all the neutralino states
$\widetilde{\chi}^0_i$, no matter which gaugino or Higgsino
component is actually produced at the interaction vertex.

The existing bounds on the $Z^{\prime}$ boson mass
\cite{langacker-review} do not necessarily imply a suppression of
the pair-production processes at the LHC energies, as this cross section
may get enhanced due to the resonance effects for the
center of mass energy
near the $Z^{\prime}$ mass. This implies that the singlino pair
production could be as strong as that involving the other two Higginos
$\widetilde{H}_{u,d}^0$.

Once produced, all neutralinos decay into isolated leptons,
hard jets (initiated by quarks or gluons), photons and the
lightest neutralino $\widetilde{\chi}_1^{0}$ (which appears as a
momentum imbalance or the missing transverse energy $\slashchar{E}_T$ in all
the SUSY processes since it is the lightest supersymmetric particle
(LSP), which is stable due to the conserved $R$
parity) via a chain of cascade decays. The decay patterns of
interest, especially those offering clean collider signatures,
are the ones which
yield isolated leptons. In this sense, a typical cascade decay
would look like
\begin{eqnarray}
\label{decay}
\mbox{(heavy ino)} \rightarrow \mbox{(lepton)}\; \mbox{(slepton)}^{\star}
\rightarrow \mbox{(lepton)}\; \mbox{(anti-lepton)}\; \mbox{(light ino)}\,,
\end{eqnarray}
where 'ino' stands for any of the neutral or charged gauginos or Higgsinos
in the model. Every cascade must necessarily end with the 'lightest ino' \ie, the LSP, and therefore,
decay chains of this sort proceed through several intermediate steps depending on
the mass and the couplings of the mother-ino.

It is highly illustrative to analyze these cascade decays in
the Lagrangian basis $\widetilde{G}_a$, and we do so for the remainder of this section.  A precise analysis in the
physical basis $\widetilde{\chi}^0_i$, which takes into account the mixings in
the neutralino mass matrix (\ref{mneut}), will be given in the next section.

The cascade decays (\ref{decay}) are the key processes for
determining the sparticle properties from the decay rates and
topologies at the LHC \cite{abdullin}. In the MSSM they involve
the hypercharge and the isospin gauginos as well as the Higgsinos.
In the $U(1)^{\prime}$ model, with the addition of new neutral fermions
$\widetilde{Z}^{\prime}$ and $\widetilde{S}$, the ino decays
can acquire certain novel features not present in the MSSM. This
point can be exemplified by considering the decay
\begin{eqnarray}
\label{w3decay-MSSM}
\widetilde{W}^{3} \rightarrow \ell^+ \widetilde{\ell}^{\star -} \rightarrow \ell^{+} \ell^{-} \widetilde{B}\,,
\end{eqnarray}
which in the MSSM hardly ever extends further since
$\widetilde{W}^3$ and $\widetilde{W}^{\pm}$ are nearly
mass-degenerate. In fact, the $SU(2)_L$ breaking effects that split
them in mass turn out to be small so that
$\widetilde{\chi}^0_2$ and $\widetilde{\chi}^{\pm}_1$ have
approximately the same mass \cite{abdullin,cvetic}. Hence, in the MSSM the
decay of $\widetilde{W}^3$ dominantly gives a dilepton
signal. In contrast to this, in the $U(1)^{\prime}$ model, if
$\widetilde{Z}^{\prime}$ falls in between $\widetilde{W}^{3}$
and $\widetilde{B}$ in mass, the cascade (\ref{w3decay-MSSM})
proceeds through one more step
\begin{eqnarray}
\label{w3decay-U1p}
\widetilde{W}^{3} \rightarrow \ell^+ \widetilde{\ell}^{\star -} \rightarrow \ell^{+} \ell^{-} \widetilde{Z}^{\prime}
\rightarrow \ell^{+} \ell^{-} {\ell}^{\prime\, +} \widetilde{\ell}^{\prime\, \star -} \rightarrow
\ell^{+} \ell^{-} {\ell}^{\prime\, +} {\ell}^{\prime\, -} \widetilde{B}\,,
\end{eqnarray}
to yield a tetralepton final state. Obviously, this final
state also arises when $\widetilde{Z}^{\prime}$ is heavier than
$\widetilde{W}^{3}$.
Engineered by the $U(1)^{\prime}$ gaugino, this is one
distinctive feature that helps distinguish the
$U(1)^{\prime}$ signatures from those of the MSSM.

Unlike the $U(1)^{\prime}$ gaugino, the singlino $\widetilde{S}$,
 since it does not couple to quarks and leptons
directly, exhibits a completely different decay pattern, in that the Higgs bosons are always involved in the process. One possible
decay channel proceeds with the $U(1)^{\prime}$ gaugino
\begin{eqnarray}
\widetilde{S} \rightarrow S \widetilde{Z}^{\prime}\,,
\end{eqnarray}
where $\widetilde{Z}^{\prime}$ decays into leptons and $\widetilde{B}$ as described above, and
the singlet Higgs $S$ decays into the SM particles via the doublet Higgs fields $H_{u,d}$.
 The other channel proceeds with the Higgsinos in the decay products,
\begin{eqnarray}
\widetilde{S} \rightarrow H^{0}_{u}
 \widetilde{H}^{0}_{d}\;,\; H^{+}_{u} \widetilde{H}^{-}_{d}\,,
\end{eqnarray}
wherein the Higgs bosons and the fermions follow the usual decay chains until the leptons (possibly also quarks)
plus the $\widetilde{B}$ state are reached.

The direct pair-production mechanisms in
(\ref{direct-production}) are not the only means of producing
neutralinos; moreover, they are not necessarily the dominant
ones. Indeed, neutralinos and charginos are
produced in cascade decays of the gluinos, squarks and sleptons. As at the LHC
energies, if accessible kinematically, gluinos and squarks
possess the largest production cross section \cite{produc}
among all the sparticles, neutralinos or charginos
arising from the squark/gluino decays must be much more abundant
than from all other sources, and an analysis of these can give
critical information about the absence/presence of an extra
$U(1)$ group. However, since all the SUSY processes end with a
debris containing $\widetilde{\chi}_1^{0}$, which escapes
detection in the detector, a complete reconstruction of the
masses and couplings of the sparticles is not possible. Therefore,
observability is based on the criterion of having a significant
excess of events of a given topology over a predetermined
background \cite{reconstruct,abdullin}. For
extracting information on a possible $U(1)^{\prime}$ group,
one has to determine the squark/gluino decay channels
pertaining to the $U(1)^{\prime}$ model, and compare the signal
with the MSSM prediction, as will be done explicitly in the next section.

The gluinos, unlike the $SU(2)_L\otimes U(1)_Y\otimes U(1)^{\prime}$ gauginos, can
 be pair-produced
via the gluon exchange in the $s$-channel at the LHC energies  via
\begin{eqnarray}
\label{squark-gluino}
p\ p \rightarrow \widetilde{g} \widetilde{g}\;,\; \widetilde{q} \widetilde{q}
\;,\;\widetilde{g} \widetilde{q}\,,
\end{eqnarray}
through gluon-gluon, gluon-quark and quark-quark  scattering \cite{produc}.
 Following their production, gluinos and squarks decay further.  If the gluino is
heavier than squarks then it decays into a quark and squark
$\widetilde{q}$
\begin{eqnarray}
\widetilde{g} \rightarrow q\ \widetilde{q}\,,
\end{eqnarray}
and subsequently $\widetilde{q}$ initiates a series of cascade
decays yielding a debris containing jets, isolated leptons and
$\widetilde{\chi}^0_1$. On the other hand, if the gluino is
lighter than (some of the) squarks then the squark
$\widetilde{q}$ decays into gluino and quark, and then the
gluino decays into lighter squarks and quarks yielding
eventually a similar debris. Therefore, the essential features of
the model can be extracted by exploring the decay patterns of
the squarks. The decay patterns of sfermions, for
either chirality, are exhibited in Table~\ref{table1}, where
the channels in the MSSM and the $U(1)^{\prime}$ model are displayed in
adjacent columns for comparison. As is clear from this table, the
effect of the $U(1)^{\prime}$ group is in the opening of a new
channel
\begin{eqnarray}
\label{new-decay}
\widetilde{f}_{L,R} \rightarrow f_{L,R}\,
\widetilde{Z}^{\prime}_{R,L}\,,
\end{eqnarray}
by the emission of the $U(1)^{\prime}$ gaugino. This channel
modifies not only the branching ratios of the squarks but also the
decay topologies of certain sparticles expected in the MSSM.

\begin{table}
\begin{tabular*}{0.9\textwidth}{@{\extracolsep{\fill}} ccc}
\hline \hline Sfermion & MSSM & $U(1)^{\prime}$ Model\\\hline\hline
$\widetilde{f}_R$ & $\begin{array}{c} \widetilde{f}_R \rightarrow
f_R \widetilde{B}\\
\widetilde{f}_R \rightarrow f_L \widetilde{H}^0_f\\
\widetilde{f}_R \rightarrow f^{\prime}_L \widetilde{H}^{\pm}_{f}
\end{array}$ & $\begin{array}{cc}\widetilde{f}_R \rightarrow f_R \widetilde{B} & {}\\
\widetilde{f}_R \rightarrow f_L \widetilde{H}^0_f & \bigoplus\; \widetilde{f}_R \rightarrow f_R \widetilde{Z}^{\prime}\\
\widetilde{f}_R \rightarrow f^{\prime}_L \widetilde{H}^{\pm}_{f} &
{} \end{array}$\\ \hline $\widetilde{f}_L$ & $\begin{array}{c}
\widetilde{f}_L \rightarrow f_L\, \widetilde{B}\\
\widetilde{f}_L \rightarrow f_L\, \widetilde{W}^3 \\
\widetilde{f}_L \rightarrow f^{\prime}_L\,\widetilde{W}^{\pm}\\
\widetilde{f}_L \rightarrow f_R \widetilde{H}^0_f \\
\widetilde{f}_L \rightarrow f^{\prime}_R\, \widetilde{H}^{\pm}_{f}
\end{array}$ & $\begin{array}{cc}
\widetilde{f}_L \rightarrow f_L\, \widetilde{B} & {}\\
\widetilde{f}_L \rightarrow f_L\, \widetilde{W}^3 & {} \\
\widetilde{f}_L \rightarrow f^{\prime}_L\, \widetilde{W}^{\pm} & \bigoplus\; \widetilde{f}_L \rightarrow f_L\, \widetilde{Z}^{\prime}\\
\widetilde{f}_L \rightarrow f_R\, \widetilde{H}^0_f\\
\widetilde{f}_L \rightarrow f^{\prime}_R\, \widetilde{H}^{\pm}_{f}
& {}\end{array}$\\ \hline\hline
\end{tabular*}
\caption{\label{table1} The decay channels of the scalar fermions
$\widetilde{f}$ in the MSSM and the $U(1)^{\prime}$ model. The
couplings to Higgsinos $\widetilde{H}^{\pm}_f$ and
$\widetilde{H}^0_f$ ($\equiv \widetilde{H}^0_{u}$ for $f=u$ and
$\equiv \widetilde{H}^0_{d}$ for $f=d, \ell$) are important
only for the fermions in the third generation, in particular, the
top quark. As follows from (\ref{superpot}), the singlino
$\widetilde{S}$ does not couple to fermions directly, and thus,
the $U(1)^{\prime}$ couplings enter via the decays into
$\widetilde{Z}^{\prime}$ only.}
\end{table}

For a clearer exposition of the features added by the squark decays
into $\widetilde{Z}^{\prime}$, we elaborate on
the decay channels listed in Table~\ref{table1}.  The squarks of the first and second
generations possess the following properties: $(i)$ The mass
and gauge eigenstates (especially for the scalar up and down
quarks) are identical due to their exceedingly small Yukawa
couplings, $(ii)$ the flavor and the gauge eigenstates of the scalar up
and down quarks are identical whereas the scalar strange quark
might possesses significant flavor mixing with the scalar bottom
quark, $(iii)$ they do not exhibit any appreciable coupling to the
Higgsinos but  only to the gauginos, and $(iv)$ they turn out to
be the heaviest scalars of approximately the same mass, nearly
mass degenerate with the gluino, in the minimal supergravity
\cite{cvetic}. In the light of these features, these squarks
provide a perfect playground for probing the gaugino sector
(and hence the extended gauge structures) with a conservative
number of SUSY parameters (no direct dependence on the $\mu$
parameter and trilinear couplings, and a weak dependence on
$\tan\beta$ via $D$--term contributions).

In contrast to the squarks in the first and second generations, the
squarks of the third generation exhibit non-negligible
couplings to Higgs bosons and fermions, and hence, all the
decay modes in Table~\ref{table1} become relevant for them.
Besides, they necessarily exhibit sizable left-right mixings
causing mass eigenstate squarks to have significant mass splitting
\cite{everett}. Moreover, at least in the minimal supergravity, the
third generation squarks, especially the stops, turn out to
weigh well below the ones in the first and second generations
thanks to the counter balancing effect of the rise in the
squark mass due to the Yukawa couplings
\cite{cvetic}. Because of these features, the third generation squarks involve a larger
set of SUSY parameters than the first and second generation
ones, and therefore, they enable exploration of various
parameters, like the trilinear couplings and the $\mu$ parameter,
not possible with the first and second generation squarks.
 In this work we will not explore the third generation squarks
any further. They are in principle distinguishable by
their decay products -- the top and bottom quarks can be tagged
at the LHC experiments with good efficiency. While their
exploration would give important information about various
SUSY parameters, and especially, on the Higgs/Higgsino sectors,
 for the purpose of disentangling the imprints of the
extra gauge symmetries in experimental data, the squarks in the
first and second generations would suffice.

As a highlighting
case study, we start with the analysis of the
decay patterns of the first
or the second generation right-handed squark. From Table~\ref{table1} it is clear that,
in the MSSM, a right-handed squark $\widetilde{q}_R$, with no
gauge quantum number other than color and hypercharge,
possesses one single decay channel
\begin{eqnarray}
\label{qrdecay-MSSM}
\widetilde{q}_R \rightarrow q_R\, \widetilde{B}\,,
\end{eqnarray}
which uniquely leads to $1\, jet + 0\,lepton+\slashchar{E}_T$
signal if the bino $\widetilde{B}$ is the LSP. If bino is not the LSP,
then it further decays into $\widetilde{\chi}_1^0$ emitting at
least one dilepton $\ell^+ \ell^-$ \cite{abdullin}. In either
case, the decay mode above has $100 \%$ branching fraction as
there is no other open decay channel for the $\widetilde{q}_R$ in the MSSM.

In contrast to the MSSM decay mode
(\ref{qrdecay-MSSM}), the right-handed squarks
exhibit a completely new decay pattern in the $U(1)^{\prime}$
model. As seen from Table~\ref{table1},
$\widetilde{q}_R$ now decays via two distinct channels
\begin{eqnarray}
\label{qrdecay-U1p} \widetilde{q}_R \rightarrow q_R\,
\widetilde{B}\;,\;\; \widetilde{q}_R \rightarrow q_R\,
\widetilde{Z}^{\prime}\,,
\end{eqnarray}
so that the branching ratio into $\widetilde{B}$ is no longer $100 \%$. A rough estimate gives
\begin{eqnarray}
\label{branch} {\cal{B}}^{
U(1)^{\prime}}\left(\widetilde{q}_R \rightarrow q_R\,
\widetilde{B}\right) \simeq \frac{ g_Y^2 Y_{q_R}^2}{ g_Y^2
Y_{q_R}^2 + g_{Y^{\prime}}^2 Y^{\prime\, 2}_{q_R}} <
{\cal{B}}^{MSSM}\left(\widetilde{q}_R \rightarrow q_R\,
\widetilde{B}\right) = 1\,,
\end{eqnarray}
where, realistically,  gauginos are taken to be light
$m_{\widetilde{B}}, m_{\widetilde{Z}^{\prime}} \ll
m_{\widetilde{q}_R}$, and various mixings encoded in the neutralino
mass matrix (\ref{mneut}) are neglected for simplicity.
This estimate reveals that the gauge fermion
$\widetilde{Z}^{\prime}$ of the
$U(1)^{\prime}$ group  modifies the decay properties of the
right-handed squarks in a way that can be probed by a
measurement of the squark branching ratio.

However, the branching fraction is not the whole story. Indeed,
depending on the nature of the LSP, one can make further observations
which could be of crucial importance for the searches for an extra
$U(1)$ group at the LHC. Below, we elaborate on several
distinct possibilities:
\begin{itemize}
\item {\it Bino LSP:} In this case, in the MSSM, right-handed squarks  with light fermionic partners decay only hadronically as
in (\ref{qrdecay-MSSM}). The resulting  $0\, lepton + 1\, jet +
\slashchar{E}_T$ signal can be unambiguously established at
the LHC \cite{abdullin}.

The situation in the $U(1)^{\prime}$ model is strikingly
different than in the MSSM. Decays into the
$\widetilde{B}$ yield purely hadronic states as in the
MSSM. However, decays into the $\widetilde{Z}^{\prime}$ give
rise to a chain of cascade decays depending on how heavy
$\widetilde{Z}^{\prime}$ is compared to other gauginos.
While the first decay channel in
(\ref{qrdecay-U1p}) still generates a $0\, lepton + 1\, jet
+ \slashchar{E}_T$  signal of relative amount
(\ref{branch}), the second channel in (\ref{qrdecay-U1p})
gives rise to the final states containing at least two
oppositely-charged leptons. One can have dileptons
\begin{eqnarray}
\label{qrdecay-lept1} \widetilde{q}_R \rightarrow q_R\,
\widetilde{Z}^{\prime} \rightarrow q_R\, \ell^+
\widetilde{\ell}^{\star -} \rightarrow q_R\, \ell^+ \ell^-
\widetilde{B}\,,
\end{eqnarray}
or tetraleptons
\begin{eqnarray}
\label{qrdecay-lept2} \widetilde{q}_R \rightarrow q_R\,
\widetilde{Z}^{\prime} \rightarrow q_R\, \ell^+
\widetilde{\ell}^{\star -} \rightarrow q_R\, \ell^+ \ell^-
\widetilde{W}^3 \rightarrow q_R\, \ell^+ \ell^- \ell^{\prime\, +}
\widetilde{\ell^{\prime}}^- \rightarrow q_R\, \ell^+ \ell^-
\ell^{\prime\, +} \ell^{\prime\, -} \widetilde{B}\,,
\end{eqnarray}
in the final state. Sleptons in the intermediate
states couple to gauginos and leptons via the modes listed
in Table~\ref{table1}.

Thus, when the LSP is dominated by bino (which is what
happens in most of the parameter space \cite{langacker-neutral}),
a prime signature of a $U(1)^{\prime}$ extension of the MSSM is the reduction of purely
hadronic events originating from the decays (\ref{qrdecay-MSSM})
and a corresponding enhancement of the leptonic events via the decays
 (\ref{qrdecay-lept1}) and (\ref{qrdecay-lept2}). While the  rates of these decays and  the
depletion in the number of purely hadronic events depend on the masses
and couplings of the intermediate sparticles in the cascades, the
 leptonic final states stemming from the right-handed
squarks should offer sufficiently clean signatures to establish the existence of a $U(1)^{\prime}$
extension at the LHC.

\item {\it Zino-prime LSP:} In this case, mainly the roles of the
$\widetilde{B}$ and $\widetilde{Z}^{\prime}$ are interchanged in
terms of hadronic/leptonic contents of the decay products. In
particular, while the second decay channel in (\ref{qrdecay-U1p})
leads to purely hadronic events, the first one gives rise to the
leptonic final states similar to (\ref{qrdecay-lept1}) and
(\ref{qrdecay-lept2}). In this scenario, an interesting point
is that the squark decays through the $U(1)^{\prime}$ gaugino
lead to non-leptonic $1\, jet + \slashchar{E}_T$ final states.

\item {\it Oblique LSP:} In general, the LSP does not need
    to be overwhelmed by a single gaugino and Higgsino
    component. Indeed, existing bounds on the relic density
    of  dark matter particles can be satisfied with an LSP
    candidate comprised of various neutral fermions.
    While in the $U(1)^{\prime}$ model under
    study, the LSP is dominated by the bino component in
    most of the parameter space \cite{langacker-neutral},
    depending on the dominant compositions of the LSP, a
    given decay mode, as listed in Table~\ref{table1}, may
    or may not exhibit a chain of cascades ending
    preferably with leptons.

\end{itemize}
The above considerations show that the decay patterns of the right-handed squarks in the
first and the second generations would prove to be sensitive probes
of gauge extensions of the MSSM under which right-handed quark
fields are charged.

The decay characteristics of the left-handed squarks differ
from those of the right-handed squarks due to their $SU(2)_L$
quantum number. Indeed, as shown in Table~\ref{table1}, the
left-handed squarks decay not only into the bino but also into
the charged and neutral winos. Therefore, a left-handed squark,
in a bino LSP scenario, can yield a $0\, lepton + 1\, jet +
\slashchar{E}_T$ final state via its decay into $\widetilde{B}$
as in (\ref{qrdecay-MSSM}), as well as the final states with $1\,
jet + \slashchar{E}_T$ plus at least one charged lepton. The
main impact of the decays into $\widetilde{Z}^{\prime}$ depends on the
 $\widetilde{Z}^{\prime}$ mass,  increasing the length of the cascade.

 Nonetheless, even in the left-handed fermion sector,
there are still interesting patterns for which the MSSM and
the $U(1)^{\prime}$ model exhibit striking differences. For  example, consider the single lepton production
mode:
\begin{eqnarray}
\label{qLdecay-MSSM} \widetilde{q}_L \rightarrow q_{L}^{\prime}
\widetilde{W}^{\pm} \rightarrow q_L^{\prime} \ell^{\pm}
\widetilde{\nu}^{\star}_{\ell} \rightarrow q_L^{\prime} \ell^{\pm}
\overline{\nu}_{\ell} \widetilde{B}\,,
\end{eqnarray}
wherein the missing energy comprises both the bino and the neutrino
emissions. Since
$\widetilde{W}^{\pm}$ and $\widetilde{W}^3$ are nearly
degenerate in mass, this cascade hardly extends any further in the
MSSM. In the $U(1)^{\prime}$ model, however, if
$\widetilde{Z}^{\prime}$ lies below $\widetilde{W}^3$ and above
$\widetilde{B}$ then the decay chain (\ref{qLdecay-MSSM})
proceeds one step further
\begin{eqnarray}
\label{qLdecay-U1p} \widetilde{q}_L \rightarrow q_{L}^{\prime}
\widetilde{W}^{\pm} \rightarrow q_L^{\prime} \ell^{\pm}
\widetilde{\nu}^{\star}_{\ell} \rightarrow q_L^{\prime} \ell^{\pm}
\overline{\nu}_{\ell} \widetilde{Z}^{\prime} \rightarrow
q_L^{\prime} \ell^{\pm} \overline{\nu}_{\ell} \ell^{\prime\, +}
\widetilde{\ell}^{\prime\, +} \rightarrow q_L^{\prime} \ell^{\pm}
\overline{\nu}_{\ell} \ell^{\prime\, +} \ell^{\prime\, -}
\widetilde{B}\,,
\label{4amode}
\end{eqnarray}
yielding a trilepton signal. This $U(1)^{\prime}$ result is strikingly
different from the one in the MSSM where the trilepton signal is
expected to be suppressed, if not completely blocked.

If the the LSP is not the bino but the $\widetilde{Z}^{\prime}$, then
essentially the roles of (\ref{qLdecay-MSSM}) and
(\ref{qLdecay-U1p}) are interchanged. A
$\widetilde{Z}^{\prime}$ LSP has the same features mentioned
while discussing the $\widetilde{q}_R$ decays. For a Higgsino
LSP decay, (\ref{qLdecay-MSSM}) gains further steps yielding
additional lepton pairs.

Summarizing this subsection, we have investigated the collider signatures of the
$U(1)^{\prime}$ group in the cascade decays of the first and second
generations scalar quarks. This extra gauge
symmetry offers various collider signatures by modifying the
rates, topologies and and the pattern of various decay modes. The
$U(1)^{\prime}$ gaugino $\widetilde{Z}^{\prime}$ and the
singlino $\widetilde{S}$ are the avatars of the $U(1)^{\prime}$
model. The discussions have been based on the Lagrangian--basis
inos $\widetilde{G}^a$ for a clear tracking of various effects.
An accurate analysis must necessarily take into account the
physical, mass-eigenstate neutral fermions
$\widetilde{\chi}^0_i$ as well as the mass-eigenstate sfermions
(mainly the ones in the third generation). This will undertaken
in the next section in numerical studies of
the squark decays.

\section{The LHC Signals of the $U(1)^{\prime}$ Model }
In this section we perform a simulation study of the scattering
processes indicative of the additional
$U(1)^{\prime}$ group. In particular, we analyze the decay
patterns of the scalar quarks in order to determine their
rates, topologies and signatures by explicitly working with the
physical neutralinos, squarks and sleptons.

The $U(1)^{\prime}$ model consists of a number of parameters
not yet specified by experiments. In order
to make realistic numerical estimates of the processes
discussed in the previous section, one has to adopt a
set of viable parameters, compatible with
the existing bounds from various sources. To this end, the following
parameter choices will be used in the numerical analysis:
\begin{itemize}
\item The first group of unknown parameters  refers to the
    $U(1)^{\prime}$ charges of the fields. All the
    properties of the $U(1)^{\prime}$ model advocated so far
    hold for a generic charge assignment. For the
    numerical analysis, we assume the $G_{SM}\otimes
    U(1)^{\prime}$ models to be descending from SUSY GUTs which provide
     the absence of anomalies
    and several other well-studied features \cite{gut-string}.
    The breaking pattern
\begin{eqnarray}
E_6 \rightarrow SO(10)\otimes U(1)_{\psi} \rightarrow SU(5)\otimes U(1)_{\chi} \otimes U(1)_{\psi} \rightarrow
G_{SM}\otimes U(1)^{\prime}_{Y^{\prime}}\,,
\end{eqnarray}
gives rise to the $G_{SM}\otimes U(1)^{\prime}$ model of interest from the $E_6$ SUSY GUT.
 Each arrow in this
chain corresponds to spontaneous symmetry breakdown at a specific (presumably ultra high) energy
scale. Here, by construction,
\begin{eqnarray}
\label{U1ptheta}
U(1)_{Y^{\prime}} = \cos\theta_{E_6}\, U(1)_{\psi} - \sin\theta_{E_6}\, U(1)_{\chi}\,,
\end{eqnarray}
and the $U(1)^{\prime}$ invariance is broken near the ${\rm
TeV}$ scale whereas the other orthogonal combination
$U(1)^{\prime \prime}_{Y^{\prime}}=\cos\theta_{E_6}\,
U(1)_{\chi} + \sin\theta_{E_6}\, U(1)_{\psi}$ is broken at
a much higher scale, not accessible to the LHC experiments. The
angle $\theta_{E_6}$ designates the breaking direction in
$U(1)_{\chi} \otimes U(1)_{\psi}$ space and it is a
function of the gauge couplings and VEVs associated with the breaking. The $U(1)_{\chi}$ and
$U(1)_{\psi}$ charge assignments are shown in
Table~\ref{table2}. In (\ref{U1ptheta}), a
low-energy $G_{SM}\otimes$U(1)$^{\prime}$ model arises with
\begin{eqnarray}
\label{e6list}
&& Y^{\prime}_f = \cos\theta_{E_6}\, Q^f_{\psi} - \sin\theta_{E_6}\, Q^f_{\chi}\,,\nonumber\\
&& g_{Y^{\prime}} = \sqrt{\frac{5}{3}}\, g_Y\,,
\end{eqnarray}
for any field $f$ in the spectrum  with the breaking determined by the angle
$\theta_{E_6}$. It is clear that if the $U(1)^{\prime}$
model is to solve the $\mu$ problem of the MSSM, then
$Y^{\prime}_{\widehat{S}} \neq 0$, and hence, as suggested
by Table~\ref{table2}, $\theta_{E_6} = \pi/2$ should be
avoided.

\begin{table}
\begin{tabular*}{0.9\textwidth}{@{\extracolsep{\fill}} llllllllllll}
\hline \hline
$\;\;\;\;\;$ $\widehat{f}$ & $\widehat{Q}$ & $\widehat{U}$ & $\widehat{D}$ & $\widehat{L}$ &
$\widehat{E}$ & $\widehat{H}_d$ & $\widehat{H}_u$ & $\widehat{S}$ &
$\widehat{N}$ & $\widehat{D}_u$ & $\widehat{D}_d$ \\\hline\hline
$\;$ 2 $\sqrt{6}\, Q^f_{\psi}$ & 1 & 1 &  1 & 1 & 1 &  -2& -2& 4 &1 & -2& -2\\\hline
$\;$ 2 $\sqrt{10}\, Q^f_{\chi}$ & -1 & -1 & 3& 3 & -1& -2& 2& 0& -5 & 2& -2\\\hline\hline
\end{tabular*}
\caption{\label{table2} The $U(1)_{\psi}$ and $U(1)_{\chi}$
charges of the superfields. The left side of the table lists
 the particle spectrum of $G_{SM}\otimes U(1)^{\prime}$
model whereas on the right side, the chiral
fields  $\widehat{N}$, $\widehat{D}_u$
and $\widehat{D}_d$ form a sector
necessary for canceling the anomalies \cite{kane}, yet too heavy to
leave any significant impact on the LHC experiments \cite{langacker-kang}.
Clearly, $U(1)_{\psi}$ is a viable model for solving the $\mu$ problem
of the MSSM but $U(1)_{\chi}$ is not.}
\end{table}

\item The soft-breaking masses shared with the  MSSM are
    assigned the following values:
\begin{eqnarray}
\label{mssmlist}
&& m_{\widetilde{q}_L} = m_{\widetilde{q}_R} =1200\ {\rm GeV} ,\nonumber\\
&& m_{\widetilde{e}_L} = 350\ {\rm GeV} ,\; m_{\widetilde{e}_R} = 200\ {\rm GeV} ,\nonumber\\
&& M_{\widetilde{Y}} = 100\ {\rm GeV},\; M_{\widetilde{W}} = 400\ {\rm GeV},\; M_{\widetilde{g}} = 1300\ {\rm GeV}\,,
\end{eqnarray}
where $m_{\widetilde{q}_{L,R}}$ and
$m_{\widetilde{e}_{L,R}}$ stand, respectively, for the soft
masses (before  $G_{SM}\otimes U(1)^{\prime}$ breaking) of
squarks and sleptons in the first and second generations.
These parameter values, as for all others, refer to
${\rm TeV}$ scale, and no assumption is made of the
universality of gaugino and scalar masses at high scale.

\item The parameters pertaining to the $U(1)^{\prime}$
    sector are assigned the values (the value of
    $\mu_{eff}$ determines the singlet VEV and in turn it
    determines $M_{Z^{\prime}}$)
\begin{eqnarray}
h_s =0.6 ,\; \mu_{eff} = 1400\ {\rm GeV} ,\; \tan\beta = 10,\; \sin\chi = 5\times 10^{-3}
\end{eqnarray}
where the value of the kinetic mixing angle $\chi$ follows
from its radiative nature \cite{cvetic,kolda1}. The ranges
of the parameters must be such that the bound
$\left|\theta_{Z-Z^{\prime}}\right| \simlt 10^{-3}$
\cite{langacker-review} is respected.

\item Among the well-studied $E_6$ models \cite{gut-string} we
    specialize to the one defined by the mixing angle
\begin{eqnarray}
\label{thetae6}
\theta_{E_6} = \arcsin\left[\sqrt{3/8}\right]\simeq 37.76^{\circ}\,,
\end{eqnarray}
which corresponds to the $U(1)^{\prime} \equiv U(1)_{\eta}$
model. Experimentally, $M_{Z^{\prime}} \geq 933\ {\rm GeV}$,
\cite{cdf} though this bound is lower by typically 250 ${\rm GeV}$
if the decays into sparticles are taken into account
\cite{langacker-kang}.

\item For simplicity
and later convenience, we scale the gaugino mass parameters
$M_{\widetilde{Y}^\prime}$ and
$M_{\widetilde{Y}\widetilde{Y}^\prime}$ with the
hypercharge gaugino mass to define the ratios:
 \begin{eqnarray}
 \ry \equiv \frac{M_{\widetilde{Y}^\prime}}{M_{\widetilde{Y}}} \,,\;\;\;\;
\ryy \equiv \frac{M_{\widetilde{Y}\widetilde{Y}^\prime}}{M_{\widetilde{Y}}}\,,
\end{eqnarray}
the relevant values of which are sampled according to
(\ref{U1plist-small}), (\ref{U1plist-equal}) and
(\ref{U1plist-large}). In obtaining various numerical results we employ
    different possibilities for the remaining model
    parameters:
\begin{itemize}
\item {\bf Small $U(1)_Y$--$U(1)_{Y^{\prime}}$ Mixing:}
\begin{eqnarray}
\label{U1plist-small}
(\ry,\, \ryy) = (1/2, 0),\; (2, 0),\; (6, 0),\; (10, 0)\,.
\end{eqnarray}

\item {\bf Medium $U(1)_Y$--$U(1)_{Y^{\prime}}$
    Mixing:}
\begin{eqnarray}
\label{U1plist-equal}
(\ry,\, \ryy) =(0,0),\; (1/2, 1/2),\; (2, 2),\; (6, 6),\; (10, 10)\,.
\end{eqnarray}

\item {\bf Large $U(1)_Y$--$U(1)_{Y^{\prime}}$ Mixing:}
\begin{eqnarray}
\label{U1plist-large}
(\ry,\, \ryy)  = (0, 1/2),\; (0, 2),\; (0, 6),\; (0, 10)\,,
\end{eqnarray}

\end{itemize}
In each case, the
$\widetilde{Z}^{\prime}$ gaugino falls in different bands
in mass and mixing, and, depending on how they compare with
those of the electroweak gauginos, various  decay chains
can close or open, thereby leading to distinct signatures
at the LHC, as discussed in Sec. III above, and to distinct predictions
 in the figures and tables to
be given below.
\end{itemize}
The numerical analysis below will provide a generator-level
description of the LHC signals of the $U(1)^{\prime}$ model for
the parameter values specified above. The choice of the $\eta$
model is in no way better than any other model descending from the
$E_6$ SUSY GUT. Moreover, one can just adopt a low-energy
$U(1)^{\prime}$ model without resorting to the $E_6$ framework,
at the expense of a much larger set of free parameters.
Therefore, the $U(1)_{\eta}$ model adopted here can be regarded
as a prototype to get  an idea of what physics potentials such models can have
at the LHC, compared  to the MSSM.

\subsection{Branching Fractions of Squark Decay Channels}
In this section we compute the branching fractions of the various
decay channels discussed in Sec. III. The branching fractions
will eventually determine the relative populations of the
final states that constitute the signature space of events to
be searched for at the LHC.  Essentially, we analyze the decay
patterns of the squarks by considering separately the
$\widetilde{q}_R$ and $\widetilde{q}_L$ squarks in the first
and second generations (they are themselves mass and flavor
eigenstates, to an excellent approximation). We take the parameter
values from (\ref{e6list}), (\ref{mssmlist}),
(\ref{U1plist-small}),  (\ref{U1plist-equal}) and
(\ref{U1plist-large}). For each, we compute the
branching fractions in the MSSM and in the $U(1)^{\prime}$
model, and display them comparatively in the figures to follow.
The figures employ a diagrammatic display structure for a clear
understanding of the various branching illustrated by
varying $\ry$ and $\ryy$ as in (\ref{U1plist-small}),
(\ref{U1plist-equal}) and (\ref{U1plist-large}).
\begin{center}
\begin{table}[htb]
\begin{tabular}{c@{\hspace*{0.8cm}}c@{\hspace*{0.3cm}}c@{\hspace*{0.3cm}}
c@{\hspace*{0.3cm}}c@{\hspace*{0.3cm}}c@{\hspace*{0.3cm}}c}
\hline \hline
$({\cal R}_{\widetilde{Y}^{\prime}}\, , {\cal R}_{\widetilde{Y}\widetilde{Y^{\prime}}})$
& $\mnbir$ & $\mniki$ & $\mnuc$ & $\mndort$ & $\mnbes$ & $\mnalti$ \\
 \hline \hline
 MSSM & 100\gev & 398\gev & $-$ & $-$ & 1402\gev & 1405\gev  \\
\hline
 $(1/2,0)$ & 100\gev & 398\gev & 955\gev & 1007\gev & 1407\gev & 1408\gev  \\
$(2,0)$ & 97\gev & 398\gev & 885\gev & 1087\gev & 1407\gev & 1408\gev \\
$(6,0)$ & 97\gev & 398\gev & 725\gev & 1326\gev & 1407\gev & 1408\gev \\
$(10,0)$ & 97\gev & 398\gev & 600\gev & 1407\gev & 1407\gev & 1602\gev \\
\hline
 $(0,0)$ & 100\gev & 398\gev & 980\gev & 982\gev & 1407\gev & 1408\gev  \\
$(1/2,1/2)$ & 100\gev & 398\gev & 957\gev & 1008\gev & 1407\gev & 1408\gev \\
$(2,2)$ & 97\gev & 398\gev & 905\gev & 1107\gev & 1407\gev & 1408\gev \\
$(6,6)$ & 77\gev & 398\gev & 876\gev & 1405\gev & 1407\gev & 1497\gev \\
$(10,10)$ & 54\gev & 398\gev & 960\gev & 1407\gev & 1407\gev & 1998\gev \\
\hline
 $(0,1/2)$ & 100\gev & 398\gev & 982\gev & 983\gev & 1407\gev & 1408\gev  \\
$(0,2)$ & 97\gev & 398\gev & 1000\gev & 1002\gev & 1407\gev & 1408\gev \\
$(0,6)$ & 76\gev & 398\gev & 1141\gev & 1159\gev & 1407\gev & 1409\gev \\
$(0,10)$ & 53\gev & 398\gev & 1382\gev & 1391\gev & 1407\gev & 1437\gev \\
\hline \hline
\end{tabular}
\caption{\sl\small The neutralino mass spectra in the
$U(1)^{\prime}$ model for the parameter sets
(\ref{U1plist-small}), (\ref{U1plist-equal}) and
(\ref{U1plist-large}).} \label{massneut}
\end{table}
\end{center}

In Table~\ref{massneut}, we list the neutralino masses both in the MSSM
and the $U(1)^{\prime}$ model obtained for the values of ${\cal
R}_{\widetilde{Y}^{\prime}}$ and ${\cal
R}_{\widetilde{Y}\widetilde{Y^{\prime}}}$. As seen in this table, variations of
these ratios mainly modify the masses of the third and fourth
neutralinos. In other words, the MSSM mass spectrum corresponds
approximately to the states $\left\{\widetilde{\chi}_1^0,
\widetilde{\chi}_2^0, \widetilde{\chi}_5^0,
\widetilde{\chi}_6^0\right\}$; the $U(1)^{\prime}$ effects
amount  to inserting the extra states $\left\{\widetilde{\chi}_3^0,
\widetilde{\chi}_4^0\right\}$ into the mass spectrum. The
MSSM--like neutralinos are nearly immune to these ratios,
except for the the cases ${\cal R}_{\widetilde{Y}^{\prime}} = 10$
and/or ${\cal R}_{\widetilde{Y}\widetilde{Y^{\prime}}} = 10 $,
for which the mass of the $\widetilde{Z}^{\prime}$ and/or its
mixing with $\widetilde{B}$ exceed the $\widetilde{B}$ mass by
an order of magnitude. One notices that, $\mnuc$ (in small and
medium mixing regimes) and $\mnbir$ (in medium and large
mixing regimes) typically decrease with increasing ${\cal
R}_{\widetilde{Y}^{\prime}}$ and/or ${\cal
R}_{\widetilde{Y}\widetilde{Y^{\prime}}}$. This decrease in $\mnuc$ and
 $\mnbir$ is most sensitively correlated with the corresponding increase in
$\mnalti$.

The nature of a given neutralino state $\widetilde{\chi}_i^0$
is determined by its decomposition into the Lagrangian basis
$\left\{\widetilde{B}, \widetilde{W}^3, \widetilde{H}^0_d,
\widetilde{H}^0_u, \widetilde{S},
\widetilde{Z}^{\prime}\right\}$. Depicted in
Table~\ref{mixneut} are the compositions of
$\widetilde{\chi}_1^0$ (the LSP), $\widetilde{\chi}_3^0$ and
$\widetilde{\chi}_4^0$ for the parameter sets
(\ref{U1plist-small}), (\ref{U1plist-equal}) and
(\ref{U1plist-large}). As suggested by the table, the LSP is
overwhelmed by its bino component in the small mixing regime,
as in the MSSM and in accord with \cite{langacker-neutral}.
Nevertheless, its bino component become approximately equal to
its singlino component for large ${\cal
R}_{\widetilde{Y}^{\prime}}$ and/or ${\cal
R}_{\widetilde{Y}\widetilde{Y^{\prime}}}$, in the medium and
large mixing regimes. This increase in the singlino component
implies reduced couplings of the LSP to fermions and sfermions,
as discussed in Appendices A and C.

 The neutralino states $\widetilde{\chi}_{3,4}^0$ behave
differently than the LSP, as they are, as suggested by
Table~\ref{massneut}, genuine to $U(1)^{\prime}$ model. Indeed,
they are overwhelmed by $\widetilde{Z}^{\prime}$ and
$\widetilde{S}$  for all of the small, balanced and large
mixing regimes. The exceptions arise for large ${\cal
R}_{\widetilde{Y}^{\prime}}$ and/or ${\cal
R}_{\widetilde{Y}\widetilde{Y^{\prime}}}$ values for which
$\widetilde{\chi}_3^0$ develops a significant bino component,
and $\widetilde{\chi}_4^0$ changes to be Higgsino--dominated.
For the large mixing regime, however, also
$\widetilde{\chi}_4^0$ obtains a significant bino component as
${\cal R}_{\widetilde{Y}\widetilde{Y^{\prime}}}$ grows. These
compositions, as detailed in Table~\ref{mixneut}, directly
influence decay patters and products of a given neutralino: A
sizeable $\widetilde{Z}^{\prime}$ component gives rise to novel
decay patters described in Sec. III A, a sizable
$\widetilde{S}$ composition halts the cascade as it cannot
directly decay into fermions, and, similarly, a sizeable bino
component stops the cascade as it dominates
$\widetilde{\chi}_1^0$.

\renewcommand{\arraystretch}{1}
\begin{center}
\begin{longtable}[l]
{r@{\hspace*{0.5cm}}c@{\hspace*{0.5cm}}r@{\hspace*{0.3cm}}r@{\hspace*{0.3cm}}
r@{\hspace*{0.3cm}}r@{\hspace*{0.3cm}}r@{\hspace*{0.3cm}}r}
\hline \hline
$\;\;\;\;({\cal R}_{\widetilde{Y}^{\prime}}\, , {\cal R}_{\widetilde{Y}\widetilde{Y^{\prime}}})$
& $\widetilde{\chi}_{1,3,4}^0$ & $\widetilde{B}$ & $\widetilde{W}^3$ & $\widetilde{H}^0_d$ & $\widetilde{H}^0_u$ & $\widetilde{S}$ & $\widetilde{Z}^{\prime}$ \\
 \hline \hline\\[0.1mm]
\endfirsthead
\multicolumn{1}{c}{\multirow{3}{*} {\textbf{MSSM}}}
& $\left(\widetilde{\chi}_1^0\right)_{MSSM} \left(\leadsto \widetilde{\chi}_1^0\right)$ & 0.99 & $-$0.0044 & 0.019 & 0.026 & $-$ & $-$  \\[0.1mm]
& $\left(\widetilde{\chi}_3^0\right)_{MSSM} \left(\leadsto \widetilde{\chi}_5^0\right)$ & 0.032 & $-0.064$ & $-0.71$ & $-0.70$ & $-$ & $-$  \\[0.1mm]
& $\left(\widetilde{\chi}_4^0\right)_{MSSM} \left(\leadsto \widetilde{\chi}_6^0\right)$  & $-0.0084$ & 0.029 & $-0.71$ & 0.71 & $-$ & $-$  \\[0.1mm]
\hline\\[0.1mm]
\multicolumn{1}{c}{\multirow{3}{*}{\boldmath{ $(1/2,0)$}}}
& $\widetilde{\chi}_1^0$ & $-0.99$ & 0.0023 & $-0.032$ & 0.0054 & $-0.0004$ &$-0.0033$ \\[0.1mm]
& $\widetilde{\chi}_3^0$ & $-0.0023$ & 0.0038 & 0.021 & 0.067 & $-0.71$ & 0.70  \\[0.1mm]
& $\widetilde{\chi}_4^0$ & 0.0031 & $-0.0073$ & $-0.0042$ & 0.055 & $-0.70$ & $-0.71$  \\[0.1mm] 
\cline{2-8}\\[0.1mm] \multicolumn{1}{c}{\multirow{3}{*}{\boldmath{
$(2,0)$}}}
& $\widetilde{\chi}_1^0$ & 0.99 & $-0.0023$ & 0.032 & $-0.0054$ & $-0.0001$ & 0.0033 \\[0.1mm]
& $\widetilde{\chi}_3^0$  & $-0.0025$ & 0.004 & 0.019 & 0.066 & $-0.74$ & 0.67 \\[0.1mm]
& $\widetilde{\chi}_4^0$  & 0.0029 & $-0.0065$ & 0.0065 & 0.055 & $-0.67$ & $-0.74$ \\[0.1mm]
\cline{2-8}\\[0.1mm] \multicolumn{1}{c}{\multirow{3}{*}{\boldmath{
$(6,0)$}}}
& $\widetilde{\chi}_1^0$ & 0.99 & $-0.0022$ & 0.032 & $-0.0053$ & $-0.0014$ & 0.0032 \\[0.1mm]
& $\widetilde{\chi}_3^0$ & $-0.0031$ & 0.0046 & 0.015 & 0.067 & $-0.80$ & 0.59 \\[0.1mm]
& $\widetilde{\chi}_4^0$ & $-0.0033$ & $-0.0071$ & 0.037 & $-0.079$ & 0.59 & 0.80 \\[0.1mm]
\cline{2-8}\\[0.1mm] \multicolumn{1}{c}{\multirow{3}{*}{\boldmath{
$(10,0)$}}}
& $\widetilde{\chi}_1^0$ & 0.99 & $-0.0022$ & 0.032 & $-0.0052$ & $-0.0026$ & 0.0030 \\[0.1mm]
& $\widetilde{\chi}_3^0$ & $-0.0038$ & 0.0053 & 0.013 & 0.068 & $-0.85$ & 0.52 \\[0.1mm]
&  $\widetilde{\chi}_4^0$ & $-0.018$ & 0.028 & 0.71 & 0.71 & 0.063 & $-0.008$ \\[0.1mm]
\hline\\[0.1mm] \multicolumn{1}{c}{\multirow{3}{*}{\boldmath{ $(0,0)$}}}
& $\widetilde{\chi}_1^0$ & 0.99 & $-0.0023$ & 0.032 & $-0.0054$ & 0.00057 & 0.0034  \\[0.1mm]
& $\widetilde{\chi}_3^0$ & 0.0023 & $-0.0038$ & $-0.022$ & $-0.067$ & 0.70 & $-0.71$  \\[0.1mm]
& $\widetilde{\chi}_4^0$ & 0.0032 & $-0.0077$ & $-0.0036$ & 0.056 & $-0.71$ & $-0.71$  \\[0.1mm]
\cline{2-8}\\[0.1mm] \multicolumn{1}{c}{\multirow{3}{*}{\boldmath{
$(1/2,1/2)$}}}
& $\widetilde{\chi}_1^0$  & $-0.99$ & 0.0013 & $-0.032$ & 0.0016 & 0.051 & 0.0018 \\[0.1mm]
& $\widetilde{\chi}_3^0$  & $-0.035$ & 0.0037 & 0.020 & 0.066 & $-0.71$ & 0.70 \\[0.1mm]
& $\widetilde{\chi}_4^0$  & 0.036 & 0.0076 & 0.0069 & $-0.057$ & 0.70 & 0.72 \\[0.1mm]
\cline{2-8}\\[0.1mm]\multicolumn{1}{c}{\multirow{3}{*}{\boldmath{
$(2,2)$}}}
& $\widetilde{\chi}_1^0$  & 0.98 & 0.0016 & 0.032 & 0.0094 & $-0.20$ & $-0.016$ \\[0.1mm]
& $\widetilde{\chi}_3^0$ & 0.14 & $-0.0037$ & $-0.013$ & $-0.062$ & 0.73 & $-0.67$ \\[0.1mm]
& $\widetilde{\chi}_4^0$ & $-0.14$ & $-0.0076$ & $-0.020$ & 0.066 & $-0.65$ & $-0.74$ \\[0.1mm]
\cline{2-8}\\[0.1mm]\multicolumn{1}{c}{\multirow{3}{*}{\boldmath{
$(6,6)$}}}
&  $\widetilde{\chi}_1^0$ & $-0.86$ & $-0.0075$ & $-0.030$ & $-0.033$ & 0.51 & 0.036 \\[0.1mm]
&  $\widetilde{\chi}_3^0$ & $-0.38$ & 0.0032 & $-0.00064$ & 0.051 & $-0.69$ & 0.62 \\[0.1mm]
&  $\widetilde{\chi}_4^0$ & $-0.039$ & $-0.061$ & $-0.69$ & 0.70 & $-0.051$ & $-0.14$ \\[0.1mm]
\cline{2-8}\\[0.1mm]\multicolumn{1}{c}{\multirow{3}{*}{\boldmath{
$(10,10)$}}}
& $\widetilde{\chi}_1^0$ & 0.72 & 0.011 & 0.027 & 0.048 & $-0.70$ & $-0.035$ \\[0.1mm]
&  $\widetilde{\chi}_3^0$ & $-0.55$ & 0.0023 & $-0.012$ & 0.037 & $-0.60$ & 0.58 \\[0.1mm]
&  $\widetilde{\chi}_4^0$ & $-0.021$ & 0.028 & 0.71 & 0.71 & 0.054 & 0.0043 \\[0.1mm]
\hline \\[0.1mm]\multicolumn{1}{c}{\multirow{3}{*}{\boldmath{
$(0,1/2)$}}}
& $\widetilde{\chi}_1^0$ & $-0.99$ & 0.0013 & $-0.031$ & 0.0016 & 0.051 & 0.0018  \\[0.1mm]
& $\widetilde{\chi}_3^0$ & $-0.035$ & 0.0037 & 0.021 & 0.066 & $-0.70$ & 0.71  \\[0.1mm]
& $\widetilde{\chi}_4^0$ & 0.037 & 0.0079 & 0.0062 & $-0.057$ & 0.70 & 0.71  \\[0.1mm]
\cline{2-8}\\[0.1mm]\multicolumn{1}{c}{\multirow{3}{*}{\boldmath{
$(0,2)$}}}
& $\widetilde{\chi}_1^0$ & 0.98 & 0.0017 & 0.032 & 0.0097 & $-0.20$ & $-0.016$ \\[0.1mm]
& $\widetilde{\chi}_3^0$  & 0.13 & $-0.0035$ & $-0.016$ & $-0.062$ & 0.69 & $-0.71$ \\[0.1mm]
& $\widetilde{\chi}_4^0$  & $-0.15$ & $-0.0085$ & $-0.015$ & 0.063 & $-0.69$ & $-0.71$ \\[0.1mm]
\cline{2-8}\\[0.1mm]\multicolumn{1}{c}{\multirow{3}{*}{\boldmath{
$(0,6)$}}}
& $\widetilde{\chi}_1^0$ & 0.85 & 0.0078 & 0.030 & 0.035 & $-0.52$ & $-0.037$ \\[0.1mm]
& $\widetilde{\chi}_3^0$ & $-0.35$ & 0.0025 & 0.0031 & 0.048 & $-0.61$ & 0.71 \\[0.1mm]
& $\widetilde{\chi}_4^0$ & 0.39 & 0.0096 & 0.050 & $-0.087$ & 0.59 & 0.70 \\[0.1mm]
\cline{2-8}\\[0.1mm]\multicolumn{1}{c}{\multirow{3}{*}{\boldmath{
$(0,10)$}}}
& $\widetilde{\chi}_1^0$ & $-0.70$ & $-0.011$ & $-0.026$ & $-0.050$ & 0.71 & 0.035 \\[0.1mm]
& $\widetilde{\chi}_3^0$ & $0.48$ & $-0.0021$ & $0.097$ & $-0.057$ & $-0.51$ & 0.71 \\[0.1mm]
&  $\widetilde{\chi}_4^0$ & $-0.30$ & $-0.052$ & $-0.57$ & 0.59 & $-0.26$ & $-0.42$ \\[0.1mm]
\hline \hline\\*[0.2cm]
  \caption{\sl\small The components of $\widetilde{\chi}_1^0$
  (the LSP), $\widetilde{\chi}_3^0$ and $\widetilde{\chi}_4^0$ in
  the Lagrangian basis $\left\{\widetilde{B}, \widetilde{W}^3,
  \widetilde{H}^0_d, \widetilde{H}^0_u, \widetilde{S},
  \widetilde{Z}^{\prime}\right\}$ for the parameter sets
  (\ref{U1plist-small}), (\ref{U1plist-equal}) and
  (\ref{U1plist-large}).} \label{mixneut}
\end{longtable}
\end{center}

Having completed the specification of the neutralino sector, we
now turn to the analysis of the scalar quark decays.
We compute the branching ratios of the decays
\begin{eqnarray}
\mbox{squark} \rightarrow \mbox{quark} + \widetilde{\chi}_i^0\,,
\end{eqnarray}
for each quark chirality and for each of the parameter sets
(\ref{U1plist-small}), (\ref{U1plist-equal}) and
(\ref{U1plist-large}). The results are shown in Figs.
\ref{figqRbranchsmall}, \ref{figqRbranchequal},
\ref{figqRbranchlarge} for $\widetilde{q}_R$, and Figs.
\ref{figqLbranchsmall}, \ref{figqLbranchequal},
\ref{figqLbranchlarge} for $\widetilde{q}_L$.
 \mfig{htb}{figqRbranchsmall}{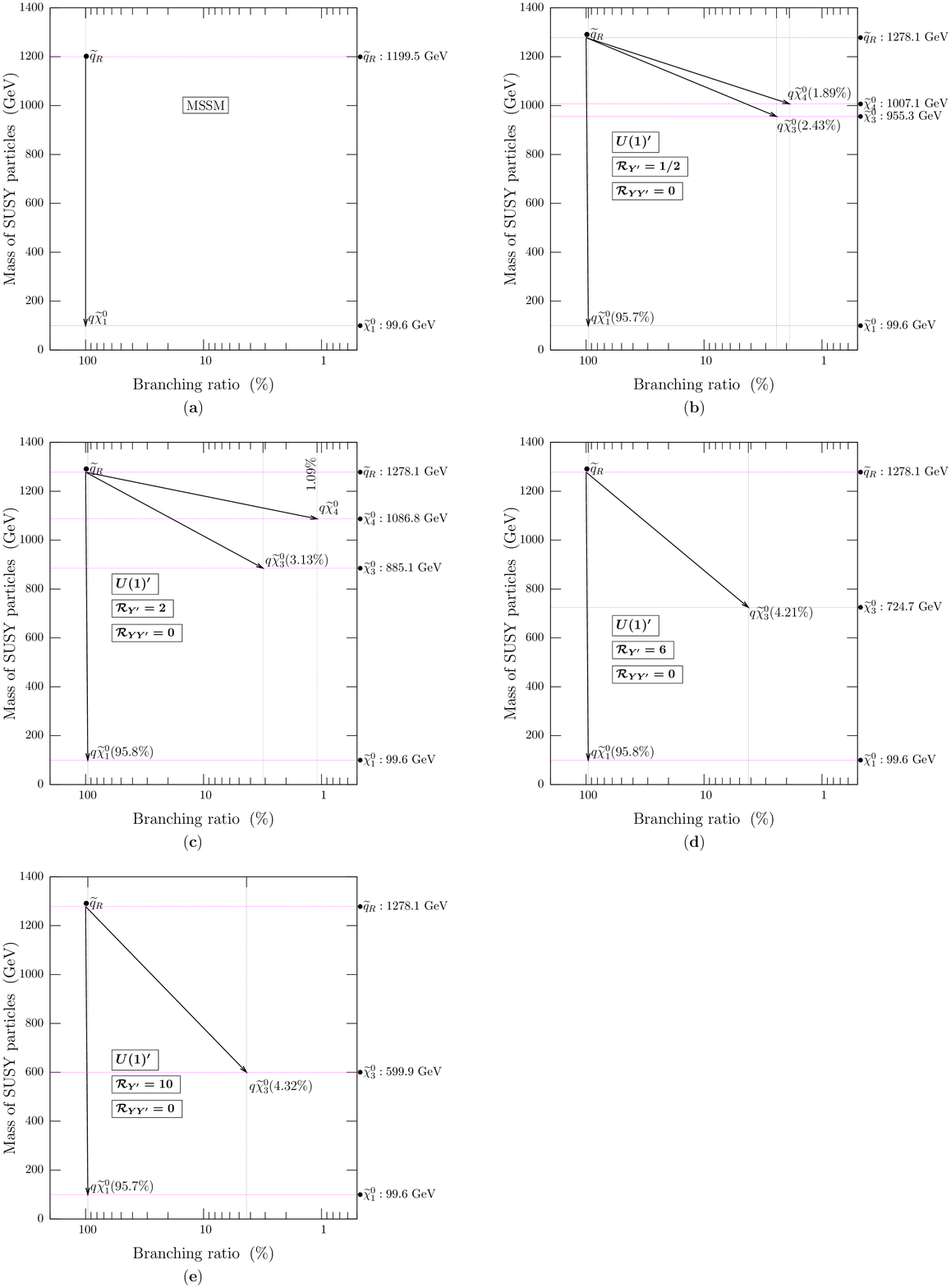}{\it The
branching fractions ($\%$) of right-handed squarks
$\widetilde{q}_R$ belonging to the first or second generation
as a function of the neutralino and chargino  masses. Shown are
branching fractions exceeding one percent level. The panel (a)
stands for the MSSM expectation while the rest correspond to
the parameter set in (\ref{U1plist-small}), that is, the small
mixing regime. The branching into $q\ \widetilde{\chi}_3^0$
grows with decreasing $\mnuc$.}
\mfig{htb}{figqRbranchequal}{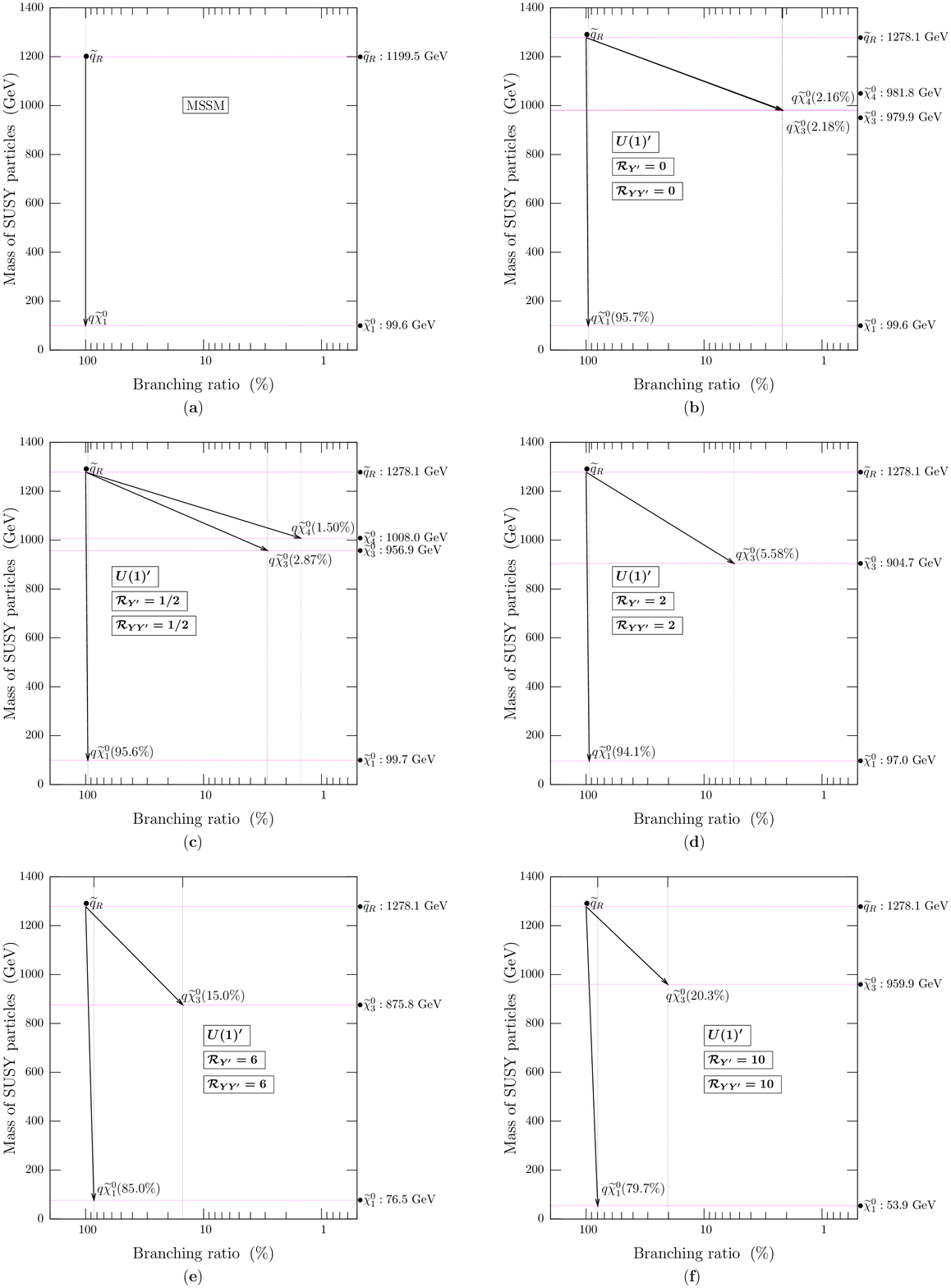}{\it The
branching fractions ($\%$) of right-handed squarks
$\widetilde{q}_R$ belonging to the first or second generation
as a function of the neutralino and chargino  masses. Shown are
branching fractions exceeding one percent level. The panel (a)
stands for the MSSM expectation while the rest correspond to
the parameter set in (\ref{U1plist-equal}), that is, the
medium mixing regime. The branching into $q\
\widetilde{\chi}_3^0$ grows with decreasing $\mnuc$, and
reaches the $20 \%$ level when ${\cal R}_{\widetilde{Y}^{\prime}} =
{\cal R}_{\widetilde{Y}\widetilde{Y^{\prime}}}= 10$.}
\mfig{htb}{figqRbranchlarge}{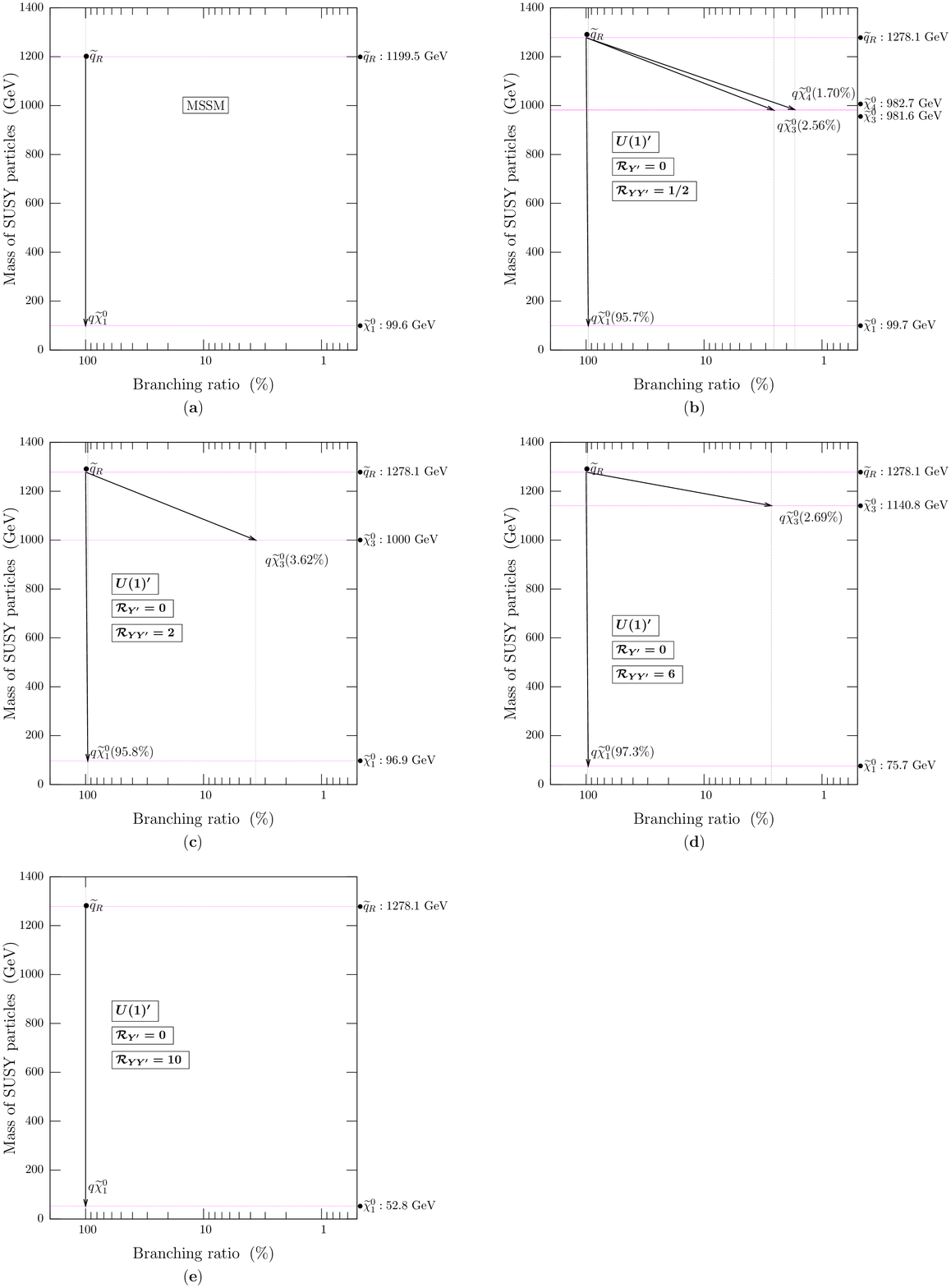}{\it The
branching fractions ($\%$) of right-handed squarks
$\widetilde{q}_R$ belonging to the first or second generation
as a function of the neutralino and chargino  masses. Shown are
branching fractions exceeding one percent level. The panel (a)
stands for the MSSM expectation while the rest correspond to
the parameter set in (\ref{U1plist-large}), that is, the large
mixing regime. The branching into $q\ \widetilde{\chi}_3^0$
decreases with increasing $\mnuc$, and is kinematically blocked
when ${\cal R}_{\widetilde{Y}\widetilde{Y^{\prime}}}= 10$. At
this extreme, the branching of the squark is
indistinguishable from the MSSM case.}

\mfig{htb}{figqLbranchsmall}{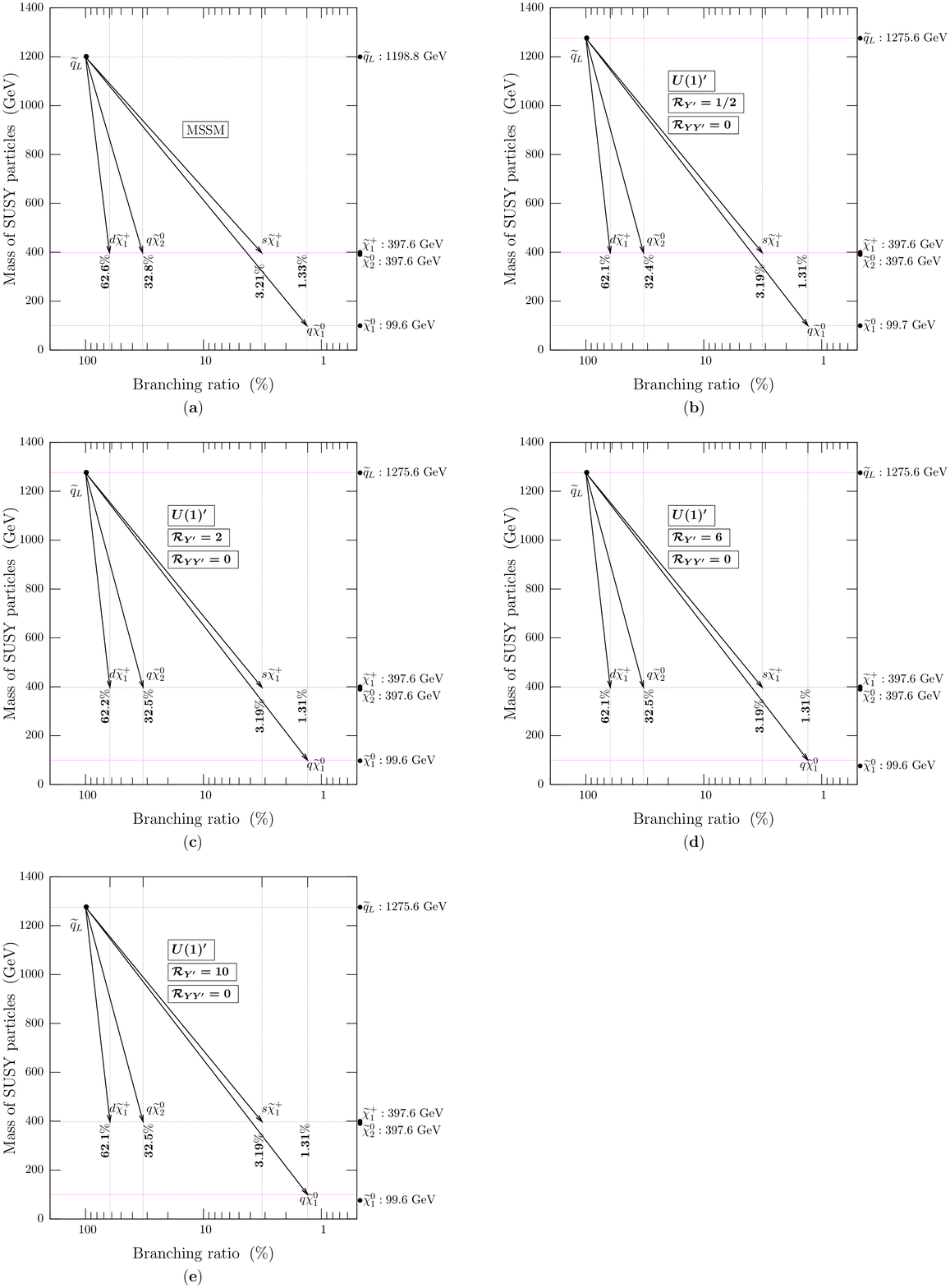}{\it The same
as in Fig. \ref{figqRbranchsmall} but for $\widetilde{q}_L$.}
\mfig{htb}{figqLbranchequal}{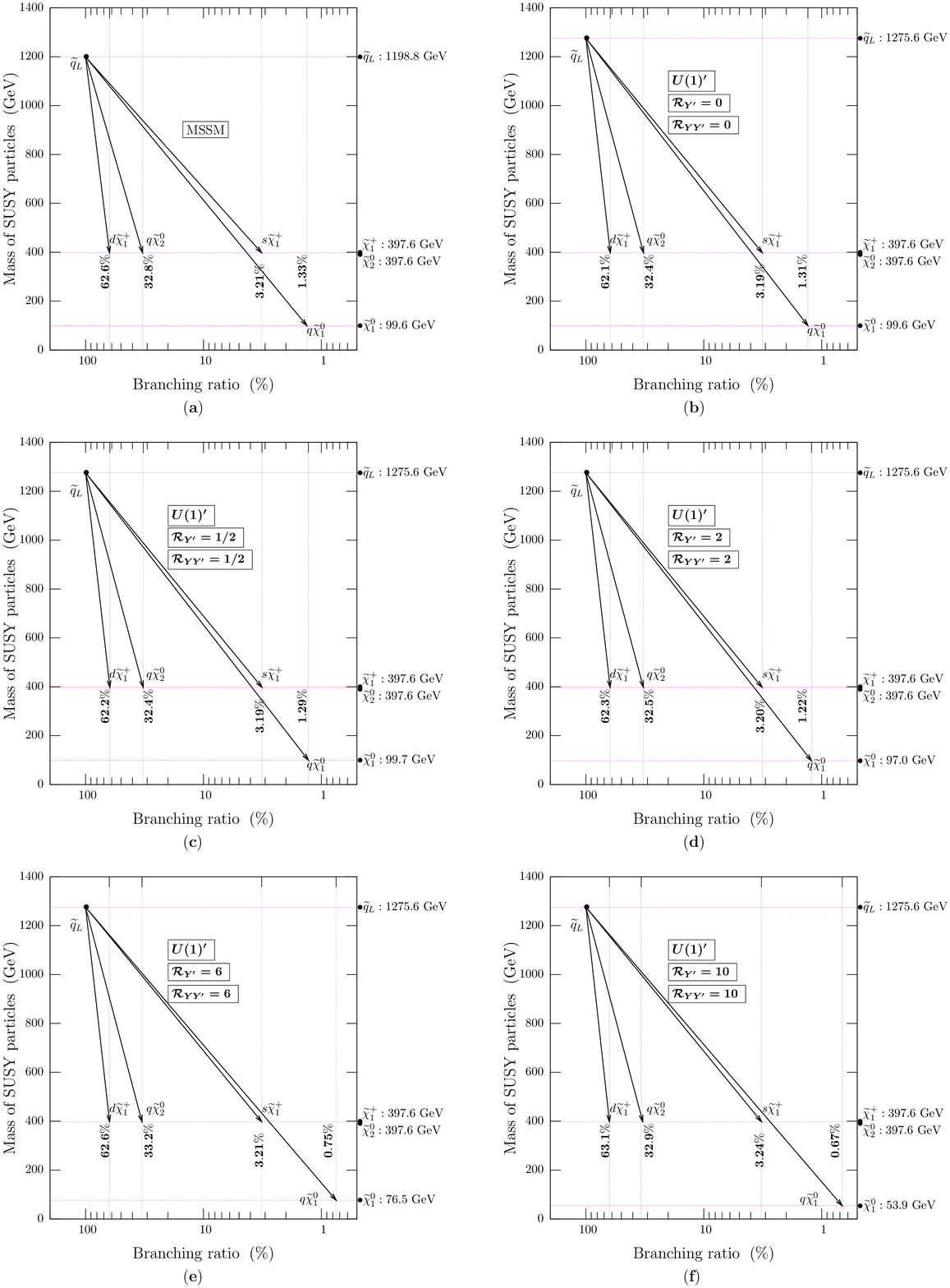}{\it The same
as in Fig. \ref{figqRbranchequal} but for $\widetilde{q}_L$.}
\mfig{htb}{figqLbranchlarge}{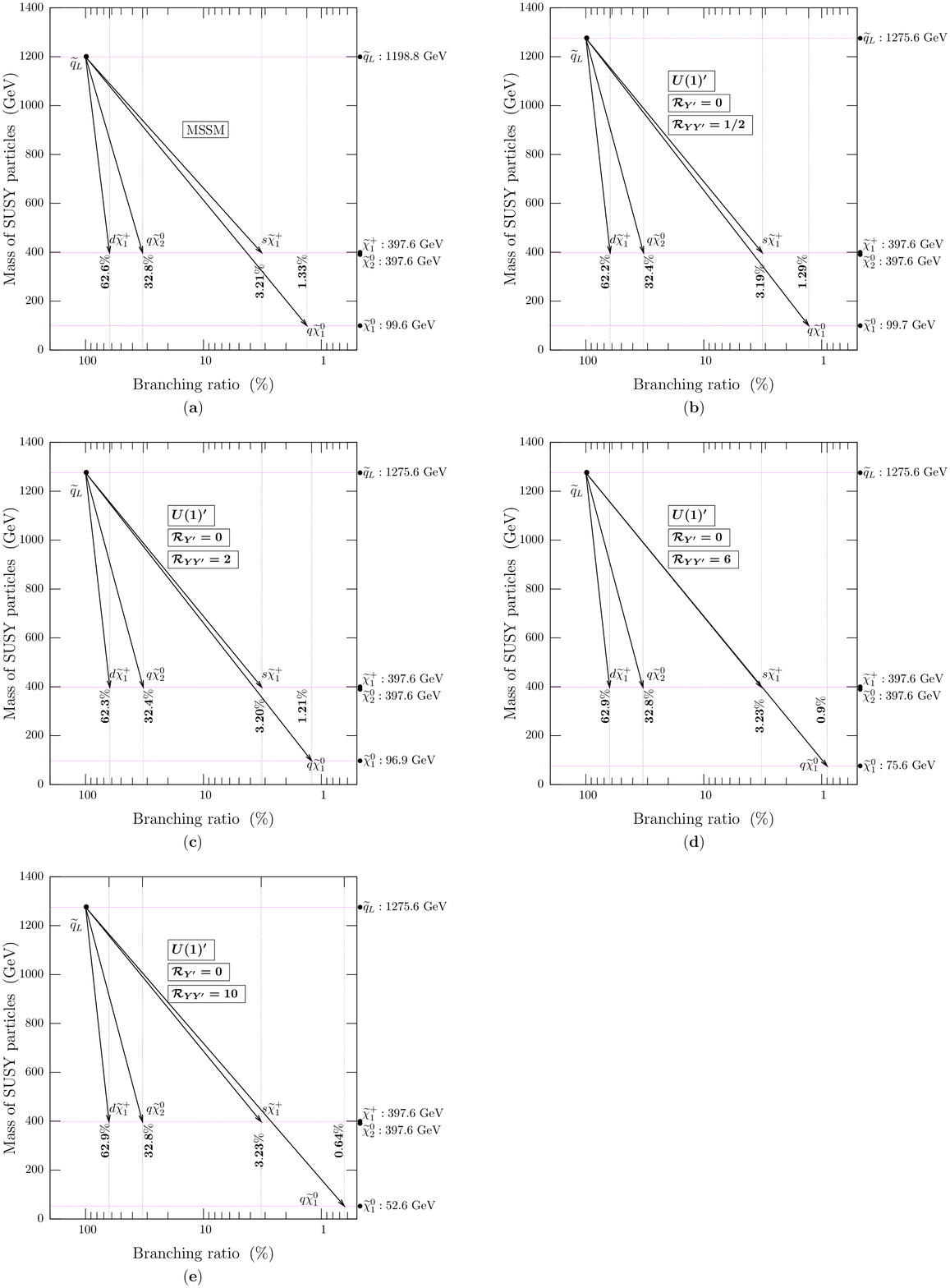}{\it The same
as in Fig. \ref{figqRbranchlarge} but for $\widetilde{q}_L$.}

As illustrated by the panels (a) of Figs.
\ref{figqRbranchsmall}, \ref{figqRbranchequal} and
\ref{figqRbranchlarge}, in the MSSM, a right-handed scalar
quark decays dominantly into the LSP since it is overwhelmingly the bino. This feature of the right-handed squarks gives rise to
$jets + \slashchar{E}_T$ signal at the LHC. By the same token,
a left-handed gluino decays into two quarks and the LSP, and it
thus causes $ 2\ jets + \slashchar{E}_T$ events at the LHC
\cite{abdullin}.

In the $U(1)^{\prime}$ model
the right-handed squarks couple to both the
$\widetilde{B}$ and $\widetilde{Z}^{\prime}$,
opening novel decay channels. These features are explicitly
depicted in Fig. \ref{figqRbranchsmall} (small mixing regime),
Fig. \ref{figqRbranchequal} (medium mixing regime), and Fig.
\ref{figqRbranchlarge} (large mixing regime). As suggested by
these figures, the right-handed squarks develop additional
decay channels with non-negligible branching fractions.

In the small mixing regime of (\ref{U1plist-small}), the
right-handed squark $\widetilde{q}_R$ decays not only into $q\
\widetilde{\chi}_1^0$ but also into $q\ \widetilde{\chi}_3^0$
(whose branching ratio increases with ${\cal
R}_{\widetilde{Y}^{\prime}}$) and $q\ \widetilde{\chi}_4^0$
(whose branching ratio decreases with ${\cal
R}_{\widetilde{Y}^{\prime}}$ as its mass grows to exceed that
of the squark).

In the medium mixing regime of (\ref{U1plist-equal}), the
right-handed squark develops a much larger branching fraction into
$q\ \widetilde{\chi}_3^0$, as shown in Fig.
\ref{figqRbranchequal}. In fact, it reaches the $20\%$ level
when ${\cal R}_{\widetilde{Y}^{\prime}} = {\cal
R}_{\widetilde{Y}\widetilde{Y^{\prime}}}= 10$. This figure is
large enough to make this parameter regime to be explored
further, as will be done in the next subsection.

For the large mixing regime of (\ref{U1plist-large}), the
branching fraction of the decays into $q\ \widetilde{\chi}_3^0$
decreases with increasing ${\cal
R}_{\widetilde{Y}\widetilde{Y^{\prime}}}$, and, as seen from
Fig. \ref{figqRbranchlarge}, eventually vanishes when the decay
channel is closed kinematically at ${\cal
R}_{\widetilde{Y}\widetilde{Y^{\prime}}}= 10$. This extreme is indistinguishable from the MSSM case, shown in
panel (a). This is expected since, all the neutralinos but
$\widetilde{\chi}_1^0$, become too heavy to be produced on-shell by the
squark decay.

These figures make it clear that, in the $U(1)^{\prime}$ model, the
right-handed squarks  can decay into neutralinos other than the LSP. This
feature guarantees that, unlike the purely hadronic events $0\
lepton + jets + \slashchar{E}_T$ expected in the MSSM, in the $U(1)^{\prime}$ model hadronic as well as leptonic events are initiated by the
right-handed squarks. This property,
which will be analyzed in detail in the next subsection, is a golden mode to discover such
extensions. One also notes that the branchings of $\widetilde{q}_R$
significantly differ from that in the MSSM only in the medium
mixing regime, that is, the parameter set
(\ref{U1plist-equal}). In addition, the large mixing regime
of (\ref{U1plist-large}), becomes indistinguishable from the
MSSM case at large ${\cal
R}_{\widetilde{Y}\widetilde{Y^{\prime}}}$.

As illustrated by the panels (a) of Figs.
\ref{figqLbranchsmall}, \ref{figqLbranchequal} and
\ref{figqLbranchlarge}, in the MSSM, a left-handed scalar quark
decays dominantly into quark plus the lighter chargino
$\widetilde{\chi}_1^{\pm}$ or quark plus the next-to-lightest
neutralino $\widetilde{\chi}_2^0$. Therefore, the left-handed
scalar quarks, as analyzed in Sec. III B and listed in
Table~\ref{table1}, give rise to leptonic final states
abundantly. The pure hadronic final states are rather rare
\cite{abdullin}.

Table~\ref{massneut} shows that the mass of $\widetilde{\chi}_2^{0}$
remains stuck to its MSSM value, to an excellent
approximation. The lighter chargino, which
is $\widetilde{W}^{\pm}$ dominated, is not expected to deviate
from its MSSM mass. Consequently, the $U(1)^{\prime}$ effects
are not expected to cause dramatic changes from the branching
fractions of $\widetilde{q}_L$ in the MSSM. This  is
seen to be the case from  Figs.
\ref{figqLbranchsmall}, \ref{figqLbranchequal} and
\ref{figqLbranchlarge} corresponding to small, medium and
large mixings among $U(1)_Y$ and $U(1)_Y^{\prime}$ gauginos, respectively,
 clearly showing  that the decay channels of the left-handed
squarks are nearly immune to the $U(1)^{\prime}$ effects.
The conclusion from this subsection is that the
$U(1)^{\prime}$-effects become visible mainly in the fermionic
decays of the right-handed scalar quarks, but not in the left-handed
ones. The medium mixing regime of (\ref{U1plist-equal})
stands as a particularly promising parameter domain for hunting
the $U(1)^{\prime}$ effects.


\subsection{The LHC Signatures of the $U(1)^{\prime}$ Model Through $jets + leptons + \slashchar{E}_T$ Events}
Having computed the squark branching ratios in the previous
section, we now turn to the analysis of various final states to be
searched for by the ATLAS and CMS experiments at the LHC. We
perform a simulation study of a number of LHC events for the
MSSM and the $U(1)^{\prime}$ model in a comparative fashion. The
scattering processes of interest have the generic form
\begin{eqnarray}
\label{process}
p\, p \rightarrow X + \mbox{SIGNAL}\,,
\end{eqnarray}
where $\mbox{SIGNAL}$ stands for the particular final state
characterizing the event. An optimal coverage of the events for
which the MSSM and the $U(1)^{\prime}$ model can exhibit striking
differences are classified in Table~\ref{table3}.
\begin{table}[htb]
\begin{tabular*}{0.9\textwidth}{@{\extracolsep{\fill}} lll}
\hline\hline $\;$ SIGNAL & FINAL STATE &  CANDIDATE PROCESSES FOR $N_{jets}=2$\\\hline\hline
$\;$ SIGNAL 1 & $ 0\, \ell +  jets + \slashchar{E}_T$ & $p\, p \rightarrow \left(\widetilde{q} \rightarrow
q\, \widetilde{\chi}_1^0\right)\, \left(\widetilde{q} \rightarrow q\, \widetilde{\chi}_1^0\right)$ \\ \hline \hline
$\;$ SIGNAL 2 & $ 1\, \ell + jets + \slashchar{E}_T$ & $p\, p \rightarrow \left(\widetilde{q} \rightarrow
q^{\prime}\, \ell \overline{\nu}_{\ell} \widetilde{\chi}_1^0\right)\, \left(\widetilde{q} \rightarrow q\, \widetilde{\chi}_1^0\right)$ \\ \hline
$\begin{array}{c}
 {\; \rm SIGNAL\;3A} \\
{\; \rm SIGNAL\;3B}\\
 \end{array}$ & $ 2\, \ell + jets + \slashchar{E}_T$ & $
\begin{array}{l} p\, p \rightarrow \left(\widetilde{q} \rightarrow
q^{\prime}\, \ell \overline{\nu}_{\ell} \widetilde{\chi}_1^0\right)\, \left(\widetilde{q} \rightarrow
q^{\prime}\, \ell \overline{\nu}_{\ell} \widetilde{\chi}_1^0\right) \\
p\, p \rightarrow \left(\widetilde{q} \rightarrow
q\, \ell^+ \ell^- \widetilde{\chi}_1^0\right)\, \left(\widetilde{q} \rightarrow q\, \widetilde{\chi}_1^0\right)
\end{array}$\\ \hline
$\begin{array}{l}
{\; \rm SIGNAL\;4A} \\
{\; \rm SIGNAL\;4B}\\
 \end{array}$
 & $ 3\, \ell + jets + \slashchar{E}_T$ & $\begin{array}{c} p\, p \rightarrow \left(\widetilde{q} \rightarrow
q^{\prime}\, \ell \overline{\nu}_{\ell} {\ell^{\prime\, +}} {\ell^{\prime\, -}} \widetilde{\chi}_1^0\right)\, \left(\widetilde{q} \rightarrow q\,
\widetilde{\chi}_1^0\right)\\
p\, p \rightarrow \left(\widetilde{q} \rightarrow
q^{\prime}\, \ell \overline{\nu}_{\ell} \widetilde{\chi}_1^0\right)\, \left(\widetilde{q} \rightarrow q\,
 {\ell^{\prime\, +}} {\ell^{\prime\, -}} \widetilde{\chi}_1^0\right)
\end{array}$ \\ \hline\hline
\end{tabular*}
\caption{\label{table3} The basic LHC signals simulated with
Monte Carlo event generators. Here $\ell = e$ or $\mu$, and
`$jets$' stands for any number of jets in the final state. Each
signal receives contributions from one or more decay processes,
the strengths of which change as one switches from the MSSM to the
$U(1)^{\prime}$ Model. The candidate processes listed here
involve only $N_{jets}=2$;  the signals started by
gluinos, which cause more jets than $N_{jets}=2$, are not
shown.}
\end{table}
We compute the
cross sections and branching ratios, and generate parton--level
events by using {\tt CalcHEP v.2.5} \cite{calchep}. We modified the package to incorporate
the  features pertaining to the $U(1)^{\prime}$ model with the help of
 {\tt LanHEP} Package \cite{Semenov:2008jy}. Hadronization (including
initial and final state radiations) and restrictions imposed by
various cuts have been achieved with {\tt PYTHIA} \cite{pythia} by using the
{\tt CalcHEP-PYTHIA} interface. The parton distributions in the
proton have been parametrized by using {\tt CTEQ6L} of {\tt
LHAPDF}. The number of events are calculated for an
integrated luminosity  ${\cal{L}}=100\
{\rm fb}^{-1}$, for which the LHC has a sensitivity to the
squark and gluino masses around $2.5\ {\rm TeV}$
\cite{abdullin}. Our goal here is to determine how the MSSM and the
$U(1)^{\prime}$ model differ in their predictions for the
signals in Table~\ref{table3}, driven by the
presence of the extra gauge and Higgs fermions. A detailed
background analysis is not warranted in this work since its
main goal is to compare the MSSM and the $U(1)^{\prime}$ model
predictions for the signal events under
consideration. Nonetheless, as a set of generic cuts for
revealing 'new physics' effects (compared to the SM ones), we
select only those events satisfying the following restrictions:
\begin{itemize}
\item Each charged lepton in the final state must have a
    transverse momentum $p_{T}^{\ell} > 15\ {\rm GeV/c}$.

\item Each jet must have a transverse momentum $p_{T}^{jet}
    > 20\ {\rm GeV/c}$.

\item The missing transverse energy must satisfy
    $\slashchar{E}_T \geq 100\ {\rm GeV}$.

\item The particles at the final state propagate in the
    transverse direction so that the pseudorapidity  stays in
    the interval $-2 \leq \eta \leq 2$.

\item The initiator energy of jets is $2\ {\rm GeV}$.

\item Two jetted showers of particles are taken to be two
    distinct jets if their spatial separation satisfies
    $\Delta R_{jj}>0.7$.
\end{itemize}

We now perform a full generator-level analysis of the events
tabulated in Table~\ref{table3} by taking into account the
generation and decays of all the squarks in the first and
second generations as well as the gluino via the $p\ p$
scatterings in (\ref{process}). We use the Feynman rules in
Appendix D, compute the populations of the events in
Table~\ref{table3}, and plot the results against various
observables of interest at the LHC. The analysis performs a
comparative study between the MSSM and the $U(1)^{\prime}$ model in
regard to their predictions for the processes in
Table~\ref{table3}. Concerning the parameter choice, we take the
$U(1)^{\prime}$ model to be in the medium mixing regime of
(\ref{U1plist-equal}), and consider the two points
\begin{eqnarray}
\label{take}
\left({\cal R}_{\widetilde{Y}^{\prime}}\, , \, {\cal
R}_{\widetilde{Y}\widetilde{Y^{\prime}}}\right) = \left(0, 0\right)\; \mbox{and}\; \left(10, 10\right)\,,
\end{eqnarray}
in all the figures that follow. These two points are picked up
on the basis of highlighting the $U(1)^{\prime}$ effects in
comparison to those of the MSSM.

Among the signals listed in Table~\ref{table3}, the signal $3\,
\ell + 2\, jet + \slashchar{E}_T$ (SIGNAL 4A), where all
leptons originate from the same branch, is not considered
further in the numerical analysis. This is due to the fact that
this signal requires a decay chain like in Eq.~(\ref{4amode})
and since we use narrow-width approximation, the scalar
neutrino $\widetilde{\nu}_\ell$ (taken to be relatively light)
has to decay through a  4-body decay  $\widetilde{\nu}_\ell\to
\bar{\nu}_\ell \ell^{\prime +}\ell^{\prime -}\widetilde{B}$
with a tiny branching ratio. Thus, the signal will be much
suppressed as compared with the others. This observation is consistent with the region of the parameter space considered here,
since for instance, scalar neutrinos heavier than
$\widetilde{\chi}_2^0$ and scalar leptons would make it
competitive with the others.

The observables with respect to which we analyze the number of
events are as follows:
\begin{itemize}
\item The number of jets $N_{jets}$ with \binsize $= 1\ {\rm
    GeV}$,

\item The transverse energy of the jets $E_{T}^{jets}$ with
    \binsize $= 3\ {\rm GeV}$,

\item The missing transverse energy $\slashchar{E}_{T}$
    with \binsize $= 20\ {\rm GeV}$,

\item The scalar sum of the transverse energies of the jets and
    leptons $E_T^{sum}$ with \binsize $= 40\ {\rm GeV}$,

\item The transverse momentum of the hardest lepton
    $p_{T}\left(\ell_{hard}\right)$ with \binsize $= 10\
    {\rm GeV}$,

\item The dilepton invariant mass $M_{\rm inv}(\ell \ell)$
    with \binsize $= 19\ {\rm GeV}$.
\end{itemize}
Distributions with respect to these variables are expected to
provide a global picture of the distinctive features of the
events in Table~\ref{table3} in regard to a comparative
analysis of the MSSM and the $U(1)^{\prime}$ models.

\begin{figure}[htb]
 \centering
 \epsfysize=2.4in
 \hspace*{-0.4in}
 \epsffile{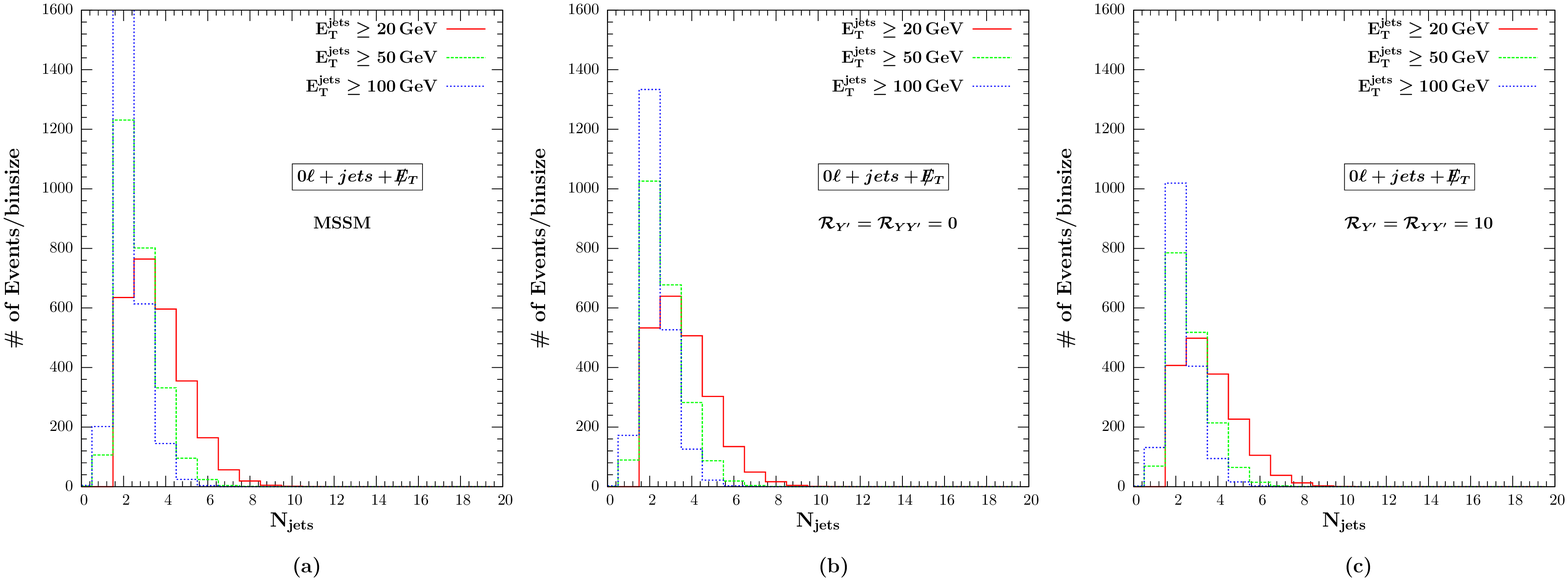}
 \caption{\it The binwise (\binsize = $1\ {\rm GeV}$) distribution of the number of purely hadronic events (the events of the type
 $\mbox{SIGNAL 1}$ in Table~\ref{table3}) with the number of jets $N_{jets}$ for different $E_{T}^{jets}$
 ranges at an integrated luminosity of ${\cal{L}} = 100\ {\rm fb}^{-1}$ in the MSSM (panel (a))
 and in the $U(1)^{\prime}$ model with $(\ry,\ryy)=(0,0)$ (panel (b)) and $(\ry,\ryy)=(10, 10)$
 (panel (c)). The number of hadronic events, in agreement with the discussions of Sec. III B,
 are depleted in the $U(1)^{\prime}$ model compared to the MSSM. It is clear that the larger the transverse
 energy of the jets the closer the event is to dijet type.}
 \label{signal1-njets}
 \end{figure}

\begin{figure}[htb]
 \centering
 \epsfysize=2.4in
 \hspace*{-0.4in}
 \epsffile{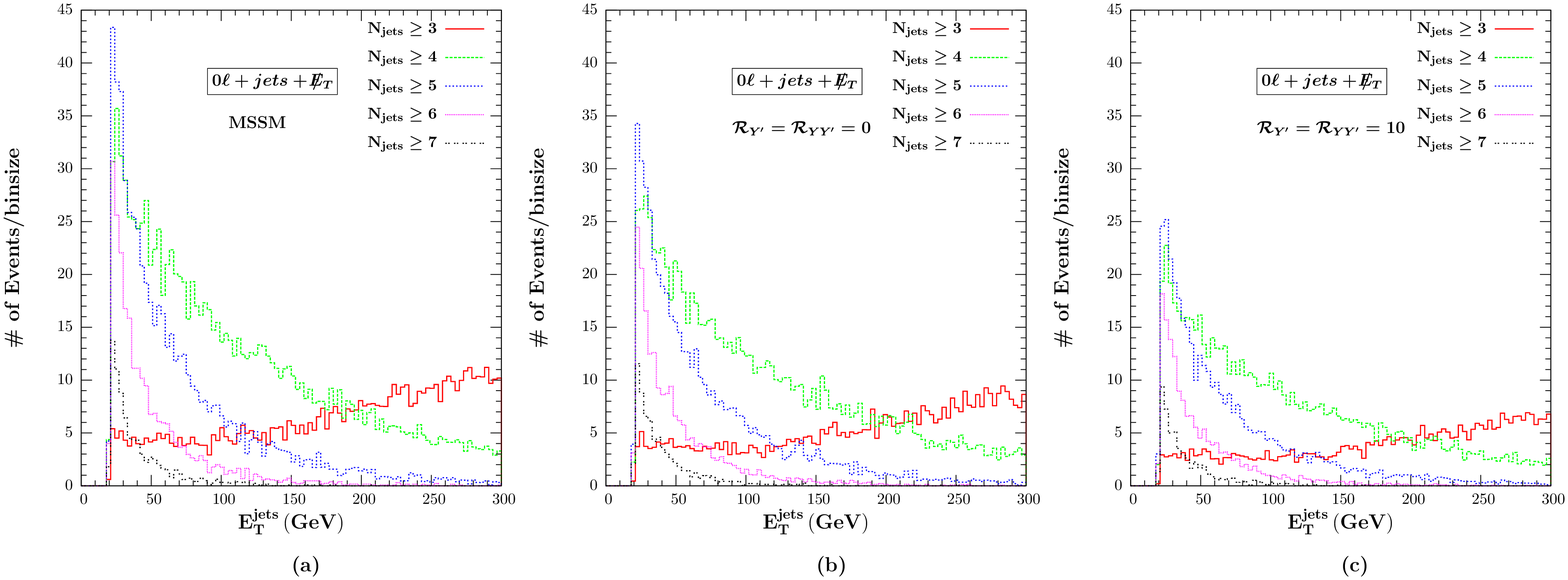}
 \caption{\it The binwise (\binsize $= 3\ {\rm GeV}$) distribution of the number of purely hadronic events (the events of the type
 $\mbox{SIGNAL 1}$ in Table~\ref{table3}) with $E_{T}^{jets}$ for different $N_{jets}$ ranges at an
 integrated luminosity of ${\cal{L}} = 100\ {\rm fb}^{-1}$ in the MSSM (panel (a))
 and in the $U(1)^{\prime}$ model with $(\ry,\ryy)=(0,0)$ (panel (b)) and $(\ry,\ryy)=(10, 10)$
 (panel (c)). The events with $N_{jets}\geq 4$ are soft (they are abundant only at low $E_{T}^{jets}$)
 and rare (they are few at large $E_{T}^{jets}$). The events with smaller numbers of jets are effective for a wide
 range of $E_{T}^{jets}$ values. In accord with Fig.~\ref{signal1-njets}, the purely hadronic events in $U(1)^{\prime}$
 model are fewer than in the MSSM, especially for the large $N_{jets}$ values.}
 \label{signal1-etjets}
 \end{figure}

\begin{figure}[htb]
 \centering
 \epsfysize=2.6in
 \hspace*{-0.4in}
 \epsffile{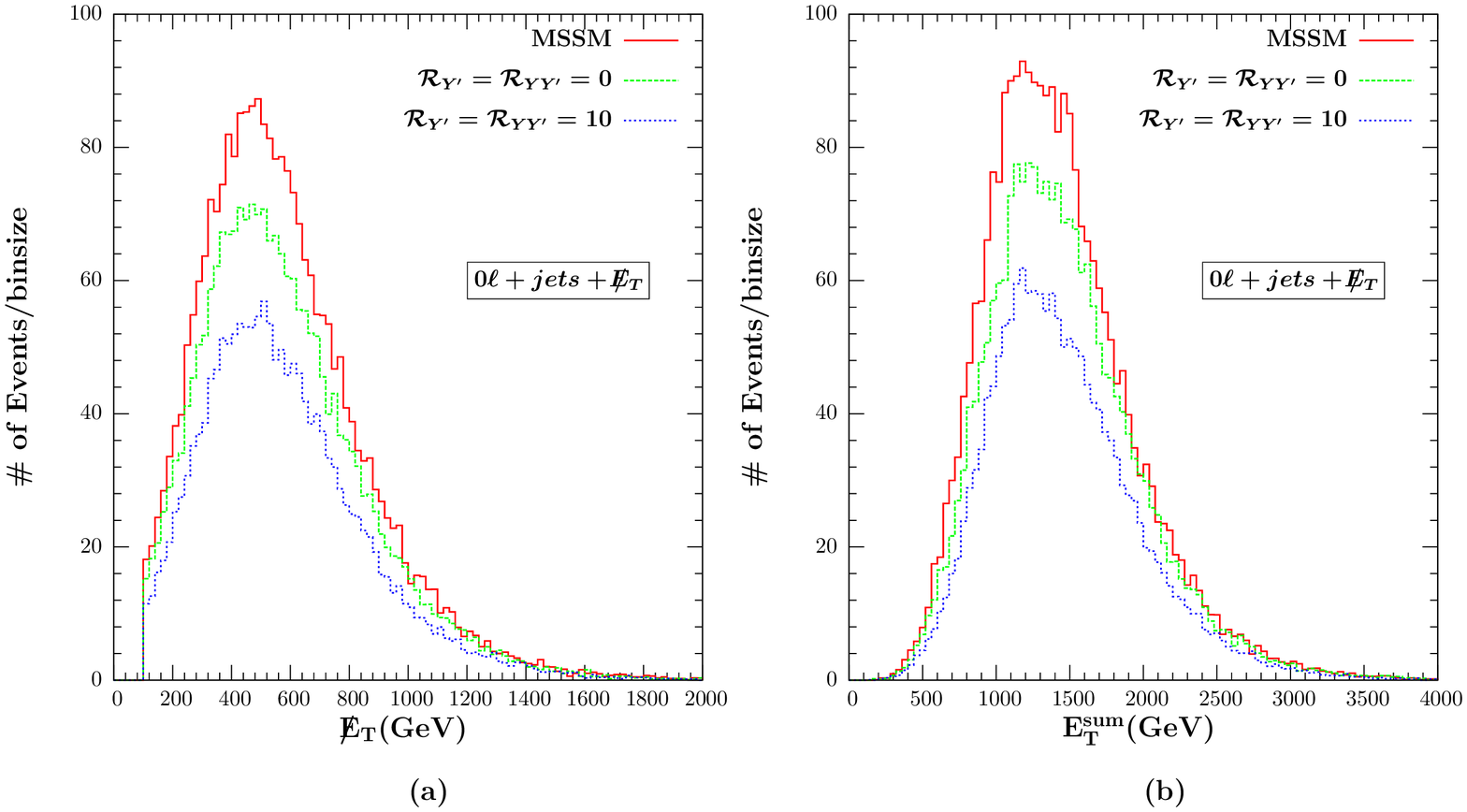}
 \caption{\it The binwise distribution of the number of purely hadronic events (the events of the type
 $\mbox{SIGNAL 1}$ in Table~\ref{table3}) with $\slashchar{E}_{T}$ (panel (a),
\binsize $= 20\ {\rm GeV}$) and $E_{T}^{sum}$ (panel (b),
 \binsize $= 40\ {\rm GeV}$ ) at an integrated luminosity of ${\cal{L}} = 100\ {\rm fb}^{-1}$ in the
 MSSM and the $U(1)^{\prime}$ model. The
 central values of the distributions vary little from model to model. Nevertheless, the number of
 events are fewer in the $U(1)^{\prime}$ model than in the MSSM. This feature is
 in accordance with the discussion in Sec. III B and with
 Figs.~\ref{signal1-njets} and \ref{signal1-etjets}.}
 \label{signal1-rest}
 \end{figure}

\begin{figure}[htb]
 \centering
 \epsfysize=2.4in
 \hspace*{-0.4in}
 \epsffile{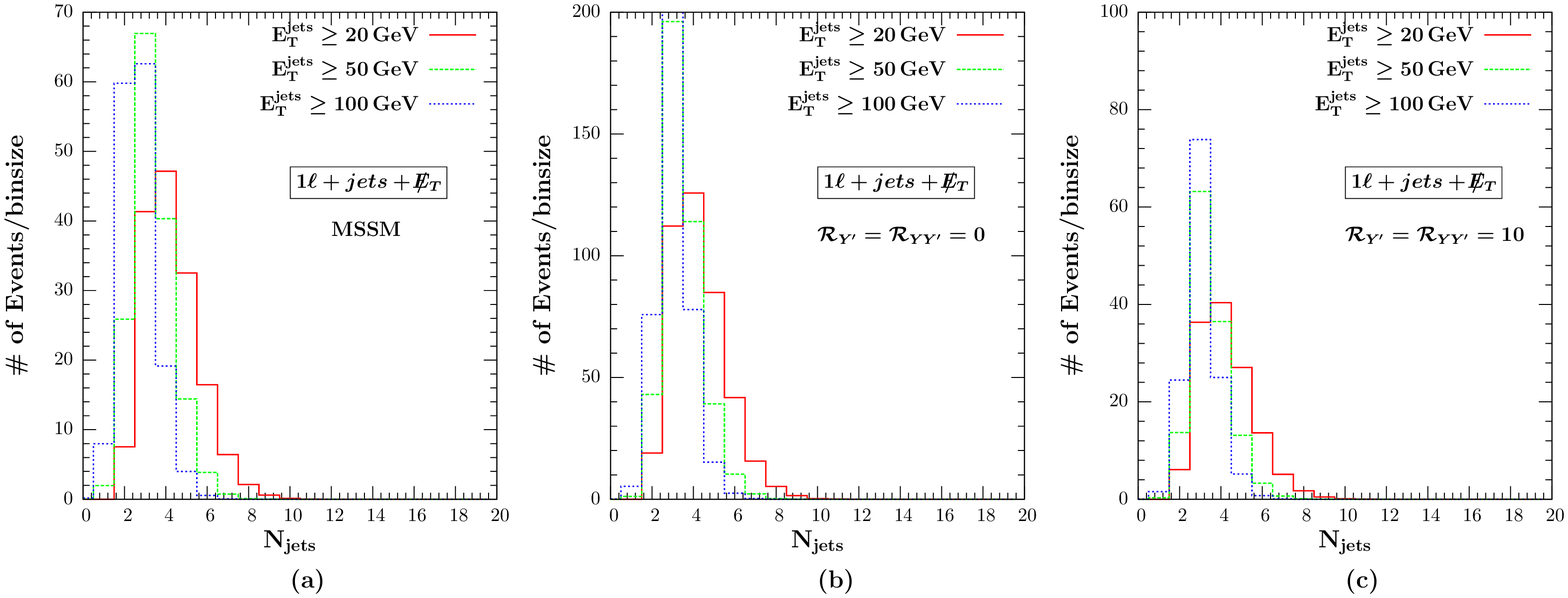}
 \caption{\it The same as in Fig.~\ref{signal1-njets} but for the single-lepton
 events (i.e.,  events of the type  $\mbox{SIGNAL 2}$ in Table~\ref{table3}).}
 \label{signal2-njets}
 \end{figure}

\begin{figure}[htb]
 \centering
 \epsfysize=2.4in
 \hspace*{-0.4in}
 \epsffile{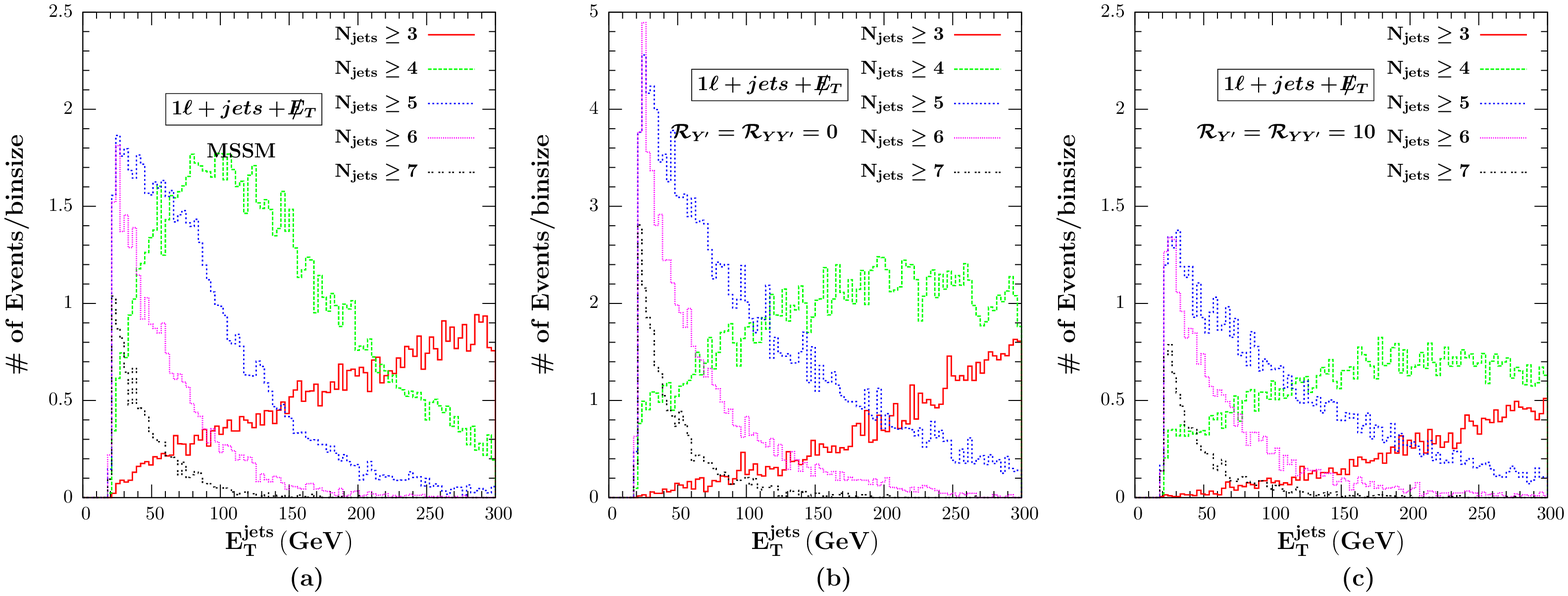}
 \caption{\it \it The same as in Fig.~\ref{signal1-etjets} but for the single-lepton
 events (i.e., events of the type  $\mbox{SIGNAL 2}$ in Table~\ref{table3}).}
 \label{signal2-etjets}
 \end{figure}

\begin{figure}[htb]
 \centering
 \epsfysize=2.4in
 \hspace*{-0.4in}
 \epsffile{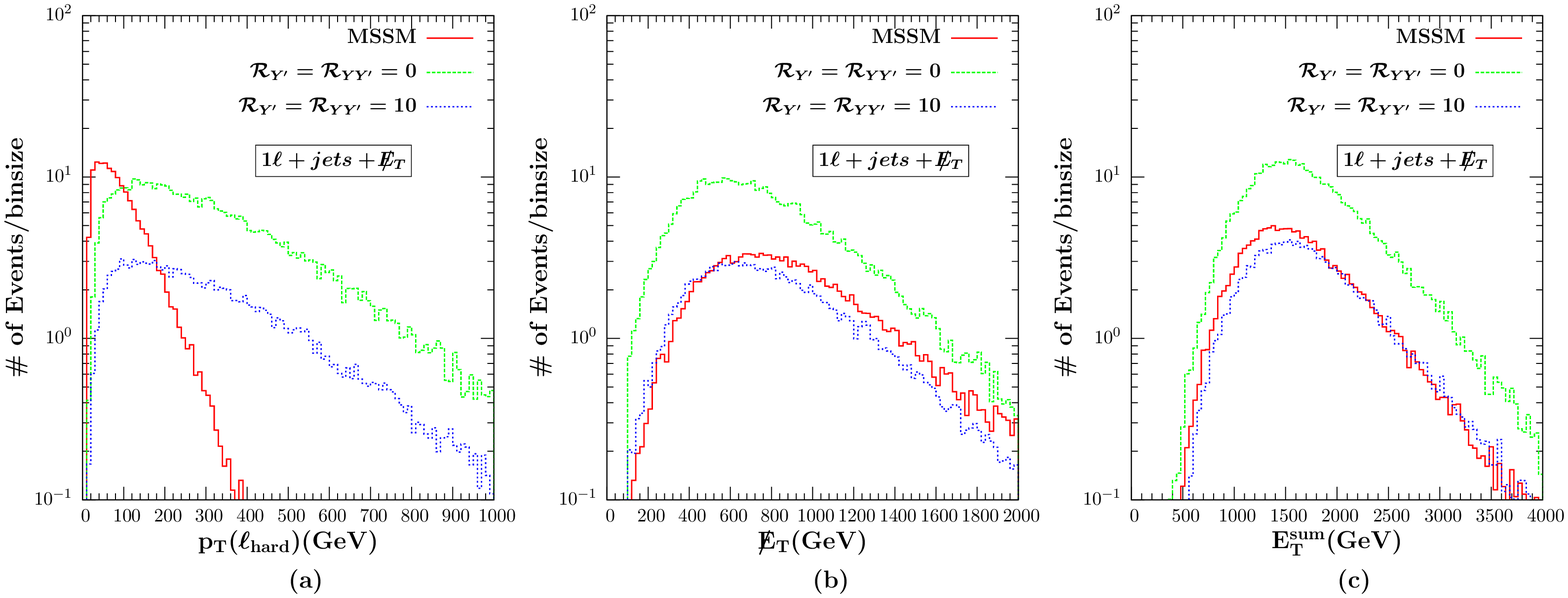}
 \caption{\it The same as in Fig.~\ref{signal1-rest} but for the single-lepton
 events (i.e., events of the type  $\mbox{SIGNAL 2}$ in Table~\ref{table3}). The
 panel (a) (\binsize $= 10\ {\rm GeV}$) is new; it describes the distribution with respect to the transverse
 momentum of the emitted lepton.}
 \label{signal2-rest}
 \end{figure}

\begin{figure}[htb]
 \centering
 \epsfysize=2.4in
 \hspace*{-0.4in}
 \epsffile{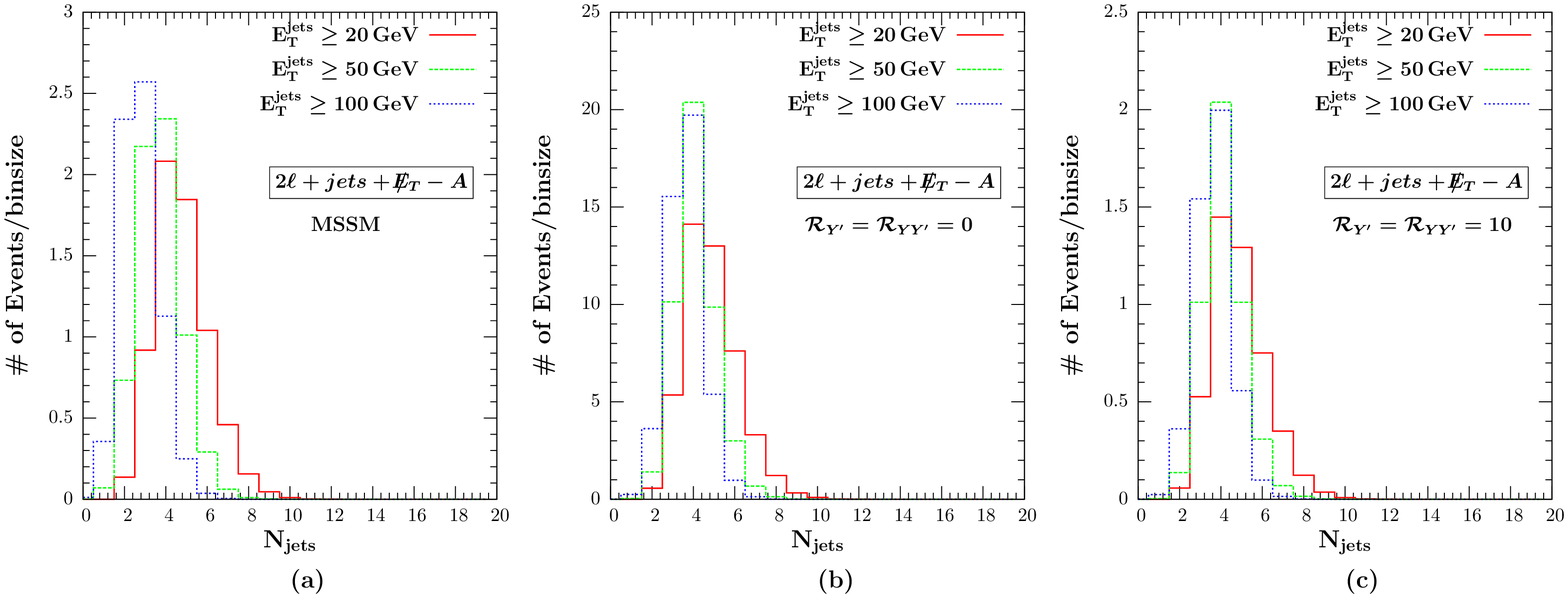}
 \caption{\it The same as in Fig.~\ref{signal1-njets} but for the dilepton
 events (i.e., events of the type  $\mbox{SIGNAL 3A}$ in Table~\ref{table3}).}
 \label{signal3a-njets}
 \end{figure}

\begin{figure}[htb]
 \centering
 \epsfysize=2.4in
 \hspace*{-0.4in}
 \epsffile{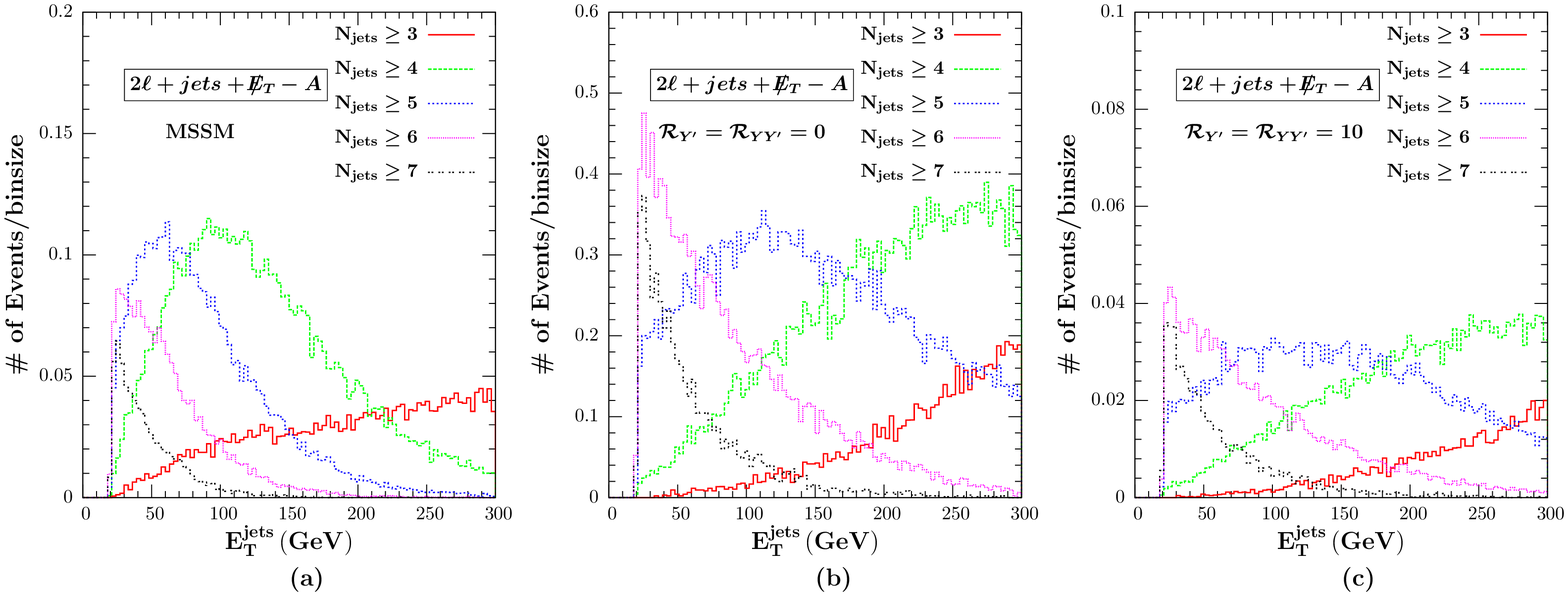}
 \caption{\it The same as in Fig.~\ref{signal1-etjets} but for the dilepton
 events (i.e., events of the type  $\mbox{SIGNAL 3A}$ in Table~\ref{table3}).}
 \label{signal3a-etjets}
 \end{figure}

\begin{figure}[htb]
 \centering
 \epsfysize=7.0in
 \hspace*{-0.4in}
 \epsffile{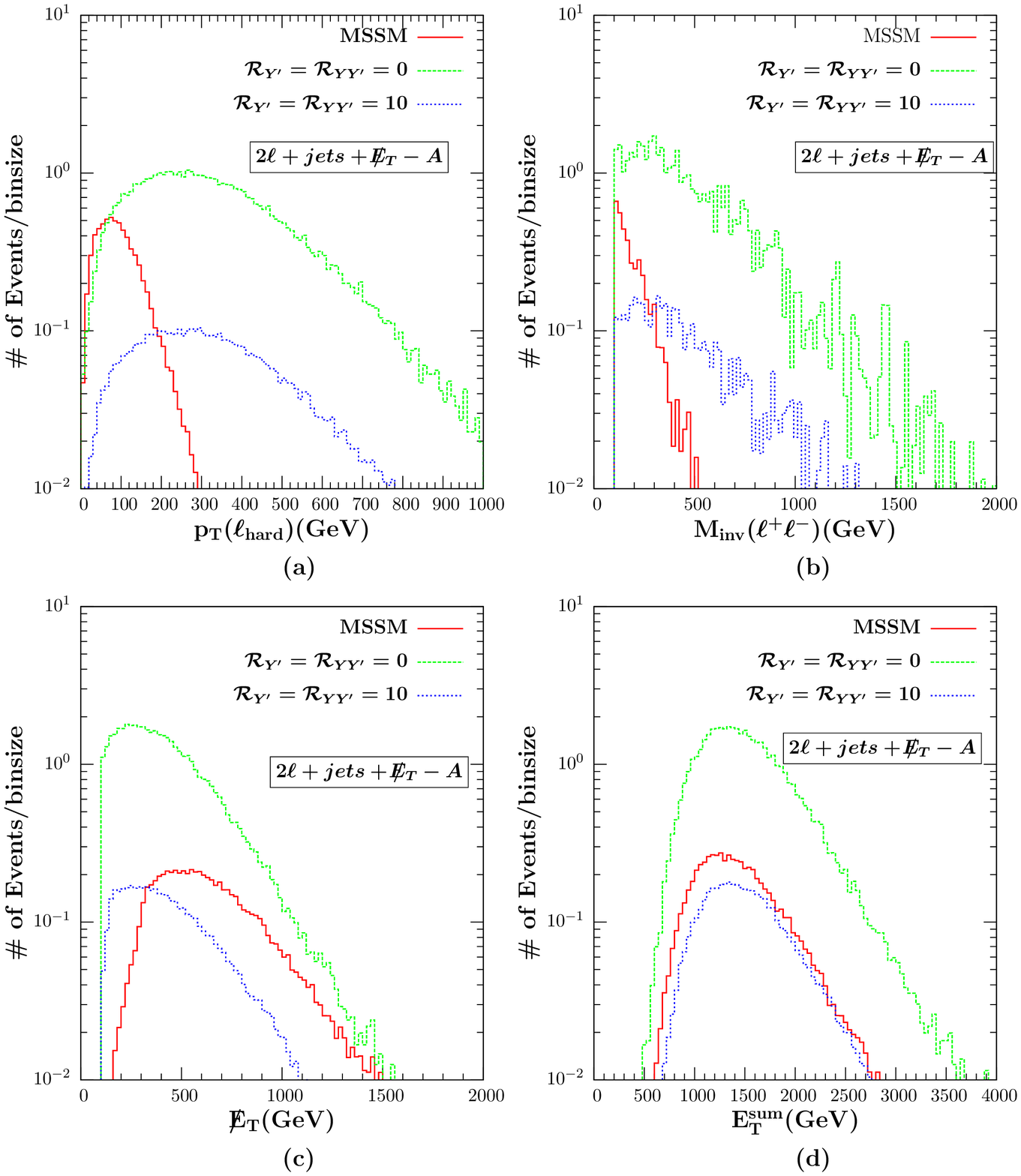}
 \caption{\it The same as in Fig.~\ref{signal2-rest} but for the dilepton
 events (i.e., events of the type  $\mbox{SIGNAL 3A}$ in Table~\ref{table3}). The
 new features compared to those in Fig. \ref{signal2-rest} are as follows:
 The panel (a) describes the distribution with respect to the transverse
 momentum of the hardest lepton, $p_T\left(\ell_{hard}\right)$. The panel
 (b) (\binsize $= 19\ {\rm GeV}$) is new; it describes the distribution with respect to the invariant
 mass of the two emitted leptons, $M_{\rm inv}\left(\ell^+ \ell^-\right)$.}
 \label{signal3a-rest}
 \end{figure}

\begin{figure}[htb]
 \centering
 \epsfysize=2.4in
 \hspace*{-0.4in}
 \epsffile{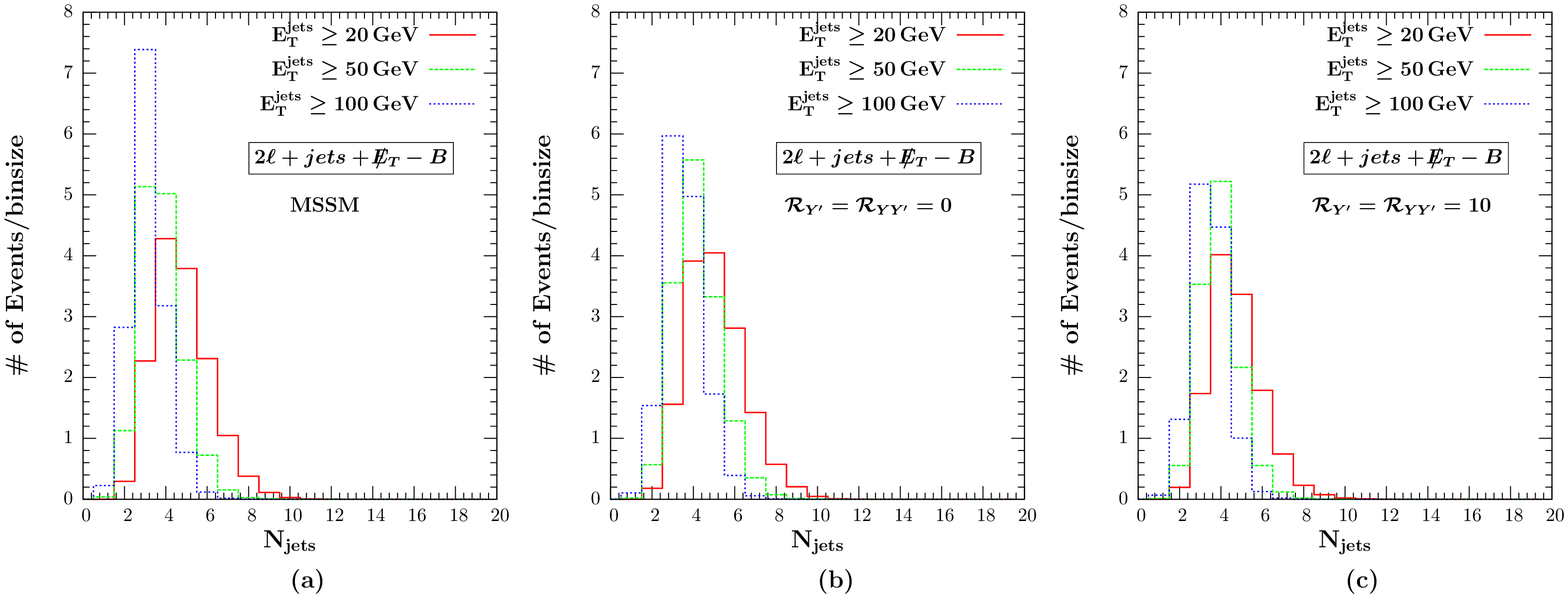}
 \caption{\it The same as in Fig.~\ref{signal1-njets} but for the dilepton
 events (i.e., events of the type  $\mbox{SIGNAL 3B}$ in Table~\ref{table3}).}
 \label{signal3b-njets}
 \end{figure}

\begin{figure}[htb]
 \centering
 \epsfysize=2.4in
 \hspace*{-0.4in}
 \epsffile{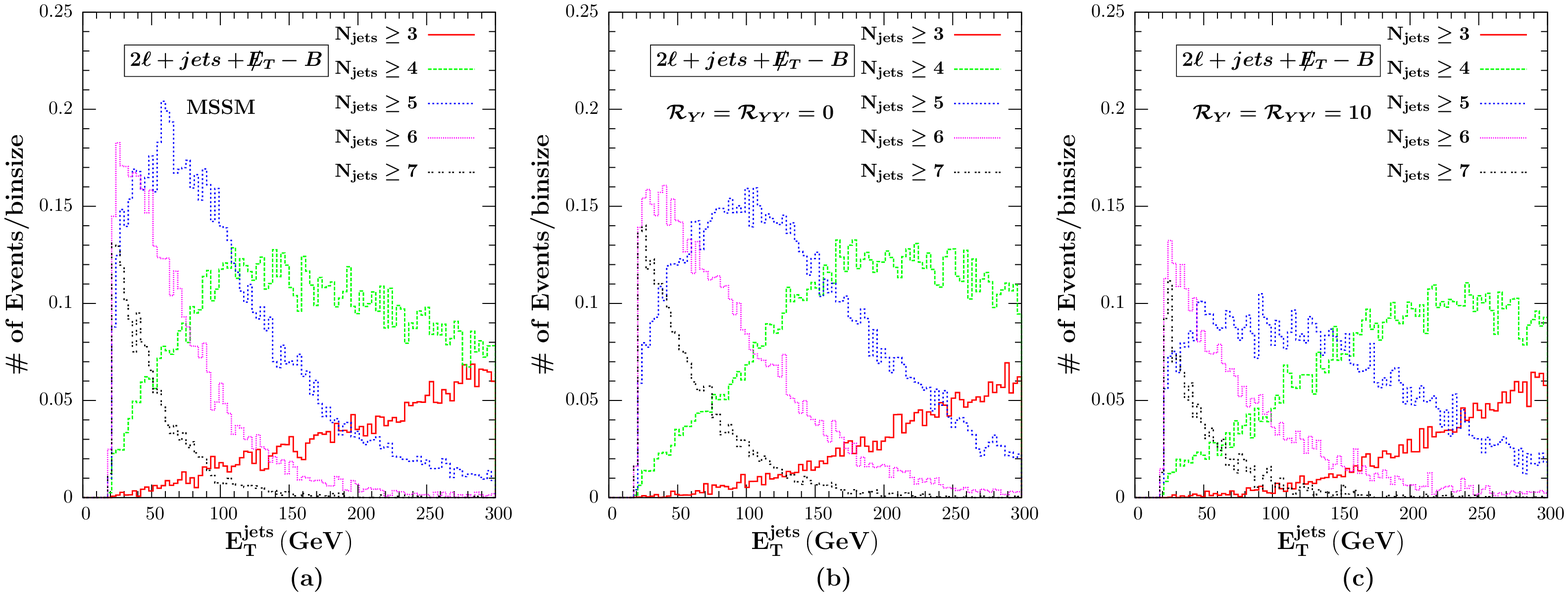}
 \caption{\it The same as in Fig.~\ref{signal1-etjets} but for the dilepton
 events (i.e., events of the type  $\mbox{SIGNAL 3B}$ in Table~\ref{table3}).}
 \label{signal3b-etjets}
 \end{figure}

\begin{figure}[htb]
 \centering
 \epsfysize=7.0in
 \hspace*{-0.4in}
 \epsffile{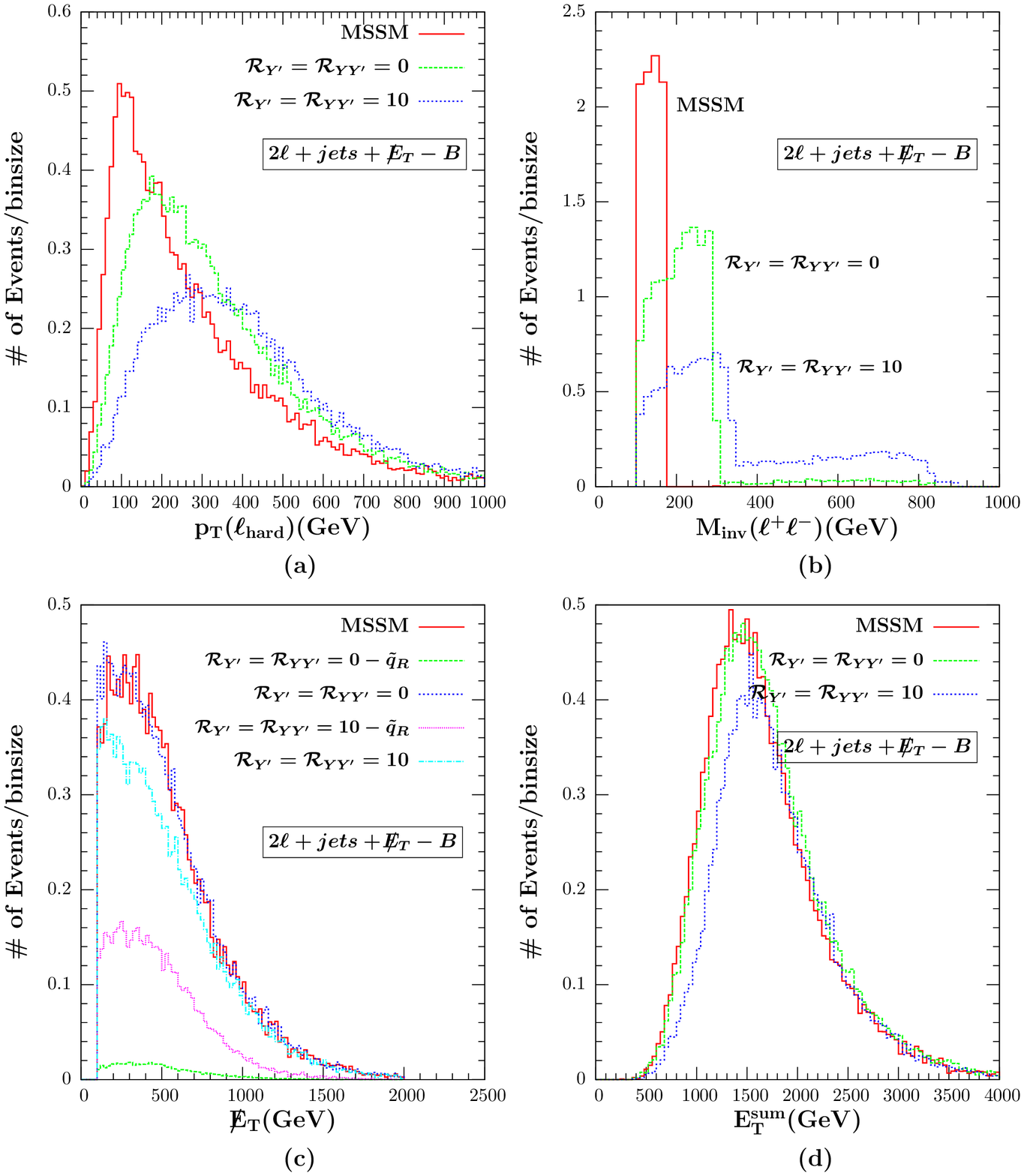}
 \caption{\it The same as in Fig.~\ref{signal3a-rest} but for the dilepton
 events (i.e., events of the type  $\mbox{SIGNAL 3B}$ in Table~\ref{table3}).}
 \label{signal3b-rest}
 \end{figure}

\begin{figure}[htb]
 \centering
 \epsfysize=2.4in
 \hspace*{-0.4in}
 \epsffile{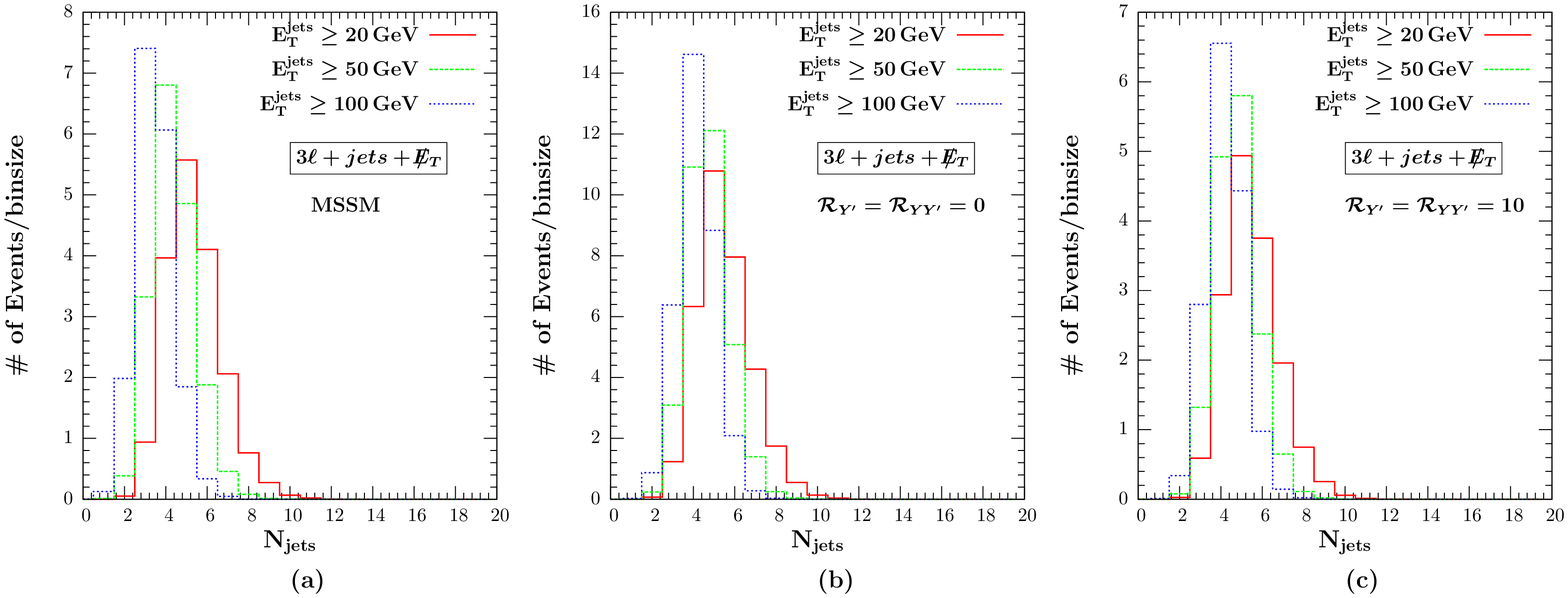}
 \caption{\it The same as in Fig.~\ref{signal1-njets} but for the trilepton
 events (i.e., events of the type  $\mbox{SIGNAL 4B}$ in Table~\ref{table3}).}
 \label{signal4b-njets}
 \end{figure}

\begin{figure}[htb]
 \centering
 \epsfysize=2.4in
 \hspace*{-0.4in}
 \epsffile{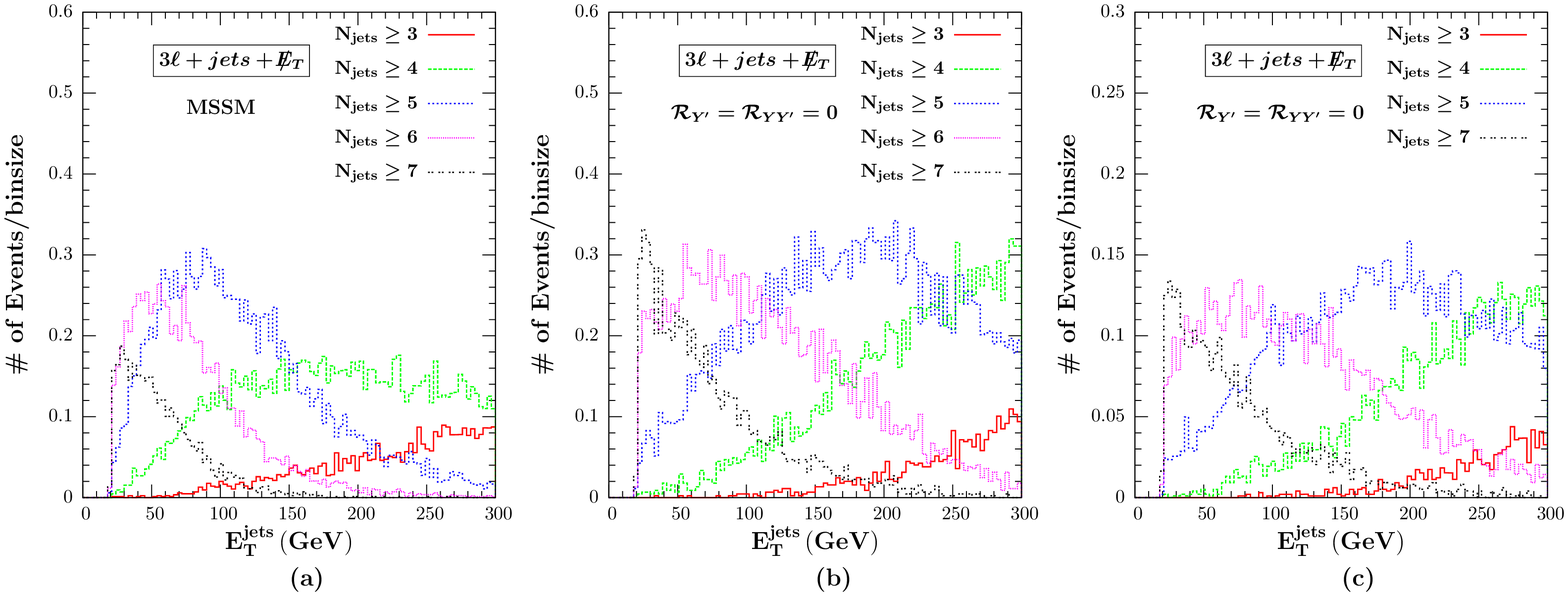}
 \caption{\it The same as in Fig.~\ref{signal1-etjets} but for the trilepton
 events (i.e., events of the type  $\mbox{SIGNAL 4B}$ in Table~\ref{table3}).}
 \label{signal4b-etjets}
 \end{figure}

\begin{figure}[htb]
 \centering
 \epsfysize=7.0in
 \hspace*{-0.4in}
 \epsffile{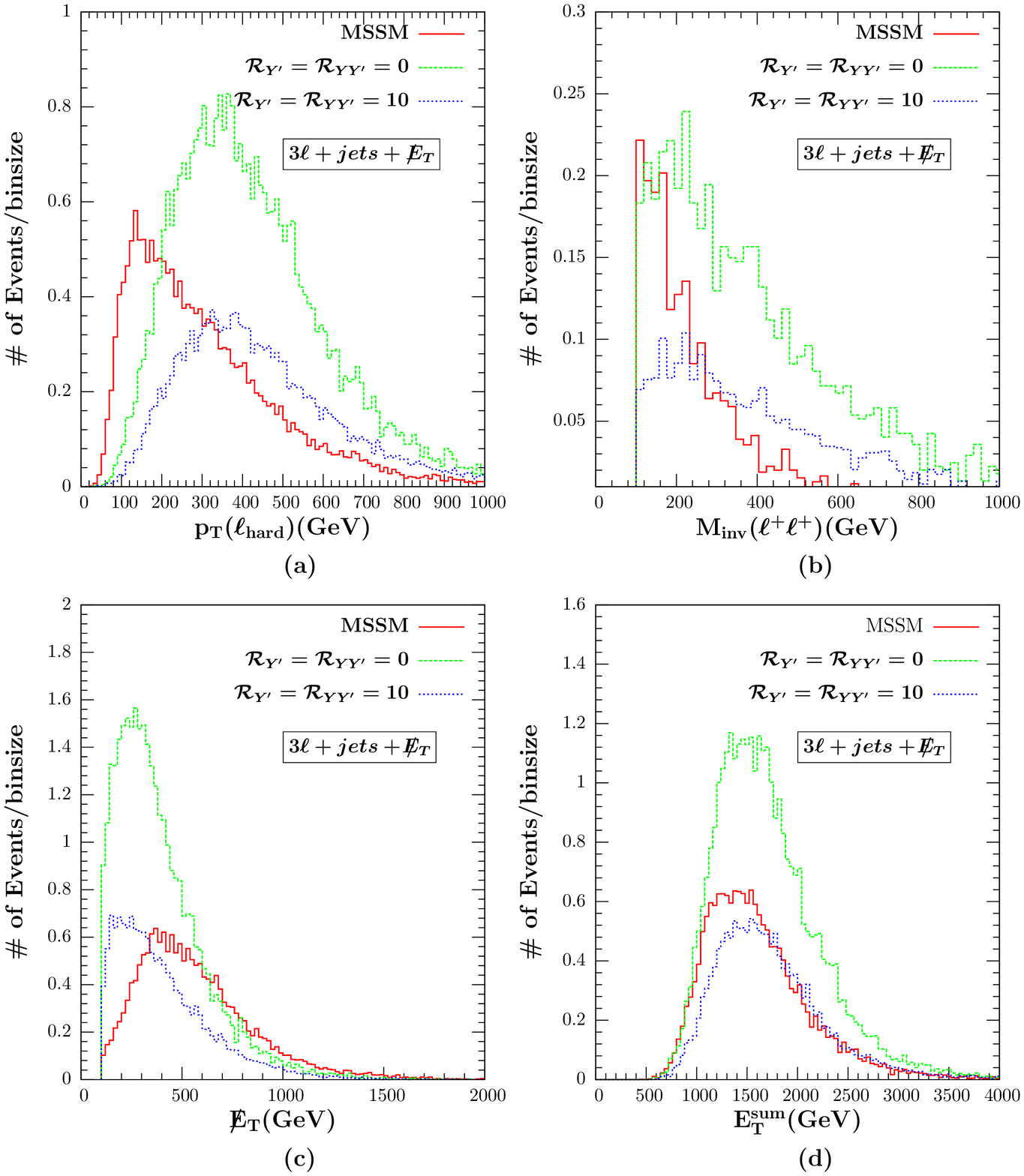}
 \caption{\it The same as in Fig.~\ref{signal3a-rest} but for the trilepton
 events (i.e., events of the type  $\mbox{SIGNAL 4B}$ in Table~\ref{table3}). The
 panel (b) is different than those in Figs.~\ref{signal3a-rest} and \ref{signal3b-rest} in that
 it describes the distribution with respect to the invariant
 mass of the two same-charge leptons, $M_{\rm inv}\left(\ell^+ \ell^+\right)$.}
 \label{signal4b-rest}
 \end{figure}

The SIGNAL 1 in Table~\ref{table3} is analyzed in Figs.
\ref{signal1-njets}, \ref{signal1-etjets} and
\ref{signal1-rest}. Similarly, SIGNAL 2 is analyzed in Figs.
\ref{signal2-njets}, \ref{signal2-etjets} and
\ref{signal2-rest}, SIGNAL 3A in Figs. \ref{signal3a-njets},
\ref{signal3a-etjets} and \ref{signal3a-rest}, SIGNAL 3B in
Figs. \ref{signal3b-njets}, \ref{signal3b-etjets} and
\ref{signal3b-rest}, and finally SIGNAL 4B in Figs.
\ref{signal4b-njets}, \ref{signal4b-etjets} and
\ref{signal4b-rest}. We discuss these plots in terms of
their ability to discriminative between the MSSM and $U(1)^{\prime}$
models. In these plots, we include contributions from all
possible squark pair-production channels:
$\widetilde{q}_R\,\widetilde{q}_R$,
$\widetilde{q}_L\,\widetilde{q}_L$, and
$\widetilde{q}_L\,\widetilde{q}_R$. In addition, we include the
effects of the pair-production of the gluinos $\widetilde{g}\,
\widetilde{g}$ as well as the associated production of the gluinos
and squarks, $\widetilde{g}\,\widetilde{q}_{L,R}$.  We combine contributions from all light quarks (the
ones in the first and second generations) as jets in the final
state without distinguishing quarks and anti-quarks.

Figs. \ref{signal1-njets}--\ref{signal1-rest} depict the
number of purely hadronic events (SIGNAL 1 in
Table~\ref{table3}) as functions of the variables listed above.
Fig. \ref{signal1-njets} shows how the number of purely
hadronic events vary with the number and transverse energy
threshold of the jets. It is seen that, the low-energy jets
$E_T^{jets} > 20\ {\rm GeV}$ exhibit a broad distribution over
$N_{jets} = 2$ (from the squark pair production), $N_{jets} = 3$
(from the gluino-squark associated production), $N_{jets} = 4$
(from the gluino pair production), and $N_{jets}\geq 5$ (from
various multiple production and decay processes). As the
transverse jet energy increases, the distribution becomes less
broad. In fact, for $E_{T}^{jets} \geq 100\ {\rm GeV}$, the
events are nearly pure dijet events induced by pair-production
of squarks. The three panels, panels (a), (b) and (c), differ
mainly by the overall change in the number of events as one
switches from the MSSM to the $U(1)^{\prime}$ model. Indeed, purely
hadronic events are depleted in number in the $U(1)^{\prime}$ model
compared to the MSSM, and the depletion is strongest for
$\left({\cal R}_{\widetilde{Y}^{\prime}}\, , \, {\cal
R}_{\widetilde{Y}\widetilde{Y^{\prime}}}\right) = (10, 10)$.

Fig. \ref{signal1-etjets} is complementary to Fig.
\ref{signal1-njets}, depicting the
variation of the number of purely hadronic events (SIGNAL 1 in
Table~\ref{table3}) with the jet transverse energy for
different lower bounds on the number of jets. We see that the
events with $N_{jets}\geq 4$ are soft (they dominate only at
low $E_T^{jets}$) and rare (they rapidly decrease in number
with increasing $E_{T}^{jets}$). The main distinction between
the MSSM and the $U(1)^{\prime}$ models is the depletion of the
number of events in the latter. The panel (a) of Fig.
\ref{signal1-rest} depicts an important distribution: The
variation of the purely hadronic events with the missing
transverse energy $\slashchar{E}_T$. It is obvious that the
number of events is maximal for the MSSM and decreases gradually
in the $U(1)^{\prime}$ model as ${\cal R}_{\widetilde{Y}^{\prime}}$
and/or ${\cal R}_{\widetilde{Y}\widetilde{Y^{\prime}}}$
increase. The distribution has a sharp edge at the LSP mass,
and peaks around $500\ {\rm GeV}$ with slight shifts depending
on the details of the underlying model. The panel (b) of Fig.
\ref{signal1-rest} shows the distribution as a function of the
scalar sum of the transverse energies (missing transverse energy in
panel (a)). Again, one notices the drop in the number of
events as one switches from the MSSM to the $U(1)^{\prime}$ model.
Clearly, the $E_T^{sum}$ value at which the distribution is
maximized corresponds to the average squark/gluino masses. This
distribution, traditionally, has been utilized to  provide a
short-cut to the scale of SUSY \cite{abdullin}. It is a
sensitive variable to be searched for at the LHC.

Summarizing, the plots in Figs.
\ref{signal1-njets}--\ref{signal1-rest} show that the purely
hadronic events are more abundant in the MSSM than in the
$U(1)^{\prime}$ model. All distributions are quite similar with
fewer events for the $U(1)^{\prime}$ model case. These results
confirm the discussions in Sec. III B, and are consistent with
the fact that the branching ratios ${\cal B}(\widetilde{q}_{L,R}\to q\LSP)$ in
(\ref{branch}) are
larger in the MSSM than in the $U(1)^{\prime}$ model.

An important feature to note is that the SIGNAL 1 is the most
abundant among all the signals listed in Table~\ref{table3} and
studied in Figs. \ref{signal2-njets}--\ref{signal4b-rest}. This
purely hadronic event, with no hard muons,
can be constructed with good precision at the
LHC with optimized jet algorithms. Measurement of the
number of events for the given kinematic variables can
facilitate the decision-making about the underlying model. We emphasize that the
MSSM and the $U(1)^{\prime}$ models differ mainly by the number of
events per bin size rather than by their distribution patterns.

Depicted in Figs. \ref{signal2-njets} -- \ref{signal2-rest} are
the distributions for the single-lepton events (SIGNAL 2 in
Table~\ref{table3}). The number and patterns of the events in the
$U(1)^{\prime}$ model dominate (for  ${\cal
R}_{\widetilde{Y}^{\prime}} = {\cal
R}_{\widetilde{Y}\widetilde{Y^{\prime}}} = 0 $ ) or are
comparable (for  ${\cal R}_{\widetilde{Y}^{\prime}} = {\cal
R}_{\widetilde{Y}\widetilde{Y^{\prime}}} = 10 $) to the one in the
MSSM. This behavioral change can be ascribed to the
$\widetilde{Z}^{\prime}$ mediation, as discussed in Sec. III B.
Fig. \ref{signal2-rest}, compared to Fig.
\ref{signal1-rest}, has one added feature, namely the variation
of the numbers of events with the transverse momentum of the
emitted lepton. This plot, the panel (a) of Fig.
\ref{signal2-rest}, proves to be highly discriminative between
the MSSM and the $U(1)^{\prime}$ model as the latter offers a
much broader distribution extending to large transverse momenta
values for the lepton.

In general, for the SIGNAL 2, the $\widetilde{q}_L\,\widetilde{q}_R$
pair-production (with or without the $\widetilde{g}$ contribution)
dominates all the others. There are no events from
$\widetilde{q}_R\,\widetilde{q}_R$ since ${\cal
B}(\widetilde{q}_R\to q\LSP)\sim 10^{-6}$ in either model and
the $\widetilde{q}_L\,\widetilde{q}_L$ contribution is much
smaller than $\widetilde{q}_L\,\widetilde{q}_R$. This  is again
directly related to the fact that ${\cal B}(\widetilde{q}_L\to
q\LSP)\ll {\cal B}(\widetilde{q}_R\to q\LSP)$. Hence, the most
dominant signal proceeds through $p p \to
\widetilde{g}\widetilde{g}\to (q \widetilde{q}_L) (q^{\prime}
\widetilde{q}_R)\to q q^{\prime} (\widetilde{q}_L\to q^{\prime
\prime}\caa)(\widetilde{q}_R\to q^{\prime\prime\prime}\naa)\to
q q^{\prime} q^{\prime\prime} q^{\prime\prime\prime}
(\caa\to\ell \widetilde{\nu}_\ell) \naa
\to(\widetilde{\nu}_\ell\to \nu_l\naa)(\ell q q^{\prime}
q^{\prime\prime} q^{\prime\prime\prime} \naa)\to \ell\nu_l q
q^{\prime} q^{\prime\prime} q^{\prime\prime \prime} \naa\naa$.
This observation is confirmed by Figs. \ref{signal2-njets} and
\ref{signal2-etjets} where the event is seen to  be a 4-jet
event at high $E_{T}^{jet}$. The hardness of the lepton (the
only one for this signal) is mainly
determined by the mass difference
$m_{\caa}-m_{\widetilde{\nu}_\ell}$ which is about $50\ {\rm
GeV}$ in the MSSM but around $340\ {\rm GeV}$ in the
$U(1)^{\prime}$ model. Therefore, larger lepton $p_T$ cuts
would help distinguish the $U(1)^{\prime}$ model from the MSSM.
As mentioned before, both $E_T^{sum}$ and $\slashchar{E}_T$
distributions are dominated by the $U(1)^{\prime}$ model events (most
visibly in the $(\ry,\ryy)=(0,0)$ case).

In Figs. \ref{signal3a-njets} -- \ref{signal3a-rest} and
Figs. \ref{signal3b-njets} -- \ref{signal3b-rest}, we show the
number of events containing two charged leptons in the final state
(SIGNAL 3A and SIGNAL 3B in Table~\ref{table3}). The
distributions of these dilepton events are expected to reveal
further distinctive features of the two models. By contrasting
the distributions in Figs. \ref{signal3a-njets} and
\ref{signal3a-etjets} with those in Figs.  \ref{signal3b-njets}
and \ref{signal3b-etjets}, one finds that the SIGNAL 3A is
dominantly a 4-jet event at high $E_T^{jets}$ whereas the SIGNAL 3B
involves both 3-jet and 4-jet topologies depending on
$E_{T}^{jets}$ range. It is convenient to start the analysis
with the dilepton signal of SIGNAL 3A type. In this event, each
charged lepton originates from a different decay branch
(started by squark or gluino). The $U(1)^{\prime}$ signal again
dominates for $(\ry,\ryy)=(0,0)$ and remains comparable to
the MSSM case for $(\ry,\ryy)=(10,10)$. Unlike the SIGNAL 2
above, this  process is dominated by the
$\widetilde{q}_L\,\widetilde{q}_L$ contribution since both
squarks need to decay into a chargino. Compared to Fig.
\ref{signal2-rest}, we have one additional plot, the panel (b)
of Fig. \ref{signal3a-rest}, showing the number of events
against the dilepton invariant mass $M_{\rm inv}\left(\ell^+
\ell^-\right)$. This distribution does not reveal a sharp edge
since the leptons originate from different branches
\cite{Demir:2008wt}. As in Fig. \ref{signal2-rest}, the transverse momentum of the
hardest of the two leptons emitted
 $p_T(\ell_{hard})$  is capable of
distinguishing the two models for large lepton $p_T$ cuts.

Compared to the SIGNAL 3A, the $p_T(\ell_{hard})$ distribution
hardly changes as one switches from the MSSM to the $U(1)^{\prime}$
model, especially at large $p_T$. This feature
continues to hold for other distributions
in Fig.~\ref{signal3b-rest}, except for the dilepton invariant
mass distribution. The reason for the discriminative nature of the
$M_{\rm inv}\left(\ell^+ \ell^-\right)$ distribution is that
the two leptons originate from the same decay branch and obtain
different distribution tails for different processes. The
results are explicated in panel (b) of
Fig.~\ref{signal3b-rest}.

Looking closely,  the
$\widetilde{q}_R\,\widetilde{q}_R$ production-and-decay  is a
completely new contribution to this signal in $U(1)^{\prime}$, and the two models
would give drastically different results if other contributions
were ignored. This expectation, which follows from the discussions
in Sec. III B, is best examined by explicating the
contributions of the individual squarks/gluinos. We do this in
panel (c) of Fig. \ref{signal3b-rest} wherein the
$\widetilde{q}_L$ and $\widetilde{q}_R$ contributions are
explicated for the $\slashchar{E}_T$ distribution. The entire
signal is dominated by the $\widetilde{q}_L\,\widetilde{q}_R$
production-and-decay where
$\widetilde{q}_R$ decays to $q\naa$. Once we sum these
sub-processes, the missing energy distribution in $U(1)^{\prime}$ is either almost
the same or a little bit suppressed compared to the MSSM depending
on the $(\ry,\ryy)$ parameters.

Depicted in Figs. \ref{signal4b-njets} -- \ref{signal4b-rest}
are the distributions of the trilepton event (SIGNAL 4B in
Table~\ref{table3}). Clearly, two oppositely charged leptons
arise from one decay branch and the third one from the other
branch. As shown in the figures, the two models can be
distinguished via the number of events and their
distributions. To emphasize the trilepton nature of the event,
we plot in the panel (b) of Fig.~\ref{signal4b-rest}
the invariant mass of the two same-charge leptons which
originate from the different branches (as in SIGNAL 3A).

Examining these features in depth for the SIGNAL 4B, even though the
$\widetilde{q}_L\,\widetilde{q}_L$ contributions dominate in
both the MSSM and the $U(1)^{\prime}$ model, a new effect shows up. The MSSM distributions
receive contributions from the $\widetilde{q}_R\,\widetilde{q}_R$
production, but not the ones in the $U(1)^{\prime}$ model. The reason is
that this signal requires one squark to decay into $q\nbb$ and
the other one into $q^\prime\caa$. For $\widetilde{q}_R$ in the
MSSM, the branching fractions into $\nbb$ and $\caa$ are small
but comparable to each other, and they are the second largest
branching ratios after the $q\LSP$ mode. However, in the
$U(1)^{\prime}$, $\widetilde{q}_R$ possesses new neutral decay
modes into $q\ncc$ and $q\ndd$, the branchings of which are of
the order of $10^{-2}$. This suppresses the  $q\caa$ channel
much further. We do not see these effects in the plots since
the $\widetilde{q}_L\,\widetilde{q}_L$ decay mode dominates
over the others.

The numerical studies of the branching fractions and event
distributions convincingly prove that the MSSM
and the $U(1)^{\prime}$ model can be discriminated at the LHC
experiments. The purely hadronic events, classified as the
SIGNAL 1 in Table~\ref{table3}, turn out to be more abundant
than the leptonic ones roughly by an order of magnitude. The
analysis for confronting various
distributions in the two models has been based on basic cuts. In analyzing the experimental
data, certain signals, like the SIGNAL 3B, may require more
detailed optimization cuts beyond the basic ones to enhance the
$U(1)^{\prime}$ signal compared to  the MSSM. Nevertheless,
on general grounds, the two models behave differently in
various kinematic observables, and measurements of events with
different leptonic contents qualify to be a viable tool to
disentangle the effects of the gauge-extended models from the bulk
of data.

\section{Conclusion}

Once the LHC becomes fully functional, one of
its most important tasks would be to discover physics beyond
the Standard Model, and in particular, to look for signals of
supersymmetry, the most extensively studied scenario as such.

From previous studies it is well-known that the signature of
supersymmetry at the LHC would be fairly straightforward. One
expects large excesses of events over the ones in the standard
model with a number of characteristic signatures: for example
events with one or more isolated leptons, an excess of
trilepton events, a pattern of missing $E_T$ plus jets, and a
characteristic $l^+l^-$ invariant mass distribution.

What is not well-studied is how would one be able to
distinguish among different, realistic models of
supersymmetry. Whereas many studies of the  MSSM and mSUGRA
models exist, fewer studies are available for the extended models.
In this work, we have studied in depth the MSSM augmented by an
extra $U(1)$ gauge symmetry, the $U(1)^{\prime}$ model. This
model, devised to solve the supersymmetric $\mu$ problem, is
further justified as a ${\rm TeV}$ scale remnant of the
supersymmetric GUTs or string models. In an attempt to keep the
model as generic as possible, we have fixed some of
the model parameters  (inspired by the supersymmetric $E_6$
GUT), restricted some parameters from the available experimental bounds,
and varied the rest freely in some reasonable ranges.
In Sec. II and III, we described the $U(1)^{\prime}$ model
and the possible search strategies at hadron colliders. As an
immediate consequence of the supersymmetric setup, we
emphasized that the collider signatures of the model can be
searched for by either considering the bosonic fields or the
fermionic fields. The former
has been under both phenomenological and experimental
study, so we focused here on the effects of the fermionic fields with
regard to their potential to reveal possible gauge extensions.
As we expect that the squarks and gluinos will be abundantly
produced  at the LHC, we
look for the $U(1)^{\prime}$ effects in their decays. As
discussed in Sec. III and simulated in Sec. IV, we arrived at
novel features in the generic LHC events which reveal the effects
of gauge extension. Combined with the possible $Z^{\prime}$
discovery in Drell-Yan process, the analysis and results
of this work illustrate  other discernible effects of a
$U(1)^{\prime}$ extension.

The analysis reported here includes inherently some model and
parameter-set dependence. Nevertheless, it predicts some clear
distinguishing features of the  $U(1)^{\prime}$ model
from the MSSM. In particular, in this
model, the right-handed squarks can decay through an extra
neutral gaugino (in addition to the LSP) leading to an enhanced
signal in the events containing at least one lepton. The
difference between this model and the MSSM becomes also visible
in the invariant mass distribution of the $\ell^+\ell^-$ pair,
and in the missing $E_T$ distribution. In spite of these promising 
observables, a more general analysis involving a fine-grained
scan of a wider set of parameters (and not just the $U(1)^{\prime}$
gaugino mass and its mixing with the hypercharge gaugino,
as employed in the present work), can reveal further properties
that can be of interest at the LHC.

We summarize main findings of the simulation
studies detailed in Sec.~IV for the signals listed in Table~\ref{table3} which
have the generic form as $m\,\ell + n\, jets + \slash\!\!\!\!
\slashchar{E}_T$. The number of jets $n$ has to be at least two
but could be bigger depending on the detailed composition in the
production and/or in the cascade decays. We consider events
with up to $m=3$ leptons, and arrive at the following features
(in comparison to the MSSM):
\begin{itemize}
\item The SIGNAL 1 ({\it no-lepton event}) of
    Table~\ref{table3} consisting of purely hadronic events. As
    expected, the number of events are fewer in the $U(1)^\prime$
    model than in the MSSM. Various distributions such as
    the jet multiplicities, transverse energy of jets, missing
    transverse energy as well as the scalar sum of transverse
    energies are considered. The distributions for the two
    models are similar in topology with fewer signal
    events surviving for the $U(1)^{\prime}$ model, after
    applying the primary selection cuts. The number of
    signal events at the peak of the distributions is in
    the range of 10 to 100 but none of the distributions is
    good enough to disentangle the $U(1)^\prime$ effects
    unless some secondary selection cuts are imposed.
 \item For the SIGNAL 2 ({\it one-lepton event}) with  one
     lepton in the final state, the $U(1)^\prime$ effects
     start becoming distinguishable not only in the number of
     events but also in the event topology. In particular, the
     $p_T$ distribution of the hardest lepton, as a new
     observable in addition to the ones discussed for SIGNAL 1, turns out to be
     very useful to distinguish the $U(1)^\prime$ effects
     (mainly in the high $p_T$-tail). The distribution is
     shown in panel (a) of Fig.~\ref{signal2-rest}. Unlike
     the $U(1)^\prime$ distributions, the MSSM distribution
     dies off rapidly since the available energy for the
     lepton is around 50 GeV for the MSSM case but around
     350 GeV for the $U(1)^\prime$ case. The missing
     transverse energy and the scalar sum of the transverse
     energy distributions are also useful, and the
     $U(1)^\prime$ effects dominate over the MSSM ones for
     especially low $\ry$ and/or $\ryy$ values. The number
     of signal events at the peak of the essential
     distributions is around 10, big enough for a
     discovery.
 \item The SIGNALS 3A and 3B ({\it two-lepton events})
     involve a lepton pair where both the leptons come from
     different branches for SIGNALS 3A, and from the same
     branch for SIGNALS 3B. This is evident from the
     invariant mass distribution of the lepton pair,
     depicted in panel (b) of Figs.~\ref{signal3a-rest} and
     \ref{signal3b-rest}. While the distributions for the
     $p_T$ of the hardest lepton, the dilepton
     invariant mass,  as well as  the missing transverse
     energy and the scalar sum of the transverse energies prove
     useful to disentangle the $U(1)^\prime$ effects in the
     SIGNAL 3A case, only two of them are promising in the
     SIGNAL 3B case, as the MSSM and the $U(1)^\prime$ model lead to
     comparable contributions in the distributions of
     missing transverse energy and the scalar sum of transverse
     energies. Again, only few events at the peak of
     primary observables qualify to be signals, in each case.
  \item For the SIGNALS 4A and 4B ({\it three-lepton
      events}), there are three leptons, all coming from
      the same branch for the SIGNALS 4A.
      Thus, the SIGNAL 4A events in our parameter set requires $1\to 4$
      decays and is not considered any further. For the SIGNAL
      4B events, however, we analyze, in addition to the  others, the
      same-sign-same-flavor lepton pair invariant mass
      distribution (which is unique to the trilepton signal, in
      general). In all these distributions, the $U(1)^\prime$
      effects dominate over the MSSM but the number of signal
      events barely reaches one in some cases. This
      means that for a clear extraction of the $U(1)^{\prime}$
      effects, higher integrated luminosities (than $100 fb)^{-1}$) are needed.
\end{itemize}
One has to keep in mind that these conclusions are based on
the generator-level analysis. The next step of such an analysis
would be to have a more realistic picture of what is
experimentally feasible by implementing a full detector
analysis. This is currently being implemented in the CMSSW
analysis system of the CMS experiment \cite{work}.

If the analysis in this work, together with the close-up provided
by the simulation study in progress, has taught us anything,
it is that the search for the extra gauge interactions, in the
supersymmetric framework, must proceed through not only the
forces mediated by gauge bosons (which have been under study
both phenomenologically and experimentally
\cite{langacker-review}) but also the by the forces mediated
by the gauge fermions.  Our
analysis has been limited to the $U(1)^{\prime}$ model;
however, the discussions in Sec. III, together with the various
distributions simulated, should provide enough guidance for the 
expectations about more general models, such as the left-right
symmetric models or the $3-3-1$ models.

\section{Acknowledgments}
The work of M.F. and I.T. was supported in part by the NSERC of
Canada under the Grant No. SAP01105354. The work of D.D. was
supported in part by the Alexander von Humboldt-Stiftung through Friedrich Wilhelm Bessel-Forschungspreis,  the Turkish
Academy of Sciences via a GEBIP grant, and the Turkish Atomic
Energy Agency (TAEK) via the CERN-CMS Research Grant. We thank
Lisa Everett and Paul Langacker for reading the
manuscript and making important comments, Antonio Masiero
for highly illuminating discussions,  Orhan {\c C}ak{\i}r for
helpful communications about the {\tt Pythia} package, and  Neil
Christensen for fruitful discussions about the {\tt CalcHEP}
package. We thank Shabaan Khalil and Hiroshi Okada for
discussions about the low-scale $U(1)_{B-L}$. We also thank the
participants of the DARK 2008 Conference, the LPC Dijet Group
Meetings, the ATLAS Canada/Carleton LHC Theory Workshop, and
the TAEK Annual Meeting for comments and suggestions on
different aspects of this work.


\section*{Appendix A: The Lagrangian}
\setcounter{equation}{0}
\def\theequation{A.\arabic{equation}}
In this Appendix we provide the Lagrangian of the
$U(1)^{\prime}$ model and compare its certain features with
those of the MSSM Lagrangian. The particle spectrum of the
model with the generic $U(1)^\prime$ hypercharge assignments is
given in Table~\ref{table4}. The total Lagrangian involves
kinetic terms as well as various interaction terms among the
fields. We discuss below the distinct pieces separately.

\begin{table}
\vskip -0.4cm
\begin{center}
\begin{tabular*}{0.95\textwidth}{@{\extracolsep{\fill}} cccc}
\hline\hline
Superfields & Bosons & Fermions & $SU(3)_c\otimes SU(2)_L\otimes U(1)_Y\otimes U(1)_{Q^{\prime}}$ \\
\hline\hline
\underline{Gauge multiplets} & & & \\
$\widehat{G}^a$ & $ G^a_\mu $ & $\widetilde{g}^a$ & $\left(8,1,0,0\right)$ \\ &&& \\
$\widehat{W}^i$ & $ W^i_\mu $ & $\widetilde{W}^i$ & $\left(1,3,0,0\right)$ \\ &&& \\
$\widehat{B}$ & $B_\mu$ & $\widetilde{B}$ & $\left(1,1,0,0\right)$ \\ &&& \\
$\widehat{Z}^{\prime}$ & $Z^{\prime}_\mu$ & $\widetilde{Z}^{\prime}$ & $\left(1,1,0,0\right)$ \\
\hline
\underline{Matter multiplets} & & & \\
$\widehat{L}$ & $\widetilde{L} = \left(\begin{array}{c}\widetilde{\nu}_{\ell_L}\\ \widetilde{\ell}_L\end{array}\right)$ &
$L = \left(\begin{array}{c}\nu_{\ell_L}\\ \ell_L\end{array}\right)$ & $\left(1,2,-1,Q_L^{\prime}\right)$ \\ &&& \\
$\widehat{E}$ & $\widetilde{E} = \widetilde{\ell}^{\star}_R$ &
$\left(\ell_R\right)^C = \left(\ell^C\right)_L$ & $\left(1,1,2,Q_E^{\prime}\right)$ \\ &&& \\
$\widehat{Q}$ & $\widetilde{Q} = \left(\begin{array}{c}\widetilde{u}_L\\ \widetilde{d}_L\end{array}\right)$ &
$Q = \left(\begin{array}{c} u_L\\ d_L\end{array}\right)$ & $\left(3,2,\frac{1}{3},Q_Q^{\prime}\right)$ \\ &&& \\
$\widehat{U}$ & $\widetilde{U} = \widetilde{u}^{\star}_R$ &
$\left(u_R\right)^C = \left(u^C\right)_L$ & $\left(3,1,-\frac{4}{3},Q_U^{\prime}\right)$ \\ &&& \\
$\widehat{D}$ & $\widetilde{D} = \widetilde{d}^{\star}_R$ &
$\left(d_R\right)^C = \left(d^C\right)_L$ & $\left(3,1,\frac{2}{3},Q_D^{\prime}\right)$\\ &&& \\
$\widehat{H}_d$ & $H_d = \left(\begin{array}{c} H_d^0\\ H_d^-\end{array}\right)$ &
$\widetilde{H}_d = \left(\begin{array}{c} \widetilde{H}_d^0\\ \widetilde{H}_d^{-}\end{array}\right)$  & $\left(1,2,-1,Q_{H_d}^{\prime}\right)$ \\ &&&
\\
$\widehat{H}_u$ & $H_u = \left(\begin{array}{c} H_u^+\\ H_u^0\end{array}\right)$ &
$\widetilde{H}_u = \left(\begin{array}{c} \widetilde{H}_u^+\\ \widetilde{H}_u^{0}\end{array}\right)$  & $\left(1,2, 1,Q_{H_u}^{\prime}\right)$ \\ &&&
\\
$\widehat{S}$ & $S$ & $\widetilde{S} $ & $\left(1,1, 0, Q_{S}^{\prime}\right)$ \\
\hline\hline
\end{tabular*}
\end{center}
\caption{\it The field content of the $U(1)^{\prime}$ model
based on $G_{SM}\otimes U(1)^{\prime}$ gauge invariance. The
$U(1)^{\prime}$ charges listed here are the ones in
(\ref{Qfprime}) for which the kinetic mixing
vanishes.}\label{table4}
\end{table}

The kinetic terms of the Lagrangian are given by
\begin{eqnarray}
\label{kinetic}
{\cal{L}}^{Kinetic}_{U(1)^{\prime}} = {\cal{L}}^{Kinetic}_{MSSM} - \frac{1}{4} Z^{\prime\, \mu \nu} Z^{\prime}_{\mu \nu} +
\left({\cal{D}}_{\mu} S\right)^{\dagger} \left({\cal{D}}^{\mu}
S\right) + \widetilde{Z}^{\prime\, \dagger} i \sigma^{\mu} \partial_{\mu} \widetilde{Z}^{\prime} + \widetilde{S}^{\dagger} i \sigma^{\mu}
{\cal{D}}_{\mu} \widetilde{S}\,,
\end{eqnarray}
and the interactions of the gauge fields with the rest (fermions, sfermions, gauginos, Higgs and Higgsino fields) are contained in the piece
\begin{eqnarray}
{\cal{L}}^{gauge}_{U(1)^{\prime}} = {\cal{L}}^{gauge}_{MSSM}\left(g_Y \frac{Y_{X}}{2} B_{\mu} \rightarrow g_Y \frac{Y_{X}}{2}
B_{\mu} + g_{Y^{\prime}} Q_X^{\prime} Z^{\prime}_{\mu} \right)\,,
\end{eqnarray}
where $X$ runs over the fields charged under $U(1)_Q^{\prime}$. In (\ref{kinetic}), $Z^{\prime\, \mu \nu}$
is the field strength tensor of $Z^{\prime}_{\mu}$, and ${\cal{D}}_{\mu} S = \left(\partial_{\mu} + i g_{Y^{\prime}}
Q_{S}^{\prime} Z^{\prime}_{\mu}\right) S$.

Given the superpotential in (\ref{superpot}), part of the
$U(1)^{\prime}$ Lagrangian spanned by the $F$--terms is given
by
\begin{eqnarray}
{\cal{L}}^{F-term}_{U(1)^{\prime}} = - \sum_{i} \left| \frac{\partial W}{\partial \phi_i}\right|^2 =
{\cal{L}}^{F-term}_{MSSM}\left(\mu \rightarrow h_s S\right) -
 h_s^2 \left|H_u\cdot H_d\right|^2 \,,
\end{eqnarray}
where $\phi_i$ is the scalar component of the $i$--the chiral superfield in the superpotential.

The $D$--term contributions to the Lagrangian are given by
\begin{eqnarray}
{\cal{L}}^{D-term}_{U(1)^{\prime}} &=& - \frac{1}{2} \sum_{a} D^a D^a =
{\cal{L}}^{D-term}_{MSSM} - \frac{g_{Y^{\prime}}^2}{8} \left(Q_Q^{\prime} \widetilde{Q}^{\dagger} \widetilde{Q} + Q_U^{\prime} \widetilde{u}^{T}_R
\widetilde{u}^{\star}_R + Q_D^{\prime} \widetilde{d}^{T}_R  \widetilde{d}^{\star}_R +
\right.\nonumber\\ &+& \left.
Q_L^{\prime} \widetilde{L}^{\dagger} \widetilde{L} + Q_E^{\prime} \widetilde{E}^{T} \widetilde{E}^{\star} + Q_{H_d}^{\prime} H_d^{\dagger} H_d +
Q_{H_u}^{\prime} H_u^{\dagger} H_u +Q_S^{\prime} S^{\dagger} S \right)^2\,.
\end{eqnarray}

The soft-breaking sector of the $U(1)^{\prime}$ Lagrangian is
given by
\begin{eqnarray}
{\cal{L}}^{Soft}_{U(1)^{\prime}} &=& {\cal{L}}^{Soft}_{MSSM}\left(\mu \rightarrow 0\right) - m_{S}^2 S^{\dagger} S - \left[h_s A_s S H_u
\cdot H_d + \mbox{h.c.}\right]\nonumber\\
&+&
\frac{1}{2}\left(  M_{\widetilde{Z}^{\prime}} \widetilde{Z}^{\prime} \widetilde{Z}^{\prime} +
M_{\widetilde{Y}\, \widetilde{Z}^{\prime}} \widetilde{Y} \widetilde{Z}^{\prime} + h.c.\right)
\end{eqnarray}
where $M_{\widetilde{Y}\, \widetilde{Z}^{\prime}}$ and
$M_{\widetilde{Z}^{\prime}}$ are defined below the neutralino
mass matrix in (\ref{mneut}), and $A_s$ is the extra trilinear
soft coupling.

Finally, the part of the Lagrangian consisting of the fermion-sfermion-ino as well as
the Higgs-Higgsino-Higgsino interactions is given by
\begin{eqnarray}
{\cal{L}}^{ino-f-\phi}_{U(1)^{\prime}} &=& {\cal{L}}^{ino-f-\phi}_{MSSM}\left(\mu \rightarrow 0\right) +
i\sqrt{2} g_{Y^{\prime}} \left[ Q_Q^{\prime}  {Q}^{\dagger} \widetilde{Z}^{\prime}  \widetilde{Q} +
Q_U^{\prime} {u}_{R}^{\dagger} \widetilde{Z}^{\prime} \widetilde{u}_R +
Q_D^{\prime} {d}_R^{\dagger}  \widetilde{Z}^{\prime} \widetilde{d}_R\right.\nonumber\\  &+& \left.
Q_L^{\prime} {L}^{\dagger} \widetilde{Z}^{\prime} \widetilde{L} + Q_E^{\prime} {\ell}_R^{\dagger} \widetilde{Z}^{\prime}
\widetilde{\ell}_R + Q_{H_d}^{\prime} \widetilde{H}_d^{\dagger} \widetilde{Z}^{\prime} H_d +
Q_{H_u}^{\prime} \widetilde{H}_u^{\dagger} \widetilde{Z}^{\prime} H_u + Q_S^{\prime} \widetilde{S}^{\dagger}
\widetilde{Z}^{\prime} S  + \mbox{h.c.}\right]\nonumber\\
&+& \left[h_s S \widetilde{H}_u\cdot \widetilde{H}_d + h_s \widetilde{S} H_u\cdot \widetilde{H}_d + h_s \widetilde{S} \widetilde{H}_u\cdot {H}_d
+ \mbox{h.c.}\right]\,.
\end{eqnarray}

All parts of the $G_{SM}\otimes U(1)^{\prime}$ Lagrangian
listed above are in the current basis. Eventually, the
fields must be transformed into the physical basis wherein each
field obtains a definite mass. The neutral gauginos and
Higgsinos form the neutralino sector whose physical states are
expressed as (\ref{neutralino-def}) after diagonalizing the
mass matrix (\ref{mneut}). The chargino sector is essentially
the same as in the MSSM with the obvious replacement $\mu
\rightarrow h_s v_s /\sqrt{2}$. The Higgs sector has been
analyzed in detail at one-loop level in \cite{everett}.

The kinetic mixing in (\ref{gauge}) can be eliminated via the transformation
\begin{eqnarray}
\label{trans1} \left(\begin{array}{c} \widehat{W}_Y \\
\widehat{W}_{Y^{\prime}}
\end{array} \right) = \left( \begin{array}{cc} 1 & - \tan \chi \\
0 & 1/\cos \chi \end{array}\right) \left(\begin{array}{c}
\widehat{W}_B \\ \widehat{W}_{Z^{\prime}} \end{array} \right)\,,
\end{eqnarray}
where the kinetic eigenstates $\widehat{W}_B$ and $\widehat{W}_{Z^{\prime}}$
couple to a matter field $f$ (with
hypercharge $Y_{f}$ and the $U(1)_{Y^{\prime}}$ charge
$Y^{\prime}_{f}$) with strengths $g_Y Y_{f}$ and
$g_{{Y^{\prime}}} Q^{\prime}_{f}$, respectively.  Consequently, the  boson sector extends
the MSSM gauge boson sector by the $Z^{\prime}$ gauge boson of the
$U(1)_{Q^{\prime}}$ group, and the Higgs sector by a new singlet field.

In the gauge boson sector, spontaneous breakdown of the product group
$SU(2)_L\otimes U(1)_Y\otimes U(1)_{Q^{\prime}}$ via the Higgs VEVs
\begin{eqnarray}
\langle H_u \rangle = \frac{1}{\sqrt{2}}\left(\begin{array}{c} 0\\
v_u\end{array}\right)\,,\;\; \langle H_d \rangle = \frac{1}{\sqrt{2}}\left(\begin{array}{c} v_d\\
0\end{array}\right)\,,\;\; \langle S \rangle =
\frac{v_s}{\sqrt{2}}\,,
\end{eqnarray}
generates one massless state (photon) and a massive state ( $Z$
boson) via two orthonormal combinations of $W^{3}_{\mu}$ and
$B_{\mu}$ gauge bosons. The $W^{1}_{\mu}$ and $W^{2}_{\mu}$
linearly combine to give $W^{\pm}_{\mu}$, as the only charged
vector bosons in the model. In contrast to the MSSM,
the $Z$ boson is not a physical state by itself since it mixes
with the $Z^{\prime}$ boson. This mass mixing arises from the fact that
the Higgs doublets $H_{u,d}$ are charged under each factor of
$SU(2)_L\otimes U(1)_Y\otimes U(1)_{Q^{\prime}}$, and the
associated mass-squared matrix is given by
\cite{cvetic,langacker-review}
\begin{eqnarray}
\label{mzzp}
M^{2}_{Z-Z^{\prime}} = \left(\begin{array}{cc} M_Z^2 & \Delta^2 \\
\Delta^2 & M_{Z^{\prime}}^2\end{array}\right)\,,
\end{eqnarray}
in the $\left(Z_{\mu}, Z^{\prime}_{\mu}\right)$ basis. Its entries are
\begin{eqnarray}
M_Z^2 &=& \frac{1}{4} G_Z^2 \left(v_u^2 + v_d^2\right),\nonumber\\
M_{Z^{\prime}}^2 &=& g_{Y^{\prime}}^2 \left( Q^{\prime\ 2}_{H_u}
v_u^2 + Q^{\prime\ 2}_{H_d} v_d^2 + Q^{\prime\ 2}_{S} v_s^2
\right)\,,\nonumber\\
\Delta^2 &=& \frac{1}{2} G_Z g_{Y^{\prime}} \left(
Q^{\prime}_{H_u} v_u^2 - Q^{\prime}_{H_d} v_d^2\right)\,,
\end{eqnarray}
where $G_Z^2 = g_2^2 + g_Y^2$. The physical neutral vector bosons, $Z_{1,2}$,
are obtained by diagonalizing $M^{2}_{Z-Z^{\prime}}$:
\begin{eqnarray}
\label{mzzp-angle}
\left(\begin{array}{c} Z_1\\Z_2 \end{array}\right) =
\left(\begin{array}{cc} \cos\theta_{Z-Z^{\prime}} & \sin\theta_{Z-Z^{\prime}} \\ -\sin\theta_{Z-Z^{\prime}} &
\cos\theta_{Z-Z^{\prime}}\end{array}\right) \left(\begin{array}{c} Z \\Z^{\prime}
\end{array}\right)\,,
\end{eqnarray}
where
\begin{eqnarray}
\theta_{Z-Z^{\prime}} = - \frac{1}{2} \arctan \left( \frac{ 2
\Delta^2}{M_{Z^{\prime}}^2 - M_Z^2}\right)\,,
\end{eqnarray}
is their mass mixing angle, and
\begin{eqnarray}
M^{2}_{Z_{1(2)}}= \frac{1}{2} \left[ M_{Z^{\prime}}^2 + M_Z^2 -
(+) \sqrt{\left(M_{Z^{\prime}}^2 - M_Z^2\right)^2 + 4
\Delta^4}\right]\,,
\end{eqnarray}
are their squared masses. The collider searches at LEP and
Tevatron plus various indirect observations require
$Z$--$Z^{\prime}$ mixing angle $\theta_{Z-Z^{\prime}}$ to be at
most a few $10^{-3}$ with an unavoidable model dependence
coming from the $Z^{\prime}$ couplings
\cite{langacker-review,LEP,indirect,collid,spin,langacker-kang}.
This bound requires either $M_{Z_2}$ to be large enough (well
in the ${\rm TeV}$ range) or $\Delta^2$ to be sufficiently
suppressed by the vacuum configuration, that is,
$\tan^2\beta\equiv v_u^2/v_d^2 \sim
Q^{\prime}_{H_d}/Q^{\prime}_{H_u}$. Which of these options is
realized depends on the $U(1)^{\prime}$ charge assignments and
the soft-breaking masses in the Higgs sector ( see \cite{secluded}
for a variant reducing the $Z$--$Z^{\prime}$ mixing).

In the Higgs sector, the
$U(1)^{\prime}$ model consists of an extra CP-even Higgs boson,
$H^{\prime}$ with a mass $m_{H^{\prime}} \sim M_{Z^{\prime}}$
stemming from the extra chiral field $\widehat{S}$, the scalar
component of which is responsible for generating the $\mu$
parameter. There is no new CP--odd scalar since the imaginary parts
of $H_u^0$, $H_d^0$ and $S$ combine to give masses to the $Z$ and
$Z^{\prime}$ bosons, leaving behind a single CP--odd Higgs
boson $A^0$ as in the MSSM. Consequently, in terms of the Higgs
boson spectrum, the $U(1)^{\prime}$ model differs from the MSSM
only in having an extra CP--even Higgs boson,
$H^{\prime}$. This feature, however, is not necessarily the
most important one given that the mass spectra of the Higgs bosons
differ significantly in the two models. Indeed, the lightest
Higgs boson $h$ in the $U(1)^{\prime}$ model weighs well above
$M_Z$ already at tree level \cite{cvetic}, and thus, large
radiative corrections (and hence large top-stop mass splitting)
are not warranted to satisfy the LEP lower bound on $m_h$
\cite{everett,higgs,higgsp}. This property can prove useful in
moderating the little hierarchy problem (especially when a set
of the MSSM singlet chiral fields are included to form a secluded
sector \cite{kane}).

\section*{Appendix B: The Scalar Fermions}
\setcounter{equation}{0}
\def\theequation{B.\arabic{equation}}

Given rather tight FCNC bounds, we neglect all the
inter-generational mixings, and consider only intra-generational
left-right mixings, though these turn out to be totally
negligible for the sfermions in the first and second
generations. The $2\times 2$ scalar fermion mixing matrix can
be written as
\begin{eqnarray}
{\cal M}^{2}_{\widetilde{f}^a}=
\left(
\begin{array}{cc}
{\cal M}^2_{\widetilde{f}^a_{LL}} & {\cal M}^2_{\widetilde{f}^{a,b}_{LR}}\\
\\
{\cal M}^{2 \dagger}_{\widetilde{f}^{a,b}_{LR}} & {\cal M}^2_{\widetilde{f}^a_{RR}}
\end{array}
\right), ~~~~~~a\ne b=u,d\,,
\end{eqnarray}
where
\begin{eqnarray}
\label{sferm-mass}
{\cal M}^2_{\widetilde{f}^a_{LL}}&=&m_{\widetilde{f}_L}^2+h_{f^a}^{2} v_a^2
+\frac{1}{2}\left(g_Y^2\,\frac{Y_{f_L^a}}{2}-\ g_2^2\, T_{3L}\right)
\left(v_u^2 - v_d^2 \right)\nonumber\\
&+&
g_{Y'}^{2} Q^\prime_{f^a_L} \left(Q_{H_u} v_u^2 + Q_{H_d} v_d^2
+ Q_s v_s^2 \right)\,,\\
{\cal M}_{\widetilde{f}^{a,b}_{LR}}^{2}&=& h_{f^a} \left(A_{f^a} v_a -h_s v_s v_b\right)\,,\\
{\cal M}^2_{\widetilde{f}^a_{RR}}&=&m_{\widetilde{f}_R}^2+h_{f^a}^{2} v_a^2
+\frac{1}{2}\left(g_Y^2\, \frac{Y_{f_L^a}}{2}\right)
\left(v_u^2 - v_d^2 \right)\nonumber\\
&+&
g_{Y'}^{2} Q^\prime_{f^a_R} \left(Q_{H_u} v_u^2 + Q_{H_d} v_d^2
+ Q_s v_s^2 \right)\,.
\end{eqnarray}
Here $m_{\widetilde{f}_{L,R}^2}$ are the soft mass-squared of the
sfermions, $v_{u,d,s}$ are the VEVs of the Higgs fields,
$Y_{f^a}(T_{3L})$ is the $U(1)_{Y}$ ($SU(2)_L$) quantum number,
$Q^\prime_{f^a}$ is the $U(1)^\prime$ charge, and $A_{f^a}$ are the
trilinear couplings. The mixing matrix can be diagonalized, in
general, by a unitary matrix $ \Gamma^f$ such that
$\Gamma^{f^a\dagger}\cdot{\cal M}^2_{\widetilde{f}^a}\cdot
\Gamma^{f^a} \equiv {\rm
Diag}(M_{\widetilde{f}^a_1}^2,M_{\widetilde{f}^a_2}^2)$.\footnote{We
note that unlike mixings in other sectors,
$\Gamma^{f^a}$ is defined differently, that is,
$(\widetilde{f}^a_{L,R})_i=\Gamma^{f^a}_{ij}\widetilde{f}^a_j$,
where $\widetilde{f}^a_j$ represent the mass eigenstates.} The
rotation matrix $\Gamma^{f^a}$ can be written for quarks and
charged leptons in the $2\times 2$ $\{\widetilde{f}^a_{L},
\widetilde{f}^a_R\}$ basis as
\begin{eqnarray}
\Gamma^{f^a}=
\left(
\begin{array}{cc}
\cos\theta_{\widetilde{f}^a} & -\sin\theta_{\widetilde{f}^a}\\
\sin\theta_{\widetilde{f}^a} & \cos\theta_{\widetilde{f}^a}
\end{array}
\right),
\end{eqnarray}
where $\displaystyle
\theta_{\widetilde{f}^a}=\frac{1}{2}\arctan 2(-2{\cal
M}^2_{\widetilde{f}^a_{LR}}, {\cal
M}^2_{\widetilde{f}^a_{RR}}-{\cal M}^2_{\widetilde{f}^a_{LL}})$
and $\arctan 2(y,x)$ is defined as
\begin{eqnarray}
{\rm \arctan 2}(y,x) = \left\{\begin{array}{ll}
                      \phi\ {\rm sign}(y),           ~~~& x>0 \\
		      \frac{\pi}{2}\ {\rm sign}(y),  ~~~& x=0 \\
                      (\pi-\phi)\ {\rm sign}(y),     ~~~& x<0 	
                     \end{array}\right.
\end{eqnarray}
with $y$ being non-zero, and $\phi$ taken in the first quadrant
such that $\tan\phi=|y/x|$.

For the sfermions in the first and second generations, the
left-right mixings are exceedingly small as they are
proportional to the corresponding fermion mass. Therefore, the
sfermion mass matrix (\ref{sferm-mass}) is automatically
diagonal. However, one has to remember  that the sfermion
masses, for fixed values of $m_{\widetilde{f}_{L,R}^2}$, are
different in the MSSM and the $U(1)^{\prime}$ models due to the
additional $D$-term contribution in the latter. This is the
reason for having different squark masses in the plots of
branching ratios in Figs. \ref{figqRbranchsmall},
\ref{figqRbranchequal} and \ref{figqRbranchlarge} for the
parameter set in (\ref{mssmlist}).

\section*{Appendix C: The Fermion-Sfermion-Neutralino Couplings}
\setcounter{equation}{0}
\def\theequation{C.\arabic{equation}}
In this Appendix we list the neutralino couplings relevant for the
production and decays of the squarks and sleptons\footnote{The
couplings of the $Z_{1,2}$ bosons to the
fermions and neutralinos as well as the couplings of the neutralinos to
the fermions and sfermions are given in Sec. IV of \cite{zerwas},
which were used for cross-checking.}. The six
physical neutralinos
$$\widetilde{\chi}_j^0 = \sum_{a} N^0_{j a} \widetilde{G}_a\,,$$ couple to
the fermions and the scalar fermions.
The neutralino-quark-scalar quark couplings read as
\begin{eqnarray}
\bar{u }^k \widetilde{\chi}_j^0 \tilde{ u}^k_{\alpha} \quad &&- i\Bigg[ \sqrt{2} \Gamma^{u_k}_{\alpha 1} \left (\frac{e}{6 \cos \theta_W} N^0_{j1}+ \frac{e}{2\sin \theta_W}N^0_{j2}  +  Q_Q^{\prime}g_Y^{\prime}N^0_{j6}
\right)+Y_{u_k} N^0_{j4}\Gamma^{u_k}_{\alpha 2}\Bigg]P_R
\nonumber\\
&&+i\Bigg[ \sqrt{2}\Gamma^{u_k}_{\alpha 2} \left( \frac{2 e}{3 \cos \theta_W}N^0_{j1} - Q_Q^{\prime}g_Y^{\prime}N^0_{j6}
 \right)-Y_{u_k} N^0_{j4}\Gamma^{u_k}_{\alpha1} \Bigg] P_L\,,
\end{eqnarray}
\begin{eqnarray}
\bar{d }^k \widetilde{\chi}_j^0 \tilde{ d}^k_{\alpha}\quad &&- i\Bigg[ \sqrt{2} \Gamma^{d_k}_{\alpha 1} \left (\frac{e}{6 \cos \theta_W} N^0_{j1}- \frac{e}{2\sin \theta_W}N^0_{j2}  +  Q_Q^{\prime}g_Y^{\prime}N^0_{j6}
\right)-Y_{d_k} N^0_{j4}\Gamma^{d_k}_{\alpha 2}\Bigg]P_R
\nonumber\\
&&+i\Bigg[ \sqrt{2}\Gamma^{d_k}_{\alpha 2} \left( \frac{- e}{3 \cos \theta_W}N^0_{j1} - Q_Q^{\prime}g_Y^{\prime}N^0_{j6}
 \right)+Y_{d_k} N^0_{j4}\Gamma^{d_k}_{\alpha 1} \Bigg] P_L\,,
\end{eqnarray}
where $\alpha=1,2$ designates the squark mass-eigenstates, $k$
is the generation label, $\Gamma^{q_k}_{\alpha i}$ are
the squark mixing matrices, assumed diagonal for the first two
generations so that $\Gamma^{u_k(d_k)}_{ij}=\delta_{ij}$ for
$k=1,2$, and finally, $Y_{q_k}$ are the quark Yukawa couplings.

The neutralino-lepton-scalar lepton couplings are given by
\begin{eqnarray}
\bar{l}^k \widetilde{\chi}_j^0 \tilde{ l}^k_{\alpha} \quad && i\Bigg[ \sqrt{2} \Gamma^{l_k}_{\alpha 1}\left (\frac{e}{\cos \theta_W} N^0_{j1}+ \frac{e}{\sin \theta_W}N^0_{j2}  -  Q_L^{\prime}g_Y^{\prime}N^0_{j6}
\right)+Y_{l_k} N^0_{j4}\Gamma^{l_k}_{\alpha 2}\Bigg]P_R
\nonumber\\
&&-i\Bigg[ \sqrt{2}\Gamma^{l_k}_{\alpha 2} \left( \frac{e}{ \cos \theta_W}N^0_{j1} + Q_E^{\prime}g_Y^{\prime}N^0_{j6}
 \right)-Y_{l_k}N^0_{j4}\Gamma^{l_k}_{\alpha 1} \Bigg] P_L\,,\\
\bar{\nu }^k \widetilde{\chi}_j^0 \tilde{ \nu}^k \quad && i\Bigg[ \sqrt{2} \left (\frac{e}{\cos \theta_W} N^0_{j1}- \frac{e}{\sin \theta_W}N^0_{j2}  -  Q_L^{\prime}g_Y^{\prime}N^0_{j6}
\right)\Bigg]P_R\,,
\end{eqnarray}
where $\Gamma^{l_k}_{ij}$, the slepton mixing matrix, is diagonal
$\Gamma^{l_k}_{ij}=\delta_{ij}$ for $k=1,2$ (corresponding
to the electron and the muon).

The charginos couple to the fermions and scalar fermions in the
same manner as in the MSSM.
\sfig{htb}{decay_fendiag}{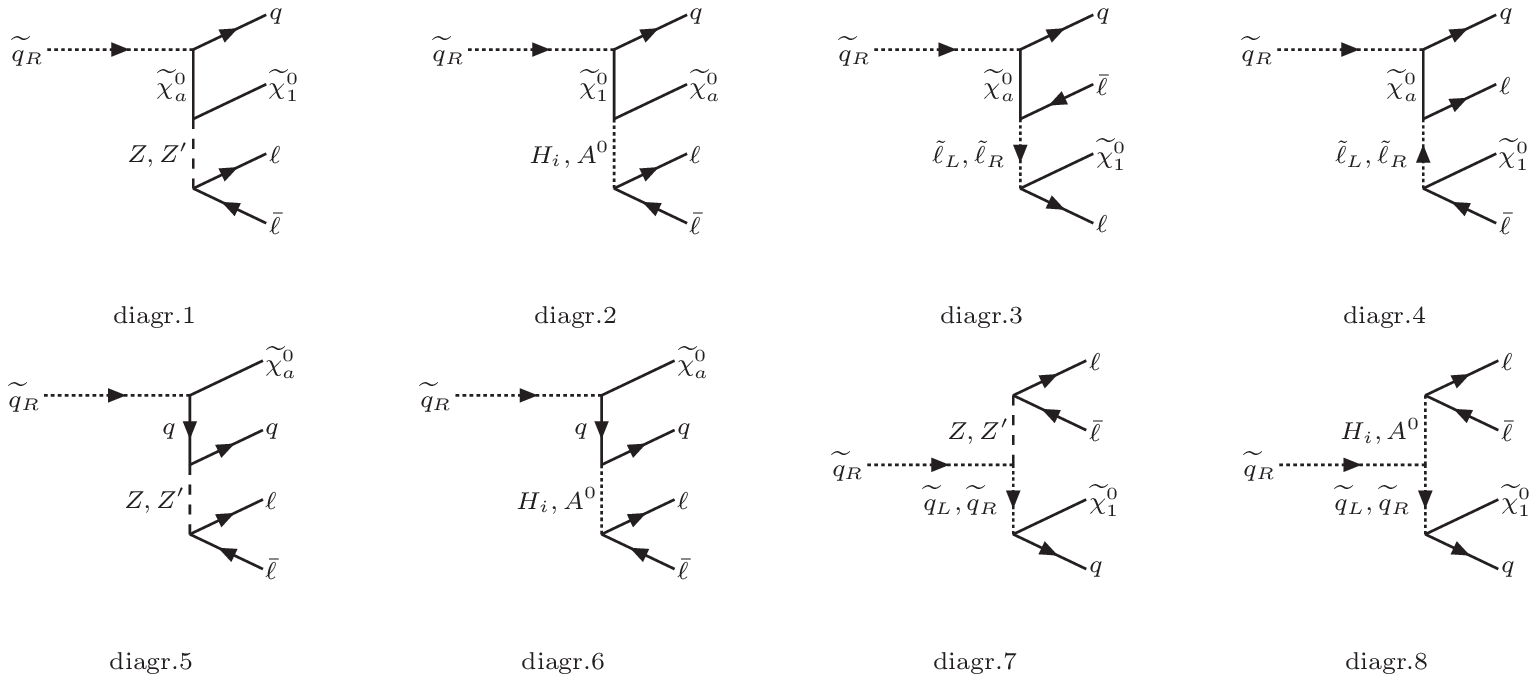}{\it The
Feynman diagrams governing the $\widetilde{q}_{R}\to
q\ell^-\ell^+\widetilde{\chi}_1^0$ decay in the $U(1)^{\prime}$
model. Here the index $a$ runs from 1 to 6 and the index $i$
from 1 to 3.}

\section*{Appendix D: An Example of  Feynman Diagrams}
\setcounter{equation}{0}
\def\theequation{D.\arabic{equation}}

In this Appendix we include, for illustration, the Feynman
diagrams contributing to the processes which have been analyzed
in the text. We have implemented the model Lagrangian and all
the information contained in the previous appendices into a
{\tt CalcHEP} code for simulation study. We illustrate the computer
code in Fig.~\ref{decay_fendiag} by picking up
$\widetilde{q}_R$ decays as an example. We note that even
though the diagrams in Fig.~\ref{decay_fendiag} are presented
as 4-body modes, we use the narrow-width approximation, and the squarks
are assumed to have a 2-body decay at first, and then, the
neutralino exhibits a 3-body decay to make up a 4-body final
state. In this respect, the diagrams 5 and 6 in
Fig.~\ref{decay_fendiag} do not contribute due to cascade
decays. For the same reason, the diagrams 7 and 8 do not
contribute either.



\begin{thebibliography}{99}

\bibitem{gut-string}
S.~Cecotti, J.~P.~Derendinger, S.~Ferrara, L.~Girardello and M.~Roncadelli,
  Phys.\ Lett.\  B {\bf 156} (1985) 318;
J.~D.~Breit, B.~A.~Ovrut and G.~C.~Segre,
  Phys.\ Lett.\  B {\bf 158} (1985) 33;
S.~M.~Barr,
  Phys.\ Rev.\ Lett.\  {\bf 55} (1985) 2778;
F.~del Aguila, G.~A.~Blair, M.~Daniel and G.~G.~Ross,
  Nucl.\ Phys.\  B {\bf 272} (1986) 413;
J.~L.~Hewett and T.~G.~Rizzo,
  Phys.\ Rept.\  {\bf 183}, 193 (1989);
M.~Cvetic and P.~Langacker,
Phys.\ Rev.\ D {\bf 54}, 3570 (1996) [arXiv:hep-ph/9511378];
G.~Cleaver, M.~Cvetic, J.~R.~Espinosa, L.~L.~Everett and
P.~Langacker,
  Phys.\ Rev.\  D {\bf 57}, 2701 (1998)
  [arXiv:hep-ph/9705391];
  Nucl.\ Phys.\  B {\bf 525}, 3 (1998)
  [arXiv:hep-th/9711178];
D.~M.~Ghilencea, L.~E.~Ibanez, N.~Irges and F.~Quevedo,
  JHEP {\bf 0208} (2002) 016
  [arXiv:hep-ph/0205083];
S.~F.~King, S.~Moretti and R.~Nevzorov,
  Phys.\ Rev.\  D {\bf 73} (2006) 035009
  [arXiv:hep-ph/0510419].


\bibitem{muprob}
J.~E.~Kim and H.~P.~Nilles,
Phys.\ Lett.\ B {\bf 138}, 150 (1984);
D.~Suematsu and Y.~Yamagishi,
Int.\ J.\ Mod.\ Phys.\ A {\bf 10}, 4521 (1995)
[arXiv:hep-ph/9411239];
M.~Cvetic and P.~Langacker,
Mod.\ Phys.\ Lett.\ A {\bf 11}, 1247 (1996)
[arXiv:hep-ph/9602424];
V.~Jain and R.~Shrock,
arXiv:hep-ph/9507238;
D.~A.~Demir,
  Phys.\ Rev.\  D {\bf 59}, 015002 (1999)
  [arXiv:hep-ph/9809358];
H.~S.~Lee, K.~T.~Matchev and T.~T.~Wang,
  Phys.\ Rev.\  D {\bf 77}, 015016 (2008)
  [arXiv:0709.0763 [hep-ph]];
D.~A.~Demir, L.~L.~Everett and P.~Langacker,
  Phys.\ Rev.\ Lett.\  {\bf 100}, 091804 (2008)
  [arXiv:0712.1341 [hep-ph]].


\bibitem{cvetic}
M.~Cvetic, D.~A.~Demir, J.~R.~Espinosa, L.~L.~Everett and
P.~Langacker,
Phys.\ Rev.\ D {\bf 56}, 2861 (1997) [Erratum-ibid.\ D {\bf 58},
119905 (1998)] [arXiv:hep-ph/9703317];
P.~Langacker and J.~Wang,
  Phys.\ Rev.\  D {\bf 58}, 115010 (1998)
  [arXiv:hep-ph/9804428].


\bibitem{kolda}
K.~S.~Babu, C.~F.~Kolda and J.~March-Russell,
Phys.\ Rev.\ D {\bf 57}, 6788 (1998) [arXiv:hep-ph/9710441];
T.~G.~Rizzo,
  Phys.\ Rev.\  D {\bf 59} (1999) 015020
  [arXiv:hep-ph/9806397].

\bibitem{zerwas}
S.~Y.~Choi, H.~E.~Haber, J.~Kalinowski and P.~M.~Zerwas,
  Nucl.\ Phys.\  B {\bf 778}, 85 (2007)
  [arXiv:hep-ph/0612218].

\bibitem{langacker-neutral}
V.~Barger, P.~Langacker and H.~S.~Lee,
  Phys.\ Lett.\  B {\bf 630} (2005) 85
  [arXiv:hep-ph/0508027];
V.~Barger, P.~Langacker and G.~Shaughnessy,
  Phys.\ Lett.\  B {\bf 644}, 361 (2007)
  [arXiv:hep-ph/0609068];
H.~S.~Lee, K.~T.~Matchev and S.~Nasri,
  Phys.\ Rev.\  D {\bf 76}, 041302 (2007)
  [arXiv:hep-ph/0702223].




\bibitem{langacker-review}
P.~Langacker,
  arXiv:0801.1345 [hep-ph].

  \bibitem{collid}
P.~Langacker, R.~W.~Robinett and J.~L.~Rosner,
  Phys.\ Rev.\  D {\bf 30} (1984) 1470;
F.~del Aguila, M.~Quiros and F.~Zwirner,
  Nucl.\ Phys.\  B {\bf 284} (1987) 530;
  Nucl.\ Phys.\  B {\bf 287} (1987) 419;
M.~Cvetic and B.~W.~Lynn,
  Phys.\ Rev.\  D {\bf 35} (1987) 51;
F.~del Aguila, M.~Cvetic and P.~Langacker,
Phys.\ Rev.\ D {\bf 48}, R969 (1993) [arXiv:hep-ph/9303299];
Phys.\ Rev.\ D {\bf 52}, 37 (1995) [arXiv:hep-ph/9501390];
F.~Del Aguila and M.~Cvetic,
Phys.\ Rev.\ D {\bf 50}, 3158 (1994) [arXiv:hep-ph/9312329];
A.~Leike,
Phys.\ Lett.\ B {\bf 402}, 374 (1997) [arXiv:hep-ph/9703263];
  Phys.\ Rept.\  {\bf 317} (1999) 143
  [arXiv:hep-ph/9805494];
T.~Appelquist, B.~A.~Dobrescu and A.~R.~Hopper,
Phys.\ Rev.\ D {\bf 68}, 035012 (2003) [arXiv:hep-ph/0212073];
M.~Dittmar, A.~S.~Nicollerat and A.~Djouadi,
  Phys.\ Lett.\  B {\bf 583} (2004) 111
  [arXiv:hep-ph/0307020];
A.~Freitas,
  Phys.\ Rev.\  D {\bf 70} (2004) 015008
  [arXiv:hep-ph/0403288];
M.~Carena, A.~Daleo, B.~A.~Dobrescu and T.~M.~P.~Tait,
  Phys.\ Rev.\ D {\bf 70}, 093009 (2004)
  [arXiv:hep-ph/0408098];
F.~Petriello and S.~Quackenbush,
  arXiv:0801.4389 [hep-ph].

\bibitem{langacker-kang}
J.~Kang and P.~Langacker,
  Phys.\ Rev.\ D {\bf 71}, 035014 (2005)
  [arXiv:hep-ph/0412190].

\bibitem{LEP}
For LEP bounds see: [LEP Collaboration],
arXiv:hep-ex/0312023;
For most recent bounds from Tevatron see:
CDF Note 9160.

\bibitem{cdf}
See the URL: http://www-cdf.fnal.gov/physics/exotic/r2a/20080306.dielectron-duke/pub25/cdfnote9160-pub.pdf

\bibitem{everett}
D.~A.~Demir and L.~L.~Everett,
Phys.\ Rev.\ D {\bf 69}, 015008 (2004) [arXiv:hep-ph/0306240].

\bibitem{higgs}
E.~Keith and E.~Ma,
  Phys.\ Rev.\  D {\bf 56} (1997) 7155
  [arXiv:hep-ph/9704441];
D.~A.~Demir and N.~K.~Pak,
Phys.\ Rev.\ D {\bf 57}, 6609 (1998) [arXiv:hep-ph/9809357];
  Phys.\ Lett.\  B {\bf 439} (1998) 309
  [arXiv:hep-ph/9809356];
H.~Amini,
New J.\ Phys.\  {\bf 5} (2003) 49 [arXiv:hep-ph/0210086];
S.~W.~Ham, E.~J.~Yoo and S.~K.~Oh,
  Phys.\ Rev.\  D {\bf 76} (2007) 015004
  [arXiv:hep-ph/0703041];
W.~Emam and S.~Khalil,
  Eur.\ Phys.\ J.\  C {\bf 522} (2007) 625
  [arXiv:0704.1395 [hep-ph]];
S.~W.~Ham, T.~Hur, P.~Ko and S.~K.~Oh,
  arXiv:0801.2361 [hep-ph].

\bibitem{higgsp}
T.~Han, P.~Langacker and B.~McElrath,
  Phys.\ Rev.\  D {\bf 70} (2004) 115006
  [arXiv:hep-ph/0405244];
V.~Barger, P.~Langacker, H.~S.~Lee and G.~Shaughnessy,
  Phys.\ Rev.\  D {\bf 73} (2006) 115010
  [arXiv:hep-ph/0603247];
V.~Barger, P.~Langacker and G.~Shaughnessy,
  Phys.\ Rev.\  D {\bf 75} (2007) 055013
  [arXiv:hep-ph/0611239];
V.~Barger, P.~Langacker, M.~McCaskey, M.~J.~Ramsey-Musolf and G.~Shaughnessy,
  Phys.\ Rev.\  D {\bf 77} (2008) 035005
  [arXiv:0706.4311 [hep-ph]].

\bibitem{pair-produce}
G.~J.~Gounaris, J.~Layssac, P.~I.~Porfyriadis and F.~M.~Renard,
  Phys.\ Rev.\  D {\bf 70} (2004) 033011
  [arXiv:hep-ph/0404162];
T.~Wohrmann and H.~Fraas,
  Phys.\ Rev.\  D {\bf 52} (1995) 78
  [arXiv:hep-ph/9501377];
H.~Baer, C.~h.~Chen, F.~Paige and X.~Tata,
  Phys.\ Rev.\  D {\bf 50} (1994) 4508
  [arXiv:hep-ph/9404212].


\bibitem{abdullin}
S.~Abdullin {\it et al.}  [CMS Collaboration],
  J.\ Phys.\ G {\bf 28}, 469 (2002)
  [arXiv:hep-ph/9806366];
A.~J.~Barr,
  ``Studies of supersymmetry models for the ATLAS experiment at the Large
  Hadron Collider,'' CERN-THESIS-2004-002, Nov 2002.

\bibitem{produc}
S.~Dawson, E.~Eichten and C.~Quigg,
  Phys.\ Rev.\  D {\bf 31} (1985) 1581;
W.~Beenakker, R.~Hopker, M.~Spira and P.~M.~Zerwas,
  Nucl.\ Phys.\  B {\bf 492} (1997) 51
  [arXiv:hep-ph/9610490];
A.~T.~Alan, K.~Cankocak and D.~A.~Demir,
  Phys.\ Rev.\  D {\bf 75} (2007) 095002
  [Erratum-ibid.\  D {\bf 76} (2007) 119903]
  [arXiv:hep-ph/0702289];
S.~Bornhauser, M.~Drees, H.~K.~Dreiner and J.~S.~Kim,
  Phys.\ Rev.\  D {\bf 76} (2007) 095020
  [arXiv:0709.2544 [hep-ph]].

\bibitem{reconstruct}
F.~E.~Paige,
  arXiv:hep-ph/9801254;
D.~R.~Tovey,
  Phys.\ Lett.\  B {\bf 498} (2001) 1
  [arXiv:hep-ph/0006276];
M.~Tytgat,
  arXiv:0710.1013 [hep-ex].

\bibitem{kane}
D.~A.~Demir, G.~L.~Kane and T.~T.~Wang,
  Phys.\ Rev.\  D {\bf 72} (2005) 015012
  [arXiv:hep-ph/0503290].

\bibitem{kolda1}
K.~S.~Babu, C.~F.~Kolda and J.~March-Russell,
  Phys.\ Rev.\  D {\bf 54} (1996) 4635
  [arXiv:hep-ph/9603212].

\bibitem{calchep}
See the URL: http://theory.sinp.msu.ru/~pukhov/calchep.html;
  A.~Pukhov,
  arXiv:hep-ph/0412191.

\bibitem{Semenov:2008jy}
  A.~Semenov,
  arXiv:0805.0555 [hep-ph];
  A.~Semenov,
  Comput.\ Phys.\ Commun.\  {\bf 115}, 124 (1998).

\bibitem{pythia}
T.~Sjostrand, S.~Mrenna and P.~Skands,
  JHEP {\bf 0605} (2006) 026
  [arXiv:hep-ph/0603175].

\bibitem{Demir:2008wt}
  D.~A.~Demir, M.~Frank, K.~Huitu, S.~K.~Rai and I.~Turan,
  Phys.\ Rev.\  D {\bf 78} (2008) 035013.

\bibitem{work} K. Cankocak, D. Demir, M. Frank and I. Turan,
    {\it Dijet Studies in SUSY with Extra U(1)}, LPC Dijet
    Group in the CMS Collaboration.

\bibitem{indirect}
U.~Amaldi {\it et al.},
  Phys.\ Rev.\  D {\bf 36} (1987) 1385;
P.~Langacker, M.~x.~Luo and A.~K.~Mann,
Rev.\ Mod.\ Phys.\  {\bf 64}, 87 (1992);
J.~Erler and P.~Langacker,
Phys.\ Rev.\ Lett.\  {\bf 84}, 212 (2000) [arXiv:hep-ph/9910315];
arXiv:hep-ph/0407097;
  Phys.\ Lett.\  B {\bf 456} (1999) 68
  [arXiv:hep-ph/9903476].

\bibitem{spin}
A.~Fiandrino and P.~Taxil,
Phys.\ Rev.\ D {\bf 44}, 3490 (1991);
P.~Taxil, E.~Tugcu and J.~M.~Virey,
Eur.\ Phys.\ J.\ C {\bf 24}, 149 (2002) [arXiv:hep-ph/0111242].

\bibitem{secluded}
J.~Erler, P.~Langacker and T.~j.~Li,
  Phys.\ Rev.\  D {\bf 66} (2002) 015002
  [arXiv:hep-ph/0205001].

\end{thebibliography}
\end{document}